\documentclass[amstex,epsf,amssymb,usenatbib]{mnras}
\usepackage{epsf,graphics,subfig}
\usepackage{amsmath,amssymb,natbib}
\usepackage{lmodern}
\usepackage{cite}

\usepackage{rotating,times,graphicx,latexsym,color,longtable,lscape} 

\usepackage{nicefrac}

\def\aj{AJ}                   
\def\apj{ApJ}                 
\def\apjl{ApJ}                
\def\apjs{ApJS}               
\def\aap{A\&A}                
\def\aaps{A\&AS}              
\def\mnras{MNRAS}             
\def\nat{Nature}              

\makeatletter

\newcommand{\Rmnum}[1]{\expandafter\@slowromancap\romannumeral #1@}
\makeatother

\title[Simulations \& signatures of protoplanetary nebulae]{Numerical simulations of wind-driven protoplanetary nebulae. \Rmnum{2}. signatures of atomic emission}
    
\author[I. Novikov \& M.D. Smith]
{Igor Novikov$^{1}$\thanks{E-mail: i-novikov@hotmail.com}
\& Michael Smith$^{1}$\thanks{E-mail: M.D.Smith@kent.ac.uk}\\
$^{1}$Centre for Astrophysics \& Planetary Science, The University of Kent, Canterbury, Kent CT2 7NH, U.K. }                                                                                                                                                             
\date{Accepted .....
      Received ..... ;
      in original form .....}

\pagerange{\pageref{firstpage}--\pageref{lastpage}}
\pubyear{2019}

\begin{document}
                                                                                                                                                             
\maketitle
                                                                                                                                                             
\label{firstpage}
                                                                                                                                                             
\begin{abstract}
We follow up on our systematic study of axisymmetric hydrodynamic simulations of protoplanetary nebula. The aim of this work is to generate the atomic analogues of the $\mathrm H_2$ near-infrared models of Paper \Rmnum{1} with the ZEUS code modified to include molecular and atomic cooling routines.
We investigate stages associated with strong $\mathrm {[Fe\,{\sc II}] \, 1.64\,\mathrm{\mu m} }$ and $\mathrm{ [S\,{\sc II}] \,6716}$ {\AA} forbidden lines, the $\mathrm{[O\,{\sc I}]\,6300}$ {\AA} airglow line, and H$\mathrm \alpha\,6563$ {\AA} emission. We simulate ($\mathrm{80\sim200\,km\,s^{-1}}$) dense ($\mathrm{\sim10^{5}\,cm^{-3}}$) outflows expanding into a stationary ambient medium.  
In the case of an atomic wind interacting with an atomic medium, a decelerating advancing turbulent shell thickens with time. This contrasts with all other cases where a shell fragments into a multitude of cometary-shaped protrusions with weak oblique shocks as the main source of gas excitation. 
We find that the atomic wind-ambient simulation leads to considerably higher excitation, stronger peak and integrated atomic emission as the nebula expands. 
The weaker emission when one component is molecular is due to the shell fragmentation into fingers so that the shock surface area is increased and oblique shocks are prevalent. 
Position-velocity diagrams indicate that the atomic-wind model may be most easy to distinguish with more emission at higher radial velocities. With post-AGB winds and shells often highly obscured and the multitude of configurations that are observed, this study suggests and motivates selection criteria for new surveys.  
\noindent 
\end{abstract} 
  
\section{Introduction} 
\label{intro}

Protoplanetary nebulae (PPN) occur  in a passage of time between the asymptotic giant branch (AGB) phase and the  planetary nebula (PN) phase. The  transformation lasts a few thousand years at most  and the PPN is largely obscured until a combination of expansion and dust expulsion brings it into view.
In this time, a usually quite spherical shape of the post-AGB envelope may distort into a dynamic display of winds, bullets and jets, creating rings, ellipses, butterflies and funnels, as summarised by  \citet{2018ApJ...865L..23F}. High-resolution images investigated by \citet{1995A&A...293..871C} show $\sim 64\%$ of outflows arrive at the PN stage as elliptical. However the shape statistics is still one of the most debated topics which is confused by several factors including projection effects. 
 
The envelope ejection and shock interaction with the ambient medium can be studied with wider implications concerning how understand molecules and dust mix with and replenish the interstellar medium. Simultaneously, we can search for clues as to the nature of the rapidly evolving stars or stellar systems \citep{2017ApJ...846...96H} which may end up driving such immaculate nebulae.

In Paper \Rmnum{1} \citep{2018MNRAS.480...75N} we began a systematic study of this transition phase via numerical simulations of molecular and atomic interactions. These are performed with molecular cooling and chemistry which turns out to be crucial to the resulting shell structure. In general, PPN are best observed in the near-infrared due to the domination of emission from radiative shocks, the dust obscuration and the absence of strong
photoionisation since the star has not turned on the ultraviolet switch \citep{2005MNRAS.360..104D}.
 We therefore focussed on the near-infrared molecular hydrogen lines in Paper \Rmnum{1}.

Our first purpose here is to explore the atomic emission line properties in the optical and near-infrared to search for potential diagnostics. We will then also investigate  the context of these chemistry/cooling simulations with previous PN simulations and so generate a complete picture of post-AGB dynamics. However, crucial limitations still remain and need to be relaxed in future research.
For example, the simulations are at this stage taken to be cylindrically symmetric, a necessity for us to achieve the resolution of the critical cooling zone in the interaction region. 


\begin{figure*}
   \subfloat[\label{genworkflow}][M2: $\rm{H_2}$ wind interacting with an $\rm{H}$ ambient medium..]{%
      \includegraphics[width=0.3\textwidth]{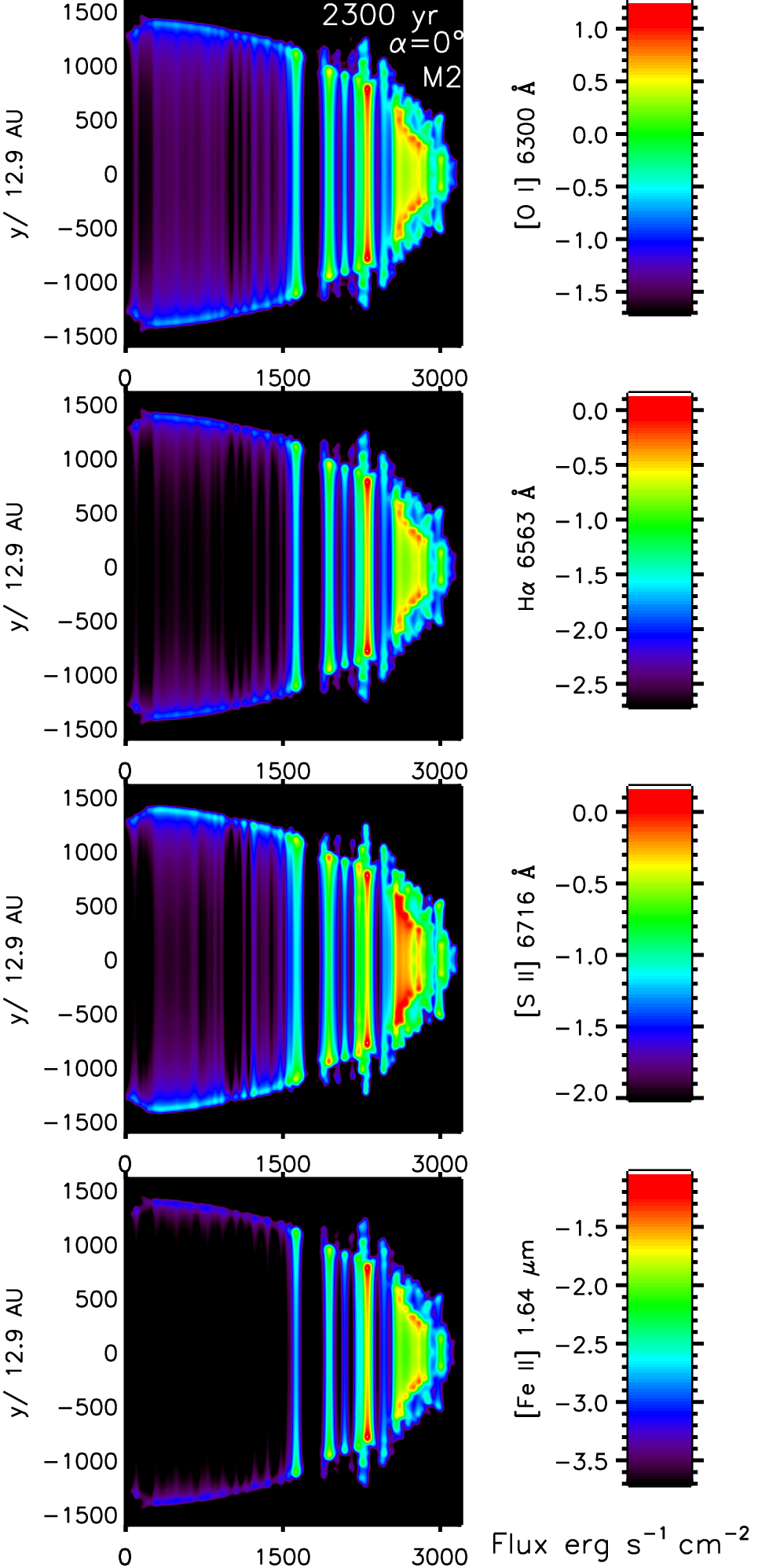}}
\hspace{\fill}
   \subfloat[\label{pyramidprocess}][M3: $\rm{H}$ wind interacting with an $\rm{H_2}$ ambient medium..]{%
      \includegraphics[width=0.3\textwidth]{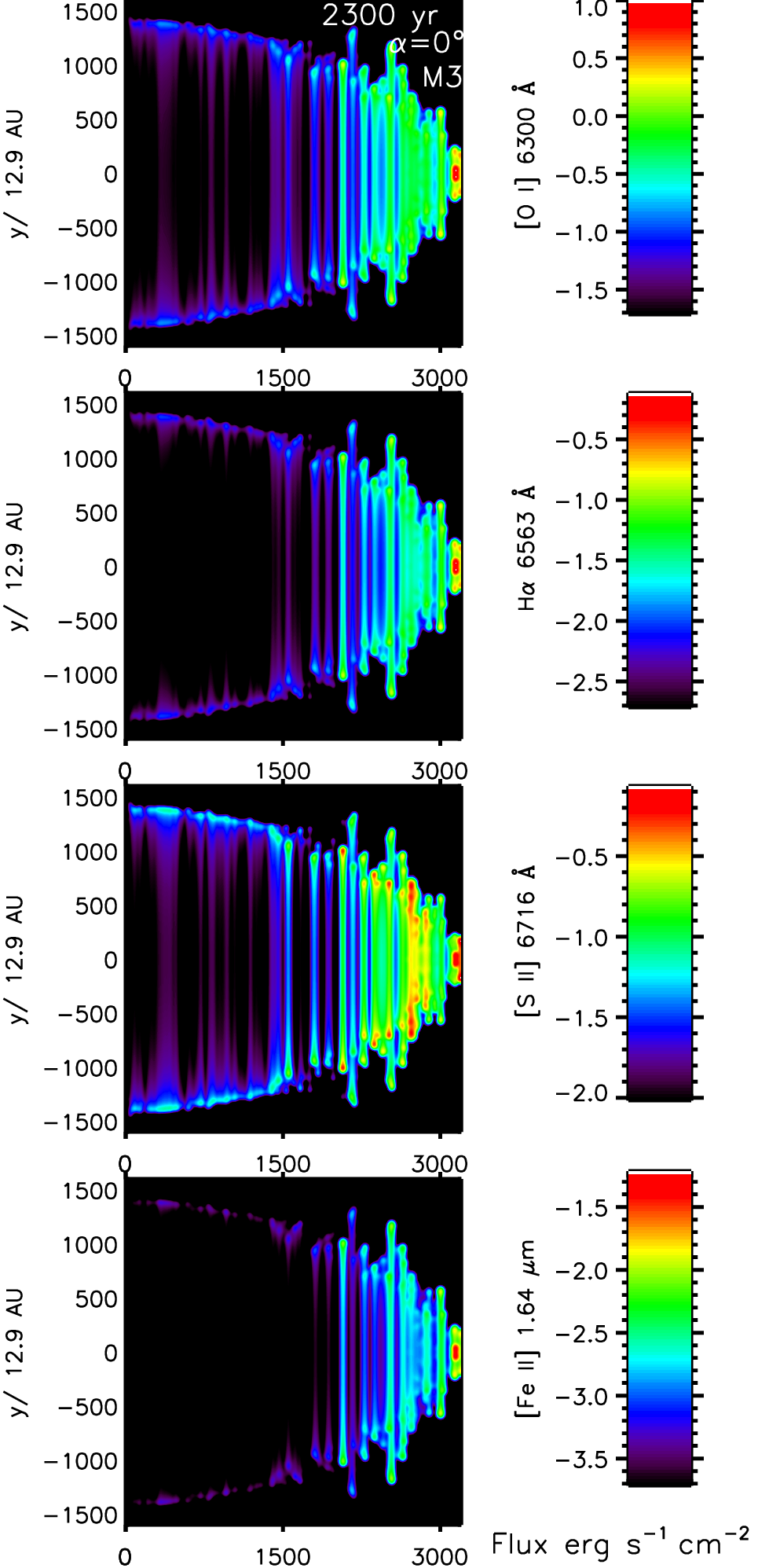}}
\hspace{\fill}
   \subfloat[\label{mt-simtask}][M4: $\rm{H}$ wind interacting with an $\rm{H}$ ambient medium.]{%
      \includegraphics[width=0.3\textwidth]{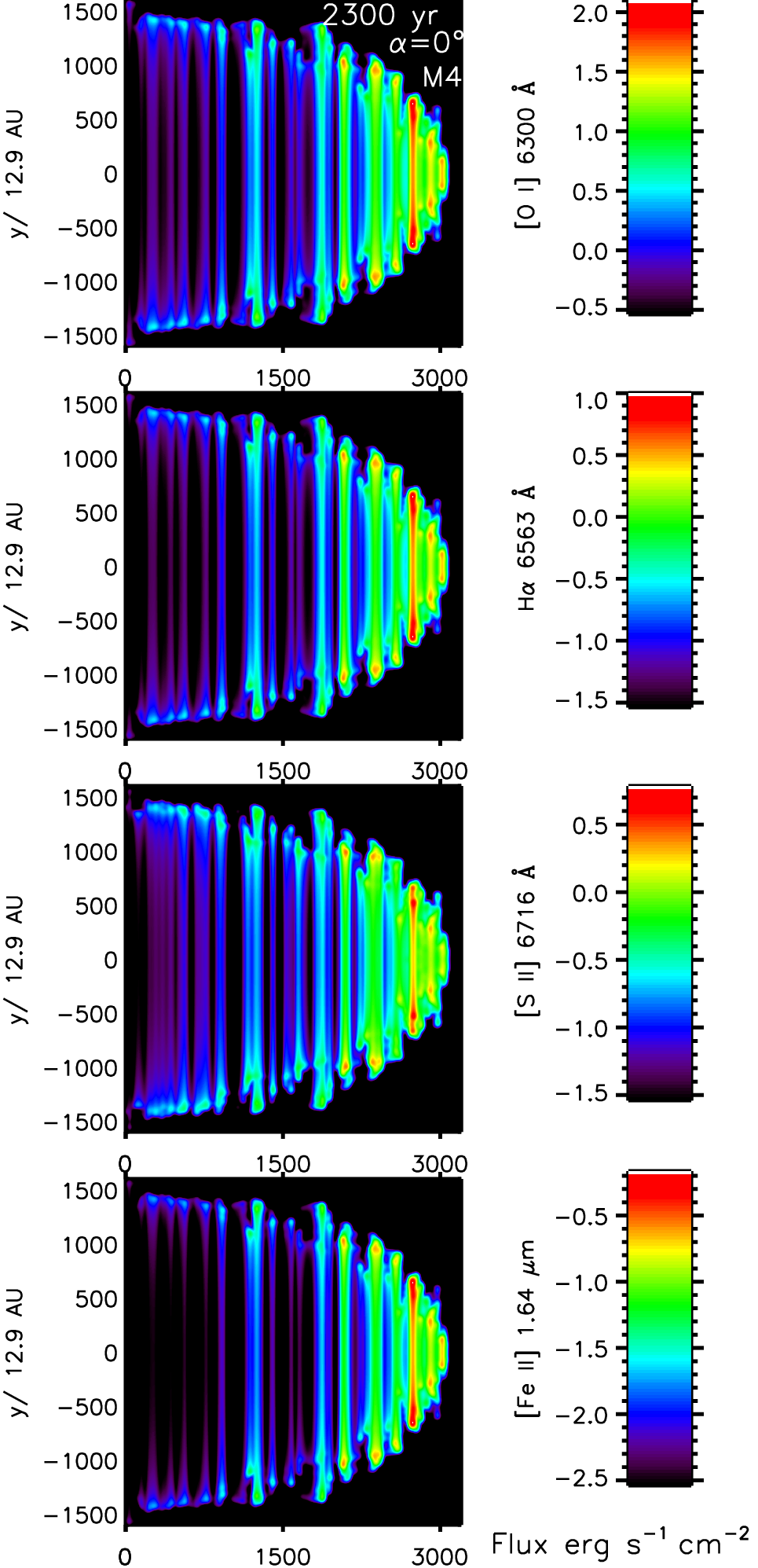}}\\
\caption{\label{2.1atEm_All}Simulated atomic lines from the 2:1 ellipsoidal wind at a late stage of expansion with axial speed of 140~km~s$^{-1}$ in the plane of the sky. The three composition models are as indicated showing shock-produced shells from high energy $\mathrm{[O\,{\sc I}]\,6300}$ {\AA}, H$\mathrm \alpha\,6563$ {\AA} and lower energy $\mathrm{ [S\,{\sc II}] \,6716}$ {\AA}, $\mathrm {[Fe\,{\sc II}] \, 1.64\,\mathrm{\mu m} }$ emission lines.}
\end{figure*}

In Paper \Rmnum{1} we followed the fraction of molecules dissociated and the fraction excited collisionally.
We identified a variety of near-infrared emission stages generated from fully molecular and atomic media. The molecular line ratio maps were used as diagnostics to differentiate between wind and envelope environments. Position-velocity diagrams completed the picture with line  widths, geometry and orientation discussed. This second work applies atomic tracers within the same programme of simulations.         
The aim of this second work is to computationally investigate whether the high velocity post-AGB winds are also capable of producing $\mathrm {[Fe\,{\sc II}] }$ and metastable $\mathrm {[O\,{\sc I}] }$ emission lines within the same collisionally-excited scenario.


\section{Method} 
\label{Method}

\subsection{The code}
The Eulerian astrophysical gas dynamics code ZEUS \citep{1992ApJS...80..753S} is extended to include molecular and atomic cooling functions. Modelling electron transitions in atoms and molecules is computationally expensive. Therefore, atomic cooling is represented macroscopically by simple functions of temperature, ion and free electron abundance.  
The modelled axisymmetric outflows were calculated on a 2--D grid, and expressed in terms of cylindrical geometry. As a result, 2--D grid values are translated onto a cylindrical grid creating a 'pseudo-3--D' projection with the line of sight as a user variable.   

\subsection{Scaling}
This work models the same hydrodynamical set up and scaling, as described in Paper \Rmnum{1} with a wind profile expanding into a stationary uniform ambient medium. We model an optically thin ambient medium with no absorption. The purely molecular wind and ambient simulation will be ignored here since little atomic emission was generated. The resultant number of models drops from the previous 36 to 27 (three speeds, three wind ellipticities and the three atomic/molecular, molecular/atomic and atomic/atomic combinations).
 
We introduce a conserved quantity of hydrogen nucleon density in the wind as $\mathrm{n_{w}} = \mathrm{\rho_{w}}/\mathrm{1.4 m_p}$ (with 10\% helium abundance $\mathrm{n(He)=0. 1n_{w}}$), set on a grid with relatively large injection radius of $\mathrm{R_{w} = 0.01 \, pc}$ chosen to resolve the input sphere surface corresponding to the inner radius of the circumstellar envelope (CSE) \citep{doi:10.1111/j.1365-2966.2004.08413.x}. Grid dimensions and scales are kept the same as in Paper \Rmnum{1} with a mass outflow rate set to $10^{-4}$~M$_\odot$~yr$^{-1}$, and employed to determine the initial set of macroscopic parameters listed in Table~\ref{partable}. The relation between the number of hydrogen atoms, molecules and total number of nucleons is
\begin{equation}
n(H) = n_w - 2 \times n(H_{2})
\end{equation}

The outflows vary in wind ellipticity through anisotropic velocity profile with planeaxis aspect ratios of 1:1 spherical, 2:1 and 4:1 with axial speed ranging from $v_w \mathrm{\sim 80 - 200\, km\, s^{-1}}$ winds. The computational grid is configured to match the ellipticity ratio, with main simulation set to 2:1 and 4:1, splitting the fixed grid into $3,200\times1,600$ and $6,400\times1,600$ zones respectively with grid zone size of $\mathrm{\sim 10^{14}\,cm}$. Atomic shocks get heated to $\mathrm{10^{5}\, K}$, at shock velocities of $\mathrm{\sim 30 - 40\, km\,s^{-1}}$, after which atomic cooling is dominant in high density thick shock front of $\mathrm{\sim 10^{6}\,cm^{-3}}$ and cooling rates of $\mathrm{\Lambda\sim10^{-13}\,erg\,s^{-1}\,cm^{-3}}$ \citep{2003MNRAS.339..133S}. The above metric constitutes to a characteristic cooling time of $\mathrm{t_{col}=nkT/\Lambda\sim10^{8}\,s}$ and cooling length of $\mathrm{\sim 10^{14}\,cm}$, same as the grid zone size and proves suitable for resolving emission in the shocked gas region. Within the post-shock zone, the cooling rate is proportional to wind density squared, resulting in $\mathrm{r_{col}\sim10^{-15}\,erg\,s^{-1}\,cm^{-3}}$, higher cooling time by a factor of 10 and a subsequent cooling length of $\mathrm{\sim 10^{15}\,cm}$. Therefore, the chosen computational domain proves to be suitable for entire shock resolution. The axial $\mathrm{v_z =  (z/R_{w}) v_{w}}$ and plane $\mathrm{v_r =  \epsilon  (r/R_{w}) v_{w}}$ speed components at $\mathrm{(z,r)}$ of inner spherical injection surface also matches the ellipticity ratio with $\mathrm{\epsilon}$ parameter equal to $1$, $0.5$ or $0.25$. Thus the speed on the surface is fixed by $\mathrm{v_{z} =  v_{w} cos\theta}$ and $\mathrm{v_{r} = \epsilon v_{w} sin\theta}$ where $\mathrm{\theta}$ is the polar angle. 

Elemental abundances are constant and, for simplicity, we take as typical of  interstellar material although strong depletions are found as elements such as iron are tied up in grains \citep{ 2015MNRAS.452.4070O}. The estimated emissivities can be scaled. The ZEUS code does not separately track the abundance of ions. Instead, it is estimated as an abundance ratio with respect to H, tabulated in Table~\ref{partable}. A free electron abundance is set to $0.1$ for simplicity. 
  
\begin{table*}
\centering
\caption[Atomic \& Molecular wind parameters]
 {Initial wind/ambient parameters. Top \& middle: Macroscopic wind/ambient parameters for 2:1, 140~km\,s$^{-1}$, $\mathrm{10^{-4} M_{\odot}yr^{-1}}$ outflows at injection radius. Bottom: Ion \& Free electron fraction by \citep{1989ApJ...345..811R}$\mathrm{^a}$,\citep{1994A&A...289..256S}$\mathrm{^b}$,\citep{1985A&AS...60..425A}$\mathrm{^c}$.  \label{partable} }
 \begin{tabular}{ | l  c  r|}
 \hline
 Property   &     Atomic & Molecular \\
\hline
Wind: \\
wind density, $\mathrm{\rm{\rho_w (gm/cm^3)}}$    & $3.76\times 10^{-20}$ & $3.76\times 10^{-20}$ \\
wind internal energy, $\mathrm{\rm{e_{w} (erg/cm^{3})}}$ & $1.83 \times 10^{-09}$ & $1.55 \times 10^{-09}$ \\
molecular fraction, $\mathrm{n(H_2)/n}$                      & $0$ & $0.5$ \\
specific heat ratio, $\mathrm{\gamma}$        & $1.66667$ & $1.42857$ \\
temperature, $\mathrm{T_{w}(K)}$      & $500$ & $500$ \\
\hline
Ambient medium: \\
 density, $\mathrm{\rho_{a} (gm/cm^3)}$          & $1.51\times 10^{-21}$  & $1.51\times 10^{-21}$ \\ 
 int. energy, $\mathrm{e_{a} (erg/cm^{3})}$ & $2.92 \times 10^{-12}$ & $2.48 \times 10^{-12}$ \\ 
 temperature, $\mathrm{T_{a}(K)}$       & $20$ & $20$ \\
 \hline 
Element &  $\mathrm{\epsilon_{ion}=n_{ion}/n_H}$ &  $\mathrm{\chi_{e}=n_e/n_H}$  \\
$\mathrm {[O\,{\sc I}]^a}$ &  $3.5(-4)$   & $0.1$  \\  
$\mathrm {[S\,{\sc II}]^c}$ &$\mathrm{\epsilon_{[S\,\sc II]}(T)}$  & $0.1$  \\  
$\mathrm {[Fe\,{\sc II}]^b}$ & $1.4(-6)$ &   $0.1$ \\
\hline
\end{tabular}

\end{table*}

Assuming stationary ambient medium, the velocity of the forward shock in terms of density ratio $\mathrm{\eta=\rho_{w(r)}/\rho_{a}}$ and wind velocity, is worked out through ram-pressure balance. The input wind follows an inverse square density profile 
\begin{equation}
\rho_{w(r)}= \rho_{w} \left( \frac{R_{w}} {r}\right) ^{2} \,R_{ppn}\geq\,r\,\geq\, R_w \,\,\,\,\,\,\,\,\, \mathrm{\bigl(cm^{-3}\bigr).}
\end{equation}
With stationary and uniform ambient medium, we have thus chosen
\begin{equation}
n_{a}=4 \times n_{w} \left( \frac{R_{w}} {R_{ppn}}\right) ^{2} \,R_{ppn}=10R_{w} \,\,\,\,\,\,\,\,\,\,\,\,\,\,\,\,\,\,\,\,\, \mathrm{\bigl(cm^{-3}\bigr).}
\label{eq:1}
\end{equation}
The chosen ambient density value creates favourable conditions for propagating forward and backwards shocks to vary in strength, as the wind expands over the defined domain. This set-up enables the reverse shock to become dominant after passing through a set distance of $\mathrm{r= 5 R_{w}}$ at the which the wind/ambient density is equated for all ellipticity cases. Factor 4 in Eq.~(\ref{eq:1}) is chosen to slow down the advancing working surface to $\mathrm{\sim \nicefrac{1}{3}\,{v_{w}}}$ for maximum reverse shock generation at $\mathrm{r=R_{ppn}}$ and $\mathrm{\eta_{w(r)}=0.25}$ in case of a 1:1 spherical outflow. Therefore, the introduced winds with particle densities of above $\mathrm{\sim 10^{4}\,cm^{-3}}$ are highly overdense with respect to the ambient medium with $\mathrm{\sim 10^{2}\,cm^{-3}}$, resulting in an initial mass density ratio of $\mathrm{\eta_o=\nicefrac{\rho_{w}}{\rho_{a}}\sim25}$ at injection radius and an initial ballistic expansion with advancing shock of $\mathrm{\sim \nicefrac{5}{6}\,{v_{w}}}$ for all ellipticity cases. For non-spherical higher ellipticity outflows, the wind's working surface expierences a longer simulation time $\mathrm{t_{lim}}$ due to $\mathrm{R_{ppn}= 20 R_{w}}$ for 2:1 and $\mathrm{R_{ppn}= 40 R_{w}}$ for 4:1 winds, setting the final working surface velocity to $\mathrm{\sim \nicefrac{1}{5}\,{v_{w}}}$ and $\mathrm{\sim \nicefrac{1}{9}\,{v_{w}}}$ respectively.    
\begin{equation}
U=\frac{\sqrt{\eta}}{1+\sqrt{\eta}}V_w \,\,\,\,\,\,\,\,\,\,\,\,\,\,\, \mathrm{\bigl(cm\,s^{-1}\bigr).}
\end{equation}
The temperature of the ambient medium is set to cold $\mathrm{20 \, K}$, and the input temperature of the wind is $\mathrm{500 \, K}$, resulting in ram-driven wind being over-pressured with respect to the ambient medium by a factor $\mathrm{\kappa_{0}\sim10^{3}}$. We do not hydrodynamicaly involve interaction of He O, S and Fe atoms, and take 10\% helium abundance for determining specific heat ratio $\mathrm{\gamma=\frac{5}{3}}$ and $\mathrm{\gamma=\frac{10}{7}}$ of atomic and molecular media respectively. The relative abundance of other atomic species shown in Table~\ref{partable} is used in emission routines to determine the line flux, without having the atoms physically present on the computational grid.
  
\subsection{Magnetic Fields}
Magnetic fields and their dynamical role are not considered in this research although PPNs investigated in CO by \citet{2001A&A...377..868B} show objects with outflow momentum excess of $\sim\,10^{3}$ between the momentum observed and values that can be attributed to stellar radiation pressure, providing indirect support for a driving mechanism different from the radiatively driven winds described by the generalized interacting Stellar Winds (GISW) model. Millimetre CO lines observations by from a group of binary post-AGB stars also reveal excess in near-infrared emission emanating from hot dust in compact possibly rotating disks  \citet{2013A&A...557A.104B}. The presence of disks and jet-like flows is also similar to the environments existent in young stellar objects (YSOs) with jets driving molecular outflows. The launching mechanism for YSO outflows are extensively studied and is believed to be magnetically driven \citep{2009ApJ...707.1485D} (and references therein). This leads to speculation that PN and PPN outflows with similar environments to YSOs are launched with magnetohydrodynamical (MHD) processes. Additionally point-symmetric flows produced by precession also suggest an MHD driver for PPN outflows \citep{2008ApJ...679.1327D} since magnetically driven models couple rotation to a magnetic field, therefore enabling jets to be bound to the rotational axis of the central star. 

Here, we assume that the wind expansion has reduced the launched magnetic flux, pressure and tension to low dynamic levels on the scales we simulate. However, there may well be an influence on the shock behaviour including cushioning through ambipolar diffusion, associated with the shell compression.

 
\begin{figure}
\includegraphics[width=0.5\textwidth,height=0.3\textheight,keepaspectratio]{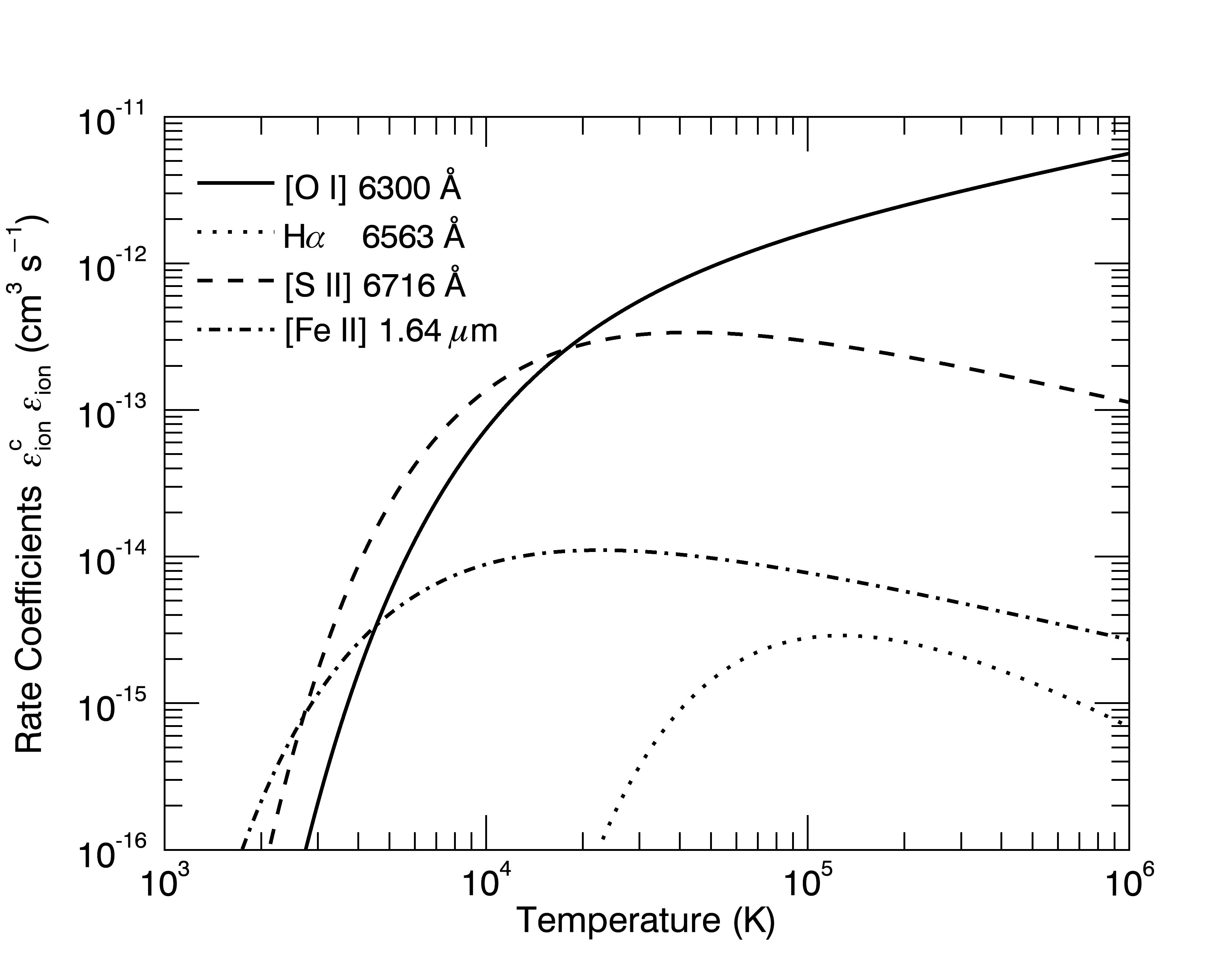}
\caption[Emission rates]
{Rate coefficients $\mathrm{\epsilon_{ion}^c(T)}$ as a function of temperature for transitions by recombination and spontaneous de-excitation multiplied by relative abundancies $\mathrm{\epsilon_{ion}}$ for: $\mathrm {[O\,{\sc I}]}$ \citep{1989ApJ...345..811R}, H$\mathrm{\alpha}$ \citep{2011ApJ...727...35D}, $\mathrm {[S\,{\sc II}]}$ \citep{1989ApJ...345..811R} and $\mathrm {[Fe\,{\sc II}]}$.\citep{1994A&A...289..256S} }  
\label{emission}
\end{figure}  

\section{Atomic emission formulae} 

The total emission element $\mathrm{dL}$ by recombination within a volume element $\mathrm{dV}$ is
\begin{equation}
d{\mathrm L} = 4 \pi j_{ion}  dV \,\,\,\,\,\,\,\,\,\mathrm{(erg\,s^{-1})}.
\end{equation}
Thus, the emissivity of atomic elements is dependent on the number densities of free electrons $\mathrm{n_{e}}$ and ions $\mathrm{n_{ion}}$ (here in units of $\mathrm{cm^{-3}}$), the temperature dependent production rate coefficients for recombination and spontaneous de-excitation, $\mathrm{\epsilon_{ion}^c(T)}$, in units of $\mathrm{cm^{3}s^{-1}}$ as presented in Fig.~\ref{emission}, and the emitted photon energy of corresponding  wavelength in $\mathrm{erg}$. The emission rate is then given by  
\begin{equation}
j_{ion} = \frac{hv}{4\pi}\, \epsilon_{\lambda} \,\,\,\,\,\,\,\mathrm{(erg\,s^{-1}\,cm^{-3}\,sr^{-1}),}
\end{equation}
where $\epsilon_{\lambda}$ is a photon emissivity of corresponding ion wavelength 
\begin{equation}
\epsilon_{\lambda} =  \epsilon_{ion}^c(T) \, n_{e}\, n_{ion} \,\,\,\, \mathrm{(cm^{-3}\,s^{-1}).}
\end{equation}
Given that we know the temperature of the medium and the number densities, we can calculate the emissivity. All four line emission models are described below in detail. The resultant radiation contribution for each shock element $\mathrm{\dot{E}=dL/dV}$ integrated along the line of sight is represented in logarithmic units of $\mathrm{erg\,s^{-1}\,cm^{-2}}$ displayed in Fig.~\ref{2.1atEm_All} and Fig.~\ref{OI-60deg-140} - \ref{FeII-60deg-140}.  
\subsection{$\mathrm{[O\,{\sc I}]\,}$ $\mathrm{630.0\,nm}$ emission}
The $\mathrm{ j_{[O\,{\sc I}]} }$ emission rate subtended over $4\pi$ steradians using the $\mathrm{6300}$ {\AA} photon transition emissivity  production by spontaneous de-excitation $\epsilon_{\lambda}$ by \citep{1989ApJ...345..811R} multiplied by emitted photon energy of corresponding  wavelength results in total radiation contribution of 
\begin{equation}
\dot{E}_{\mathrm{[O\,{\sc I}]\,}}=hv_{\mathrm{[O{\sc I}]}}\, 5.3\times 10^{-9} \frac{\Omega(T)}{T_4^{0.5}}\exp^{-2.28/T_4}n_e n_{\mathrm O^{0}}\,\mathrm{(erg\,s^{-1}\,cm^{-3})}\label{rate_OI}
\end{equation}
where $\mathrm{\Omega(T)=0.39\,T_4^{0.95}}$ is the $\mathrm{ ^3{P} \to ^1{D} }$ collision strength by \citep{1973JPhB....6L.243S} for ionised medium of $\mathrm{0.4 \, \textless\, T_4\, \textless 1.2\,K}$. With $\mathrm{T_4=T/10^4\,K}$ satisfying J-shock condition.

\subsection{$\mathrm {\textbf H \alpha}$ $\mathrm{656.28\,nm}$ emission}
We have used the $\mathrm { 6563}$ {\AA} emissivity rate of $\mathrm H\alpha$ production line by $\mathrm{H^{+}}$ and $\mathrm{e^{-}}$ recombination and spontaneous de-excitation as given by \citep{1989ApJ...345..811R}, \citep{2011ApJ...727...35D}, resulting in total emitted line flux  
\begin{multline}
\dot{E}_{\mathrm H\alpha}=hv_{h\alpha} \, 1.17\times 10^{-13}\, T_4^{-0.942-0.031\ln(T_4)}  \\ \exp^{-14.036/T_4}\, n_e n_{\mathrm H^{+}}\,\mathrm{(erg\,s^{-1}\,cm^{-3})}
\label{rate_Ha}
\end{multline}
$\mathrm{n_e = 0.1n_{H}}$ and $\mathrm{n_{ H^{+}}={n_{H} }}$ are number densities of electrons and ionised hydrogen, respectively.  

\subsection{$\mathrm {[S\,{\sc II}]}$ $\mathrm{671.6\,nm}$ emission}  
The rate coefficient by recombination and spontaneous de-excitation resulting in optical $\mathrm {[S\,{\sc II}]}$ $6716$ {\AA} transition provided by \citep{1989ApJ...345..811R} coupled with corersponding photon energy  
\begin{equation}
\dot{E}_{\mathrm{[S\,{\sc II}]\,}}=hv_{\mathrm{[S\,{\sc II}]}} \, 7.3\times 10^{-8}T_4^{-0.5}\exp^{-2.14/T_4}n_e n_{\mathrm S^{+}}\,\mathrm{(erg\,s^{-1}cm^{-3})}
\label{rate_Sii}
\end{equation}
with number density of $\mathrm{S^{+} }$ ions based on an estimated relative abundance of neutral S with respect to H of $\mathrm{\epsilon_{{[S]}} = 1.58\,\times 10^{-5} }$ provided by \citep{1989ApJ...345..811R}, along with tabulated logarithmic interpolation of ion fraction abundance $\mathrm{\epsilon_{[S\,\sc II]}(T) }$ by \citep{1985A&AS...60..425A} resulting in $\mathrm{ n_{S^{+}}=n_{w}\,\epsilon_{[S]} \,\epsilon_{[S\,\sc II]}(T) }$ and $\mathrm{ n_{w} =n_{\mathrm H} }$. 


\subsection{$\mathrm {[Fe\,{\sc II}] }$ $1.64\,\mathrm{\mu m}$ emission}
Collisional excitation rates for non-dissociative J-shock model by \citet{1994A&A...289..256S} for $1.64\,\mathrm{\mu m}$ near-infrared $\mathrm {[Fe\,{\sc II}] }$ line

\begin{multline}
\dot{E}_{\mathrm{[Fe\,{\sc II}]\,}}=hv_{\mathrm{[Fe\,{\sc II}]}} \,1.9\times 10^{-3} \\ \exp^{-1.13/T_4}/ (1+9.74^{4}\sqrt{T_4}) \, n_e n_{\mathrm Fe^{+}}\,\mathrm{(erg\,s^{-1}cm^{-3})}.
\label{rate_FeII}
\end{multline}
Strong UV radiation at later PN stage combined with heating from $\mathrm H_2$ reformation could contribute to the $\mathrm {[Fe\,{\sc II}] }$ surface brightness. Here, we use a fixed ionisation fraction of $0.1$ with $n_{\mathrm Fe^{+}}=n_\mathrm {Fe\,{\sc II}}$ notation.

\begin{figure}
\subfloat[M2: $\rm{H_2}$ wind interacting with an $\rm{H}$ ambient medium.]{
        \label{subfig:correct}
        \includegraphics[width=1.0\columnwidth,height=0.3\textheight,keepaspectratio]{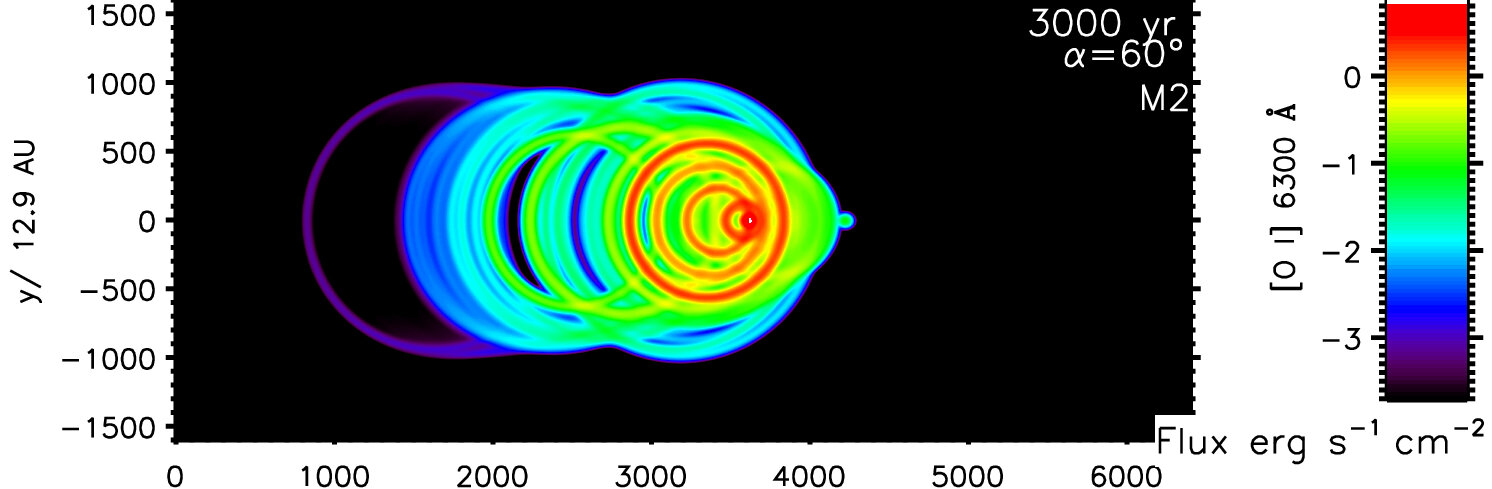} } \\
\subfloat[M3: $\rm{H}$ wind interacting with an $\rm{H_2}$ ambient medium.]{
        \label{subfig:notwhitelight}
        \includegraphics[width=1.0\columnwidth,height=0.3\textheight,keepaspectratio]{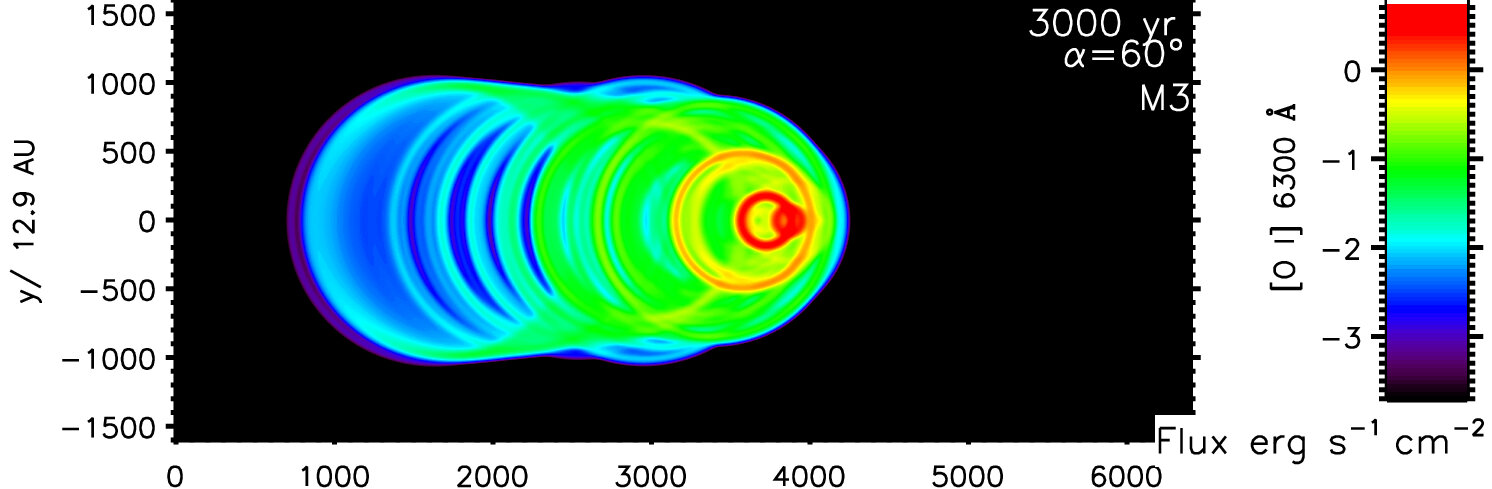} }\\
\subfloat[M4: $\rm{H}$ wind interacting with an $\rm{H}$ ambient medium.]{
        \label{subfig:nonkohler}
        \includegraphics[width=1.0\columnwidth,height=0.3\textheight,keepaspectratio]{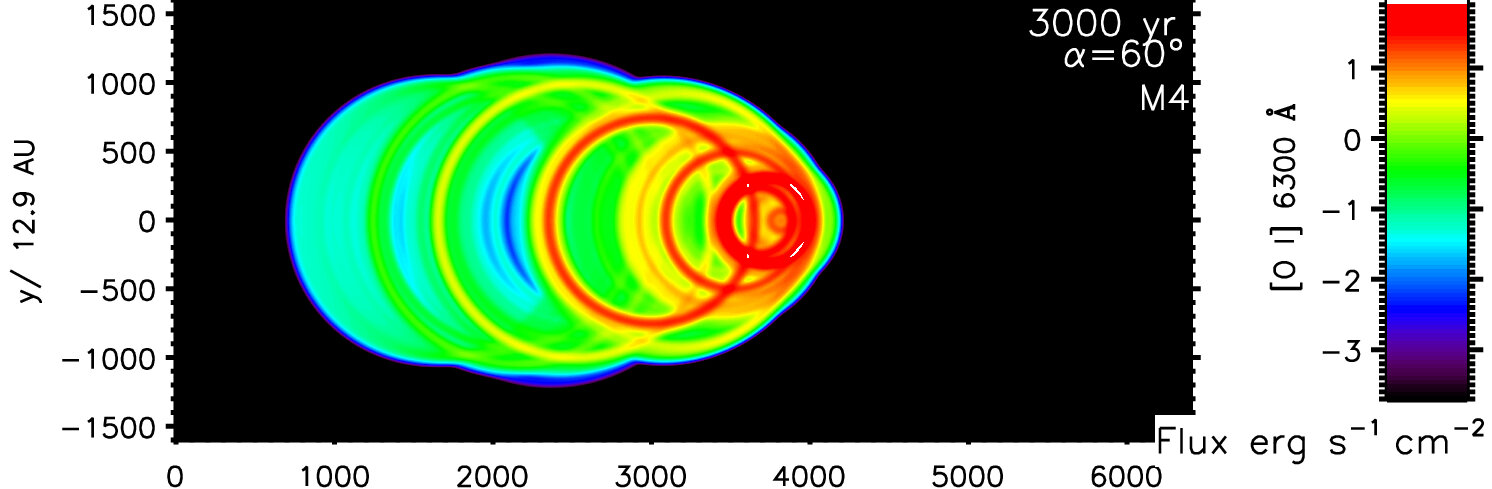}} 
        \caption[$\mathrm{[O I]\,6300}$ {\AA} airglow line flux by 4:1 outflows]
        {Simulated $\mathrm{[O\,{\sc I}]\,6300}$ {\AA} airglow line mapping of 4:1 ellipticity winds at $60^{\circ}$ to the plane of the sky and axial speed of 140~km~s$^{-1}$. The origin of the wind is at zone (1,600,0) .}
\label{OI-60deg-140}
\end{figure}

\begin{figure}
\subfloat[M2: $\rm{H_2}$ wind interacting with an $\rm{H}$ ambient medium.]{
        \label{subfig:correct}
        \includegraphics[width=1.0\columnwidth,height=0.3\textheight,keepaspectratio]{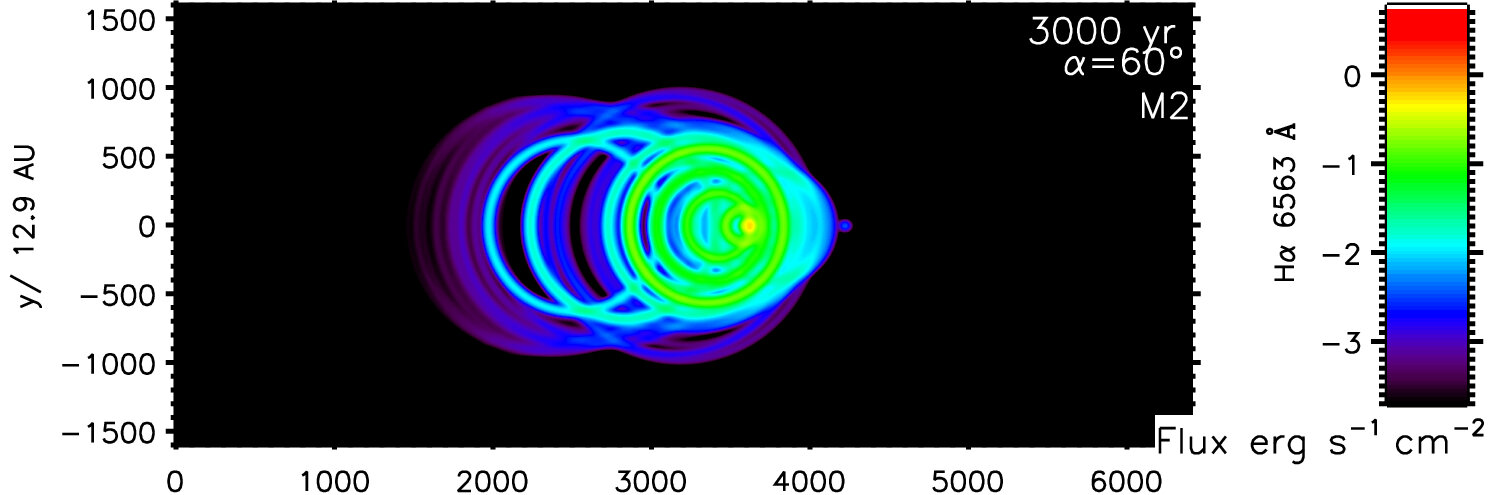} } \\
\subfloat[M3: $\rm{H}$ wind interacting with an $\rm{H_2}$ ambient medium.]{
        \label{subfig:notwhitelight}
        \includegraphics[width=1.0\columnwidth,height=0.3\textheight,keepaspectratio]{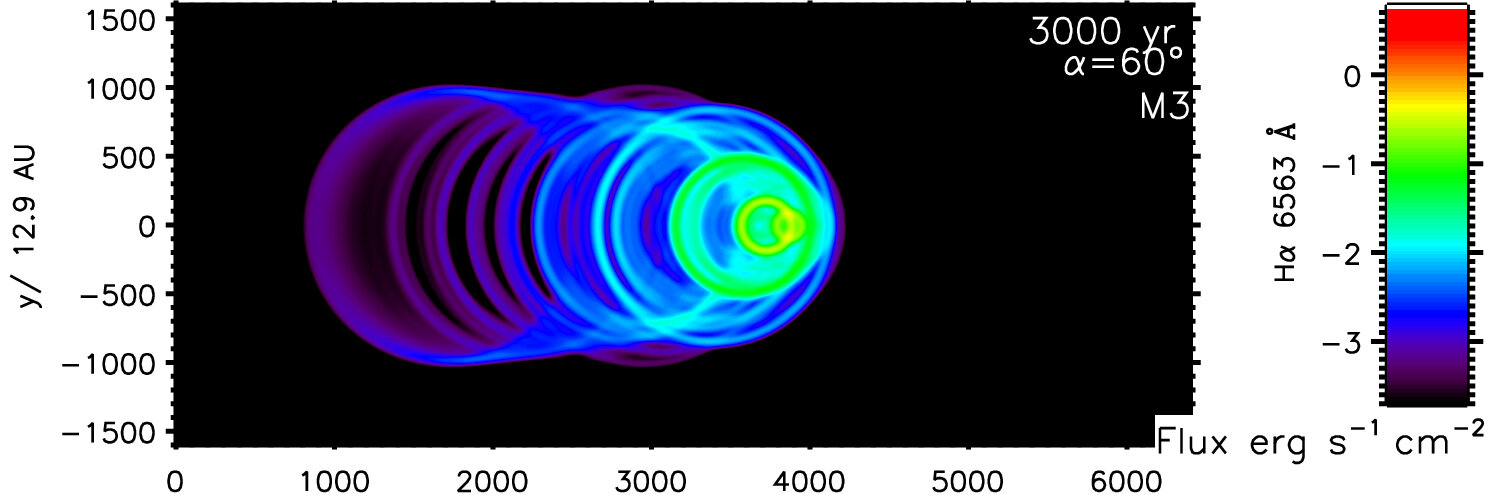} }\\
\subfloat[M4: $\rm{H}$ wind interacting with an $\rm{H}$ ambient medium.]{
        \label{subfig:nonkohler}
        \includegraphics[width=1.0\columnwidth,height=0.3\textheight,keepaspectratio]{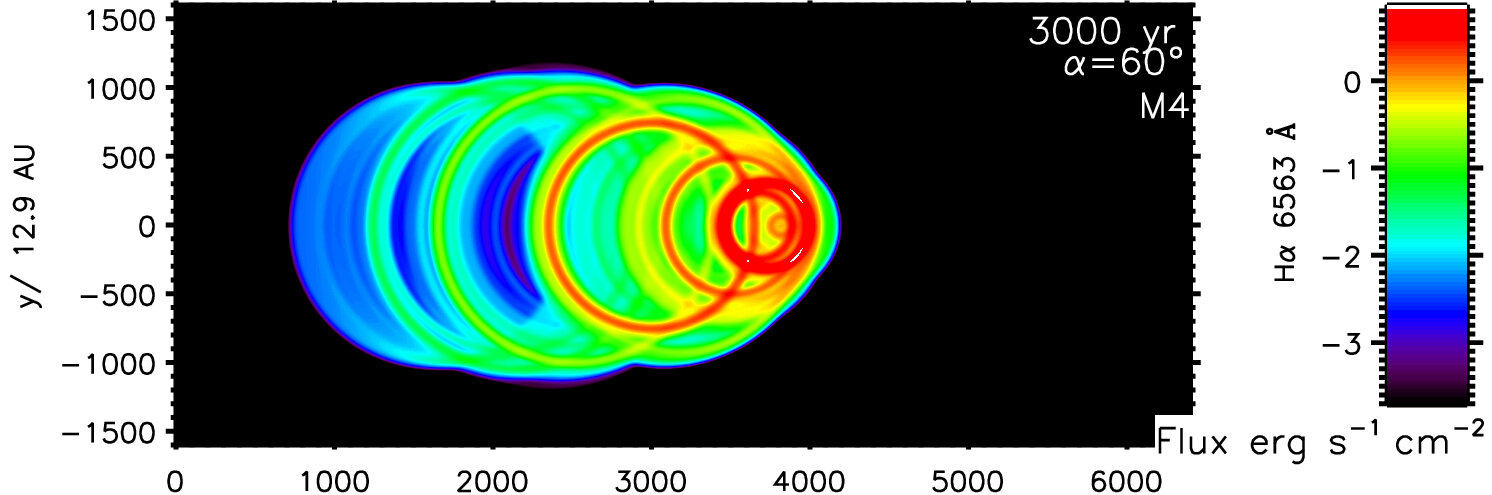}} 
        \caption[$\mathrm{[H\alpha]\,6563}$ {\AA} line flux by 4:1 outflows]
        {Simulated H$\mathrm \alpha\,6563$ {\AA} line mapping of 4:1 ellipticity winds at $60^{\circ}$ to the plane of the sky and axial speed of 140~km~s$^{-1}$. The origin of the wind is at zone (1,600,0) .}
\label{Halp-60deg-140}
\end{figure}

\begin{figure}
\subfloat[M2: $\rm{H_2}$ wind interacting with an $\rm{H}$ ambient medium.]{
        \label{subfig:correct}
        \includegraphics[width=1.0\columnwidth,height=0.3\textheight,keepaspectratio]{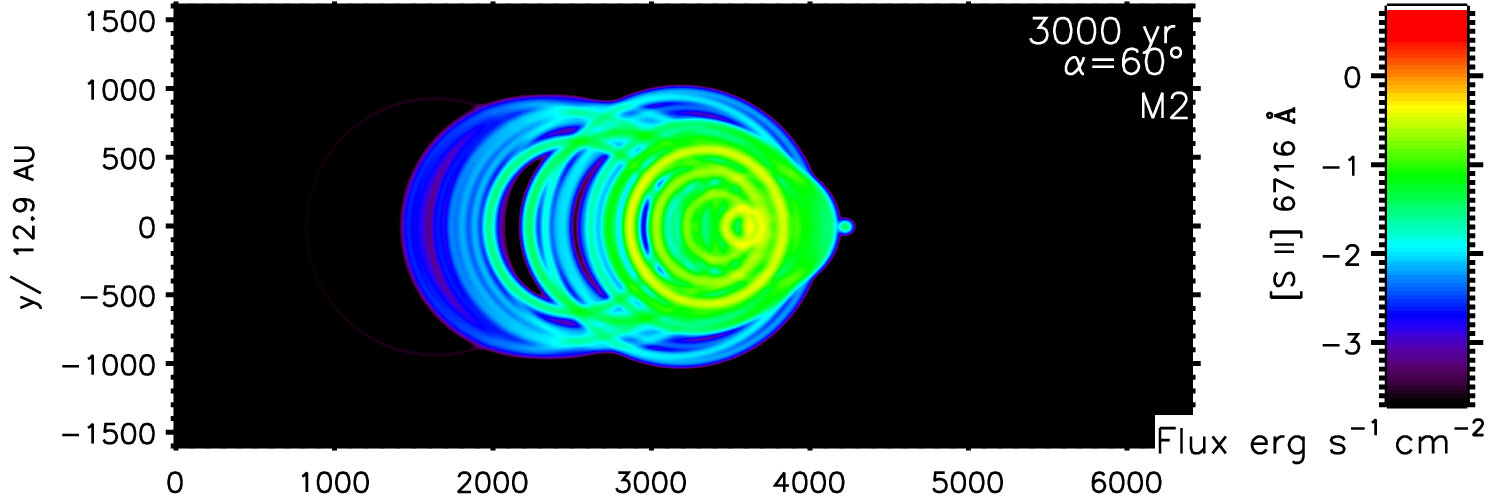} } \\
\subfloat[M3: $\rm{H}$ wind interacting with an $\rm{H_2}$ ambient medium.]{
        \label{subfig:notwhitelight}
        \includegraphics[width=1.0\columnwidth,height=0.3\textheight,keepaspectratio]{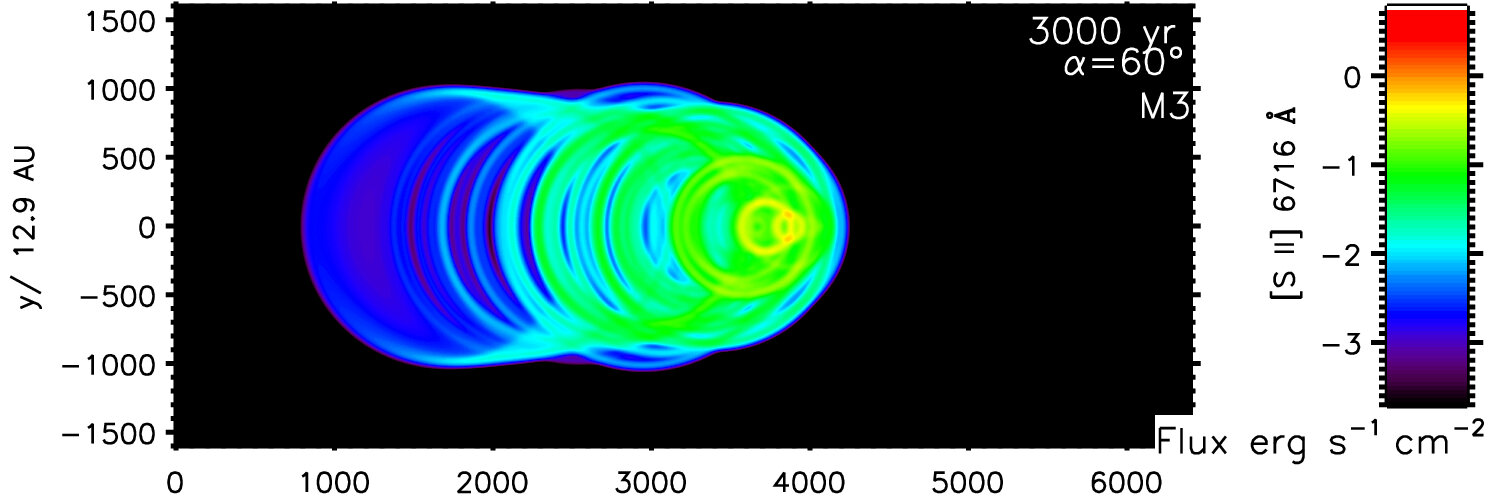} }\\
\subfloat[M4: $\rm{H}$ wind interacting with an $\rm{H}$ ambient medium.]{
        \label{subfig:nonkohler}
        \includegraphics[width=1.0\columnwidth,height=0.3\textheight,keepaspectratio]{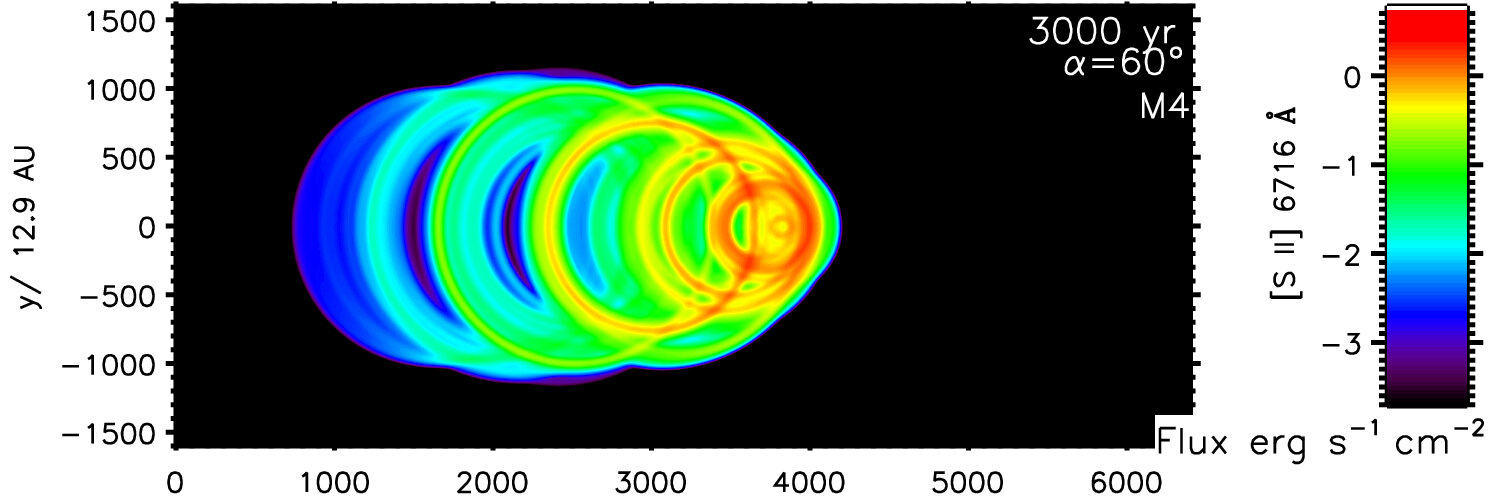}} 
        \caption[$\mathrm{[SiII]\,6716}$ {\AA} line flux by 4:1 outflows]
        {Simulated $\mathrm{ [S\,{\sc II}] \,6716}$ {\AA} line mapping of 4:1 ellipticity winds at $60^{\circ}$ to the plane of the sky and axial speed of 140~km~s$^{-1}$. The origin of the wind is at zone (1,600,0) .}
\label{SII-60deg-140}
\end{figure}

\begin{figure}
\subfloat[M2: $\rm{H_2}$ wind interacting with an $\rm{H}$ ambient medium.]{
        \label{subfig:correct}
        \includegraphics[width=1.0\columnwidth,height=0.3\textheight,keepaspectratio]{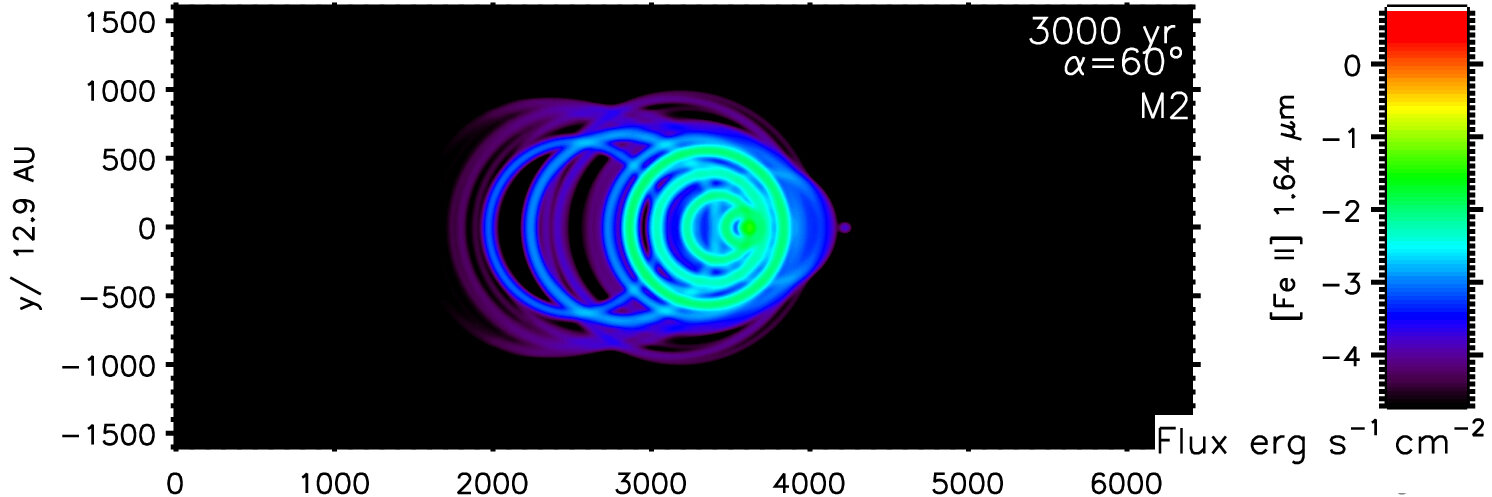} } \\
\subfloat[M3: $\rm{H}$ wind interacting with an $\rm{H_2}$ ambient medium.]{
        \label{subfig:notwhitelight}
        \includegraphics[width=1.0\columnwidth,height=0.3\textheight,keepaspectratio]{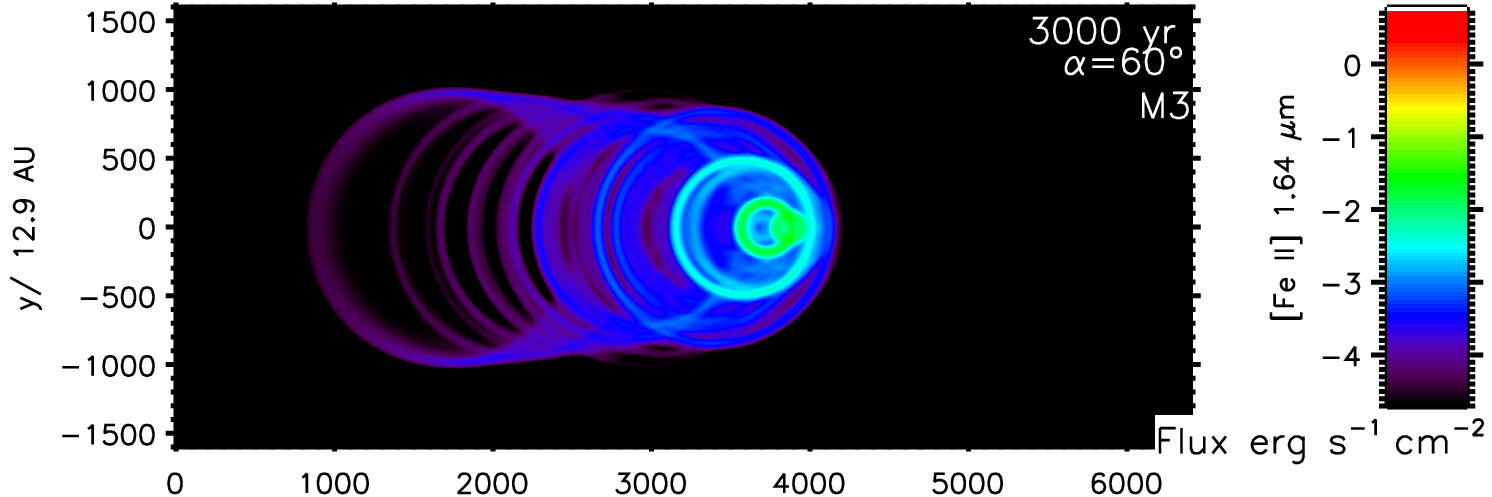} }\\
\subfloat[M4: $\rm{H}$ wind interacting with an $\rm{H}$ ambient medium.]{
        \label{subfig:nonkohler}
        \includegraphics[width=1.0\columnwidth,height=0.3\textheight,keepaspectratio]{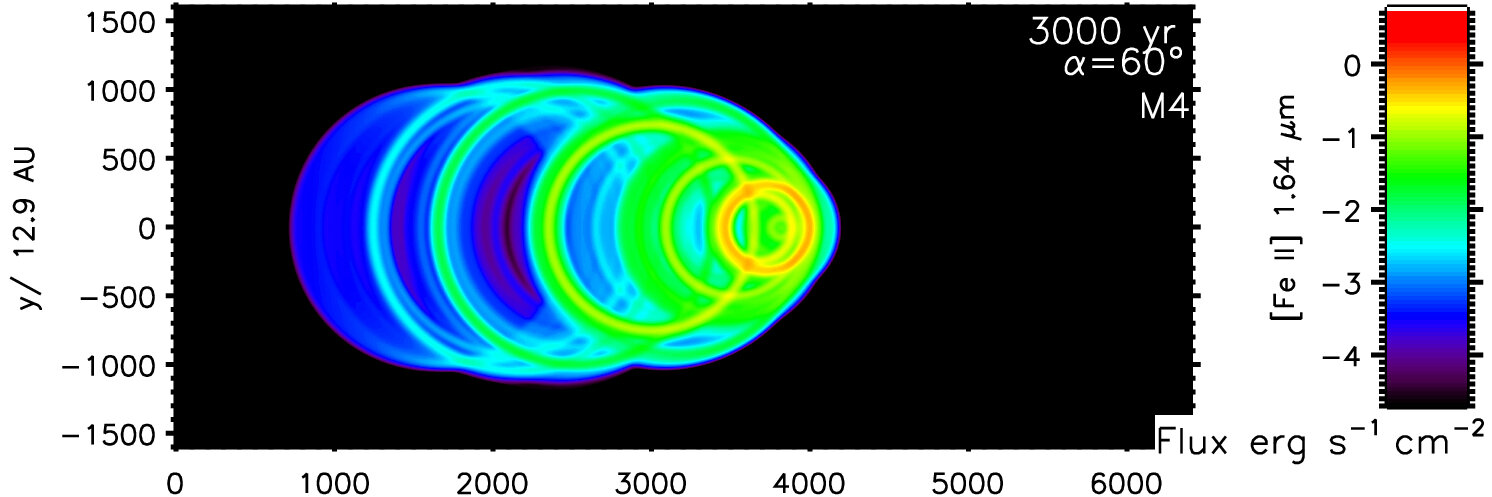}} 
        \caption[$1.64\,\mathrm{\mu m} \,\mathrm {[FeII]}$ line flux by 4:1 outflows]
        {Simulated $\mathrm {[Fe\,{\sc II}] \, 1.64\,\mathrm{\mu m} }$ line mapping of 4:1 ellipticity winds at $60^{\circ}$ to the plane of the sky and axial speed of 140~km~s$^{-1}$. The origin of the wind is at zone (1,600,0) .}
\label{FeII-60deg-140}
\end{figure}

\begin{figure*}
   \subfloat[\label{genworkflow}][M2.]{%
      \includegraphics[width=0.3\textwidth,height=0.3\textheight]{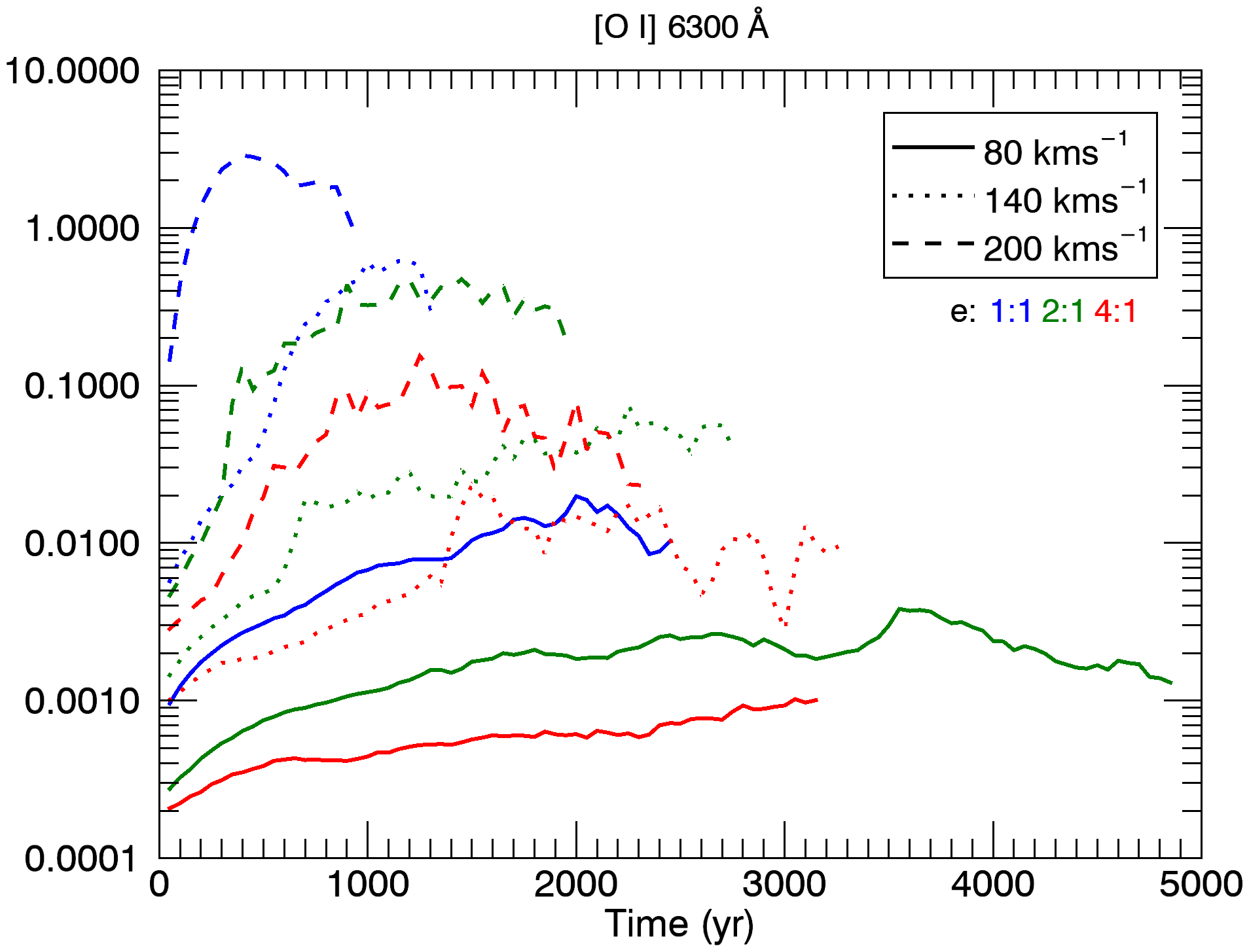}}
\hspace{\fill}
   \subfloat[\label{pyramidprocess}][M3.]{%
      \includegraphics[width=0.3\textwidth,height=0.3\textheight]{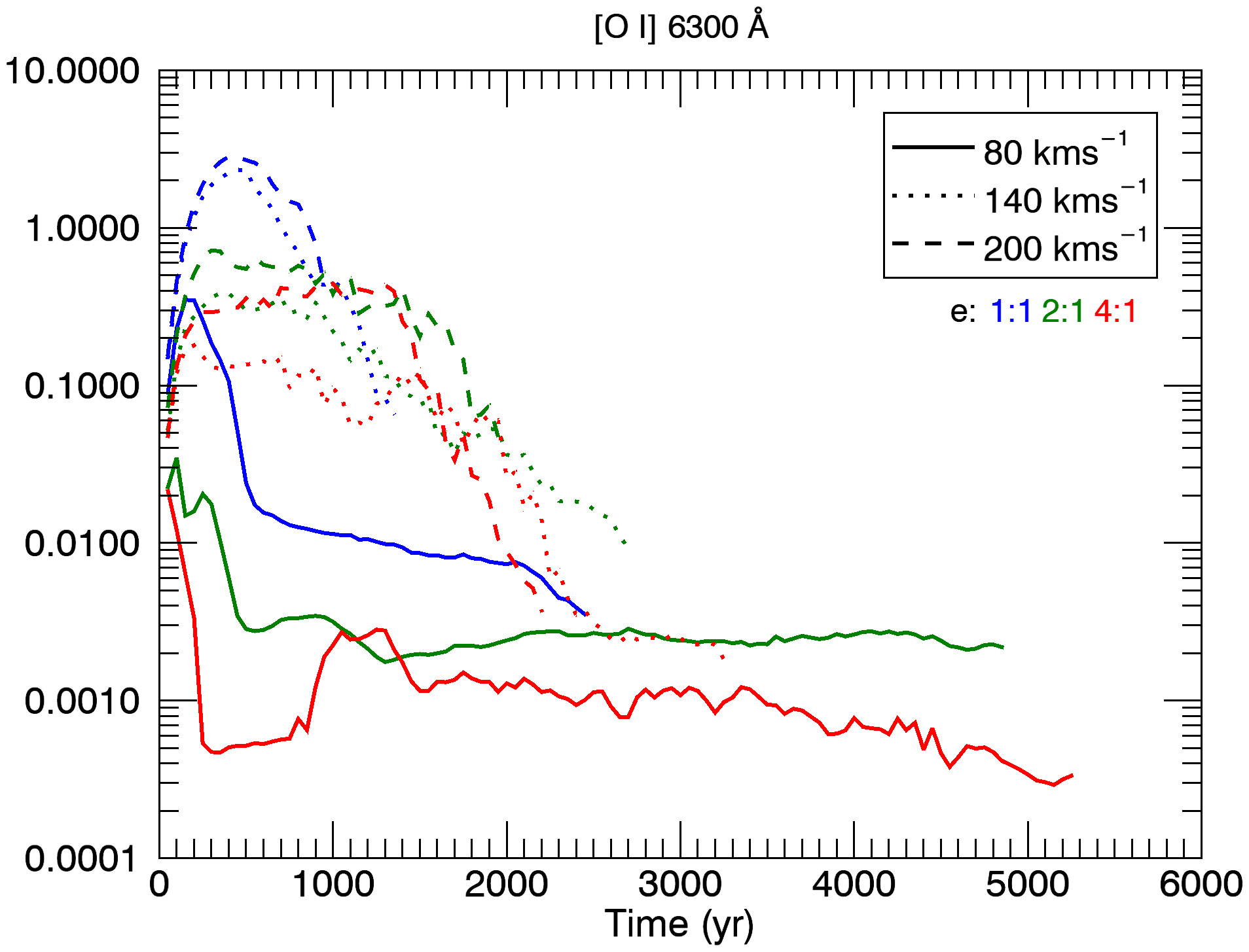}}
\hspace{\fill}
   \subfloat[\label{mt-simtask}][M4.]{%
      \includegraphics[width=0.3\textwidth,height=0.3\textheight]{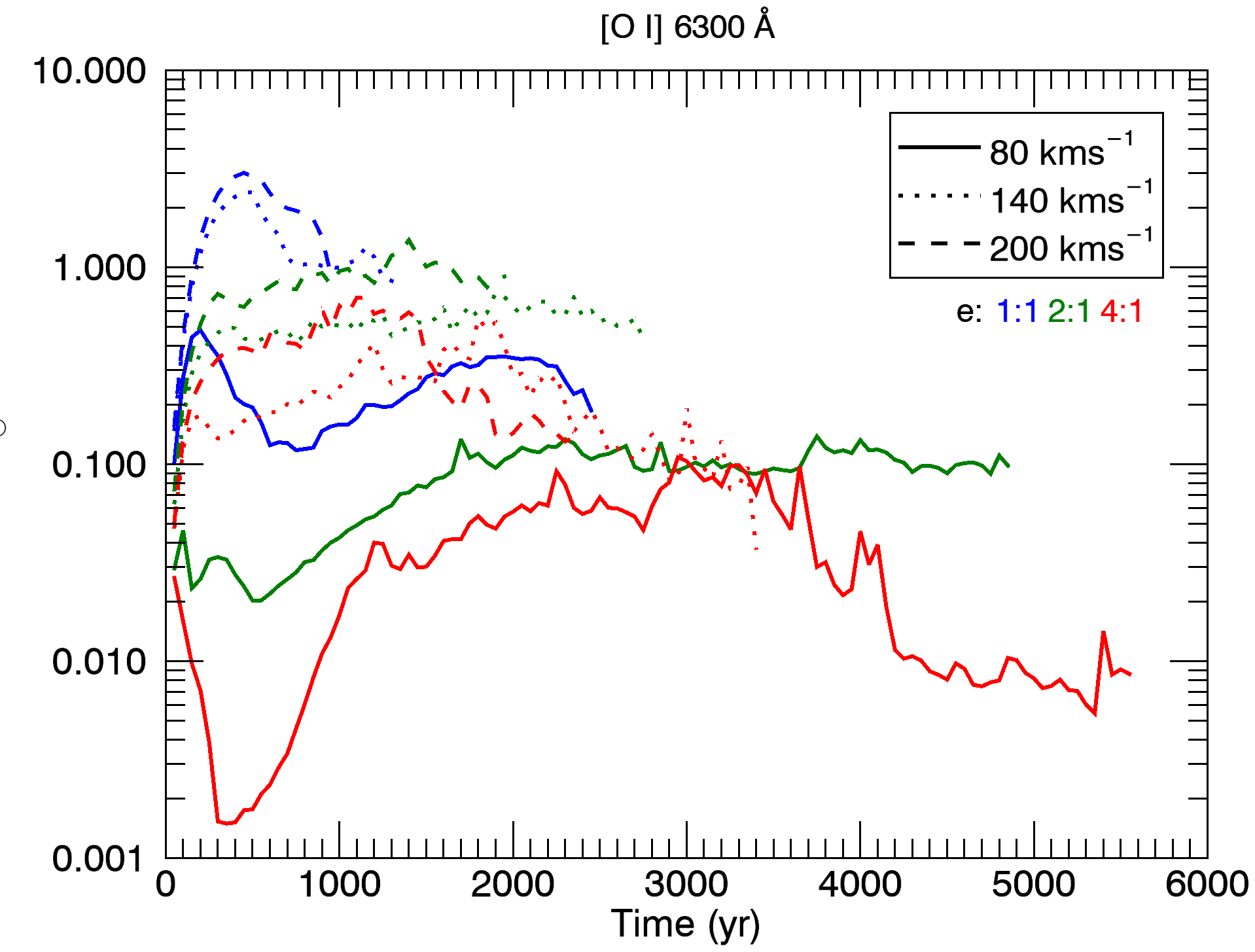}}\\
\caption[Optical $\mathrm{[O I]}$ flux variation]
{\label{atemisOI} Total $\mathrm{[O\,{\sc I}]\,6300}$ {\AA} emissivities by M2, M3 and M4 outflows at high/low wind velocities for all ellipticities.}
\end{figure*}

\begin{figure*}
   \subfloat[\label{genworkflow}][M2.]{%
      \includegraphics[width=0.3\textwidth,height=0.3\textheight]{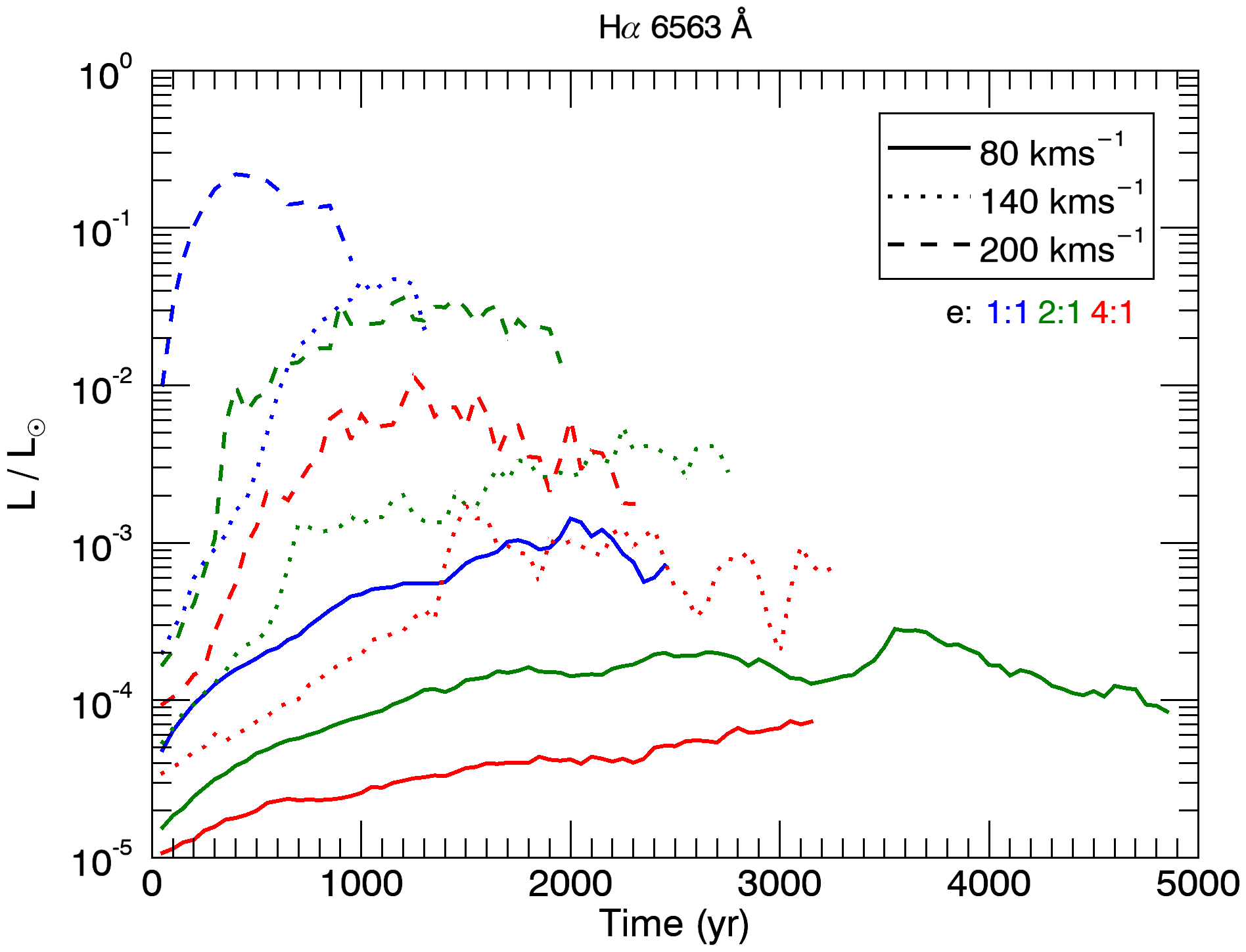}}
\hspace{\fill}
   \subfloat[\label{pyramidprocess}][M3.]{%
      \includegraphics[width=0.3\textwidth,height=0.3\textheight]{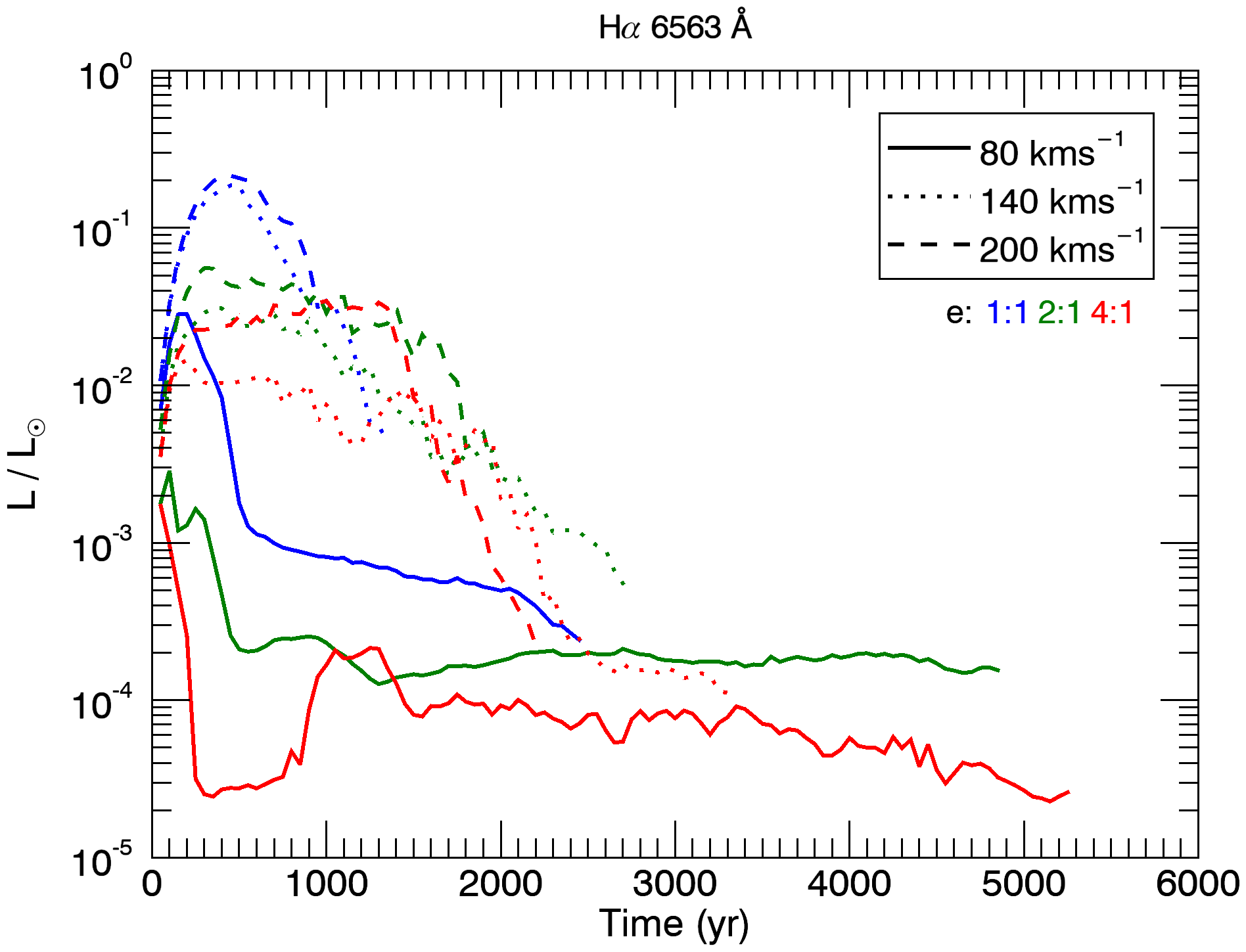}}
\hspace{\fill}
   \subfloat[\label{mt-simtask}][M4.]{%
      \includegraphics[width=0.3\textwidth,height=0.3\textheight]{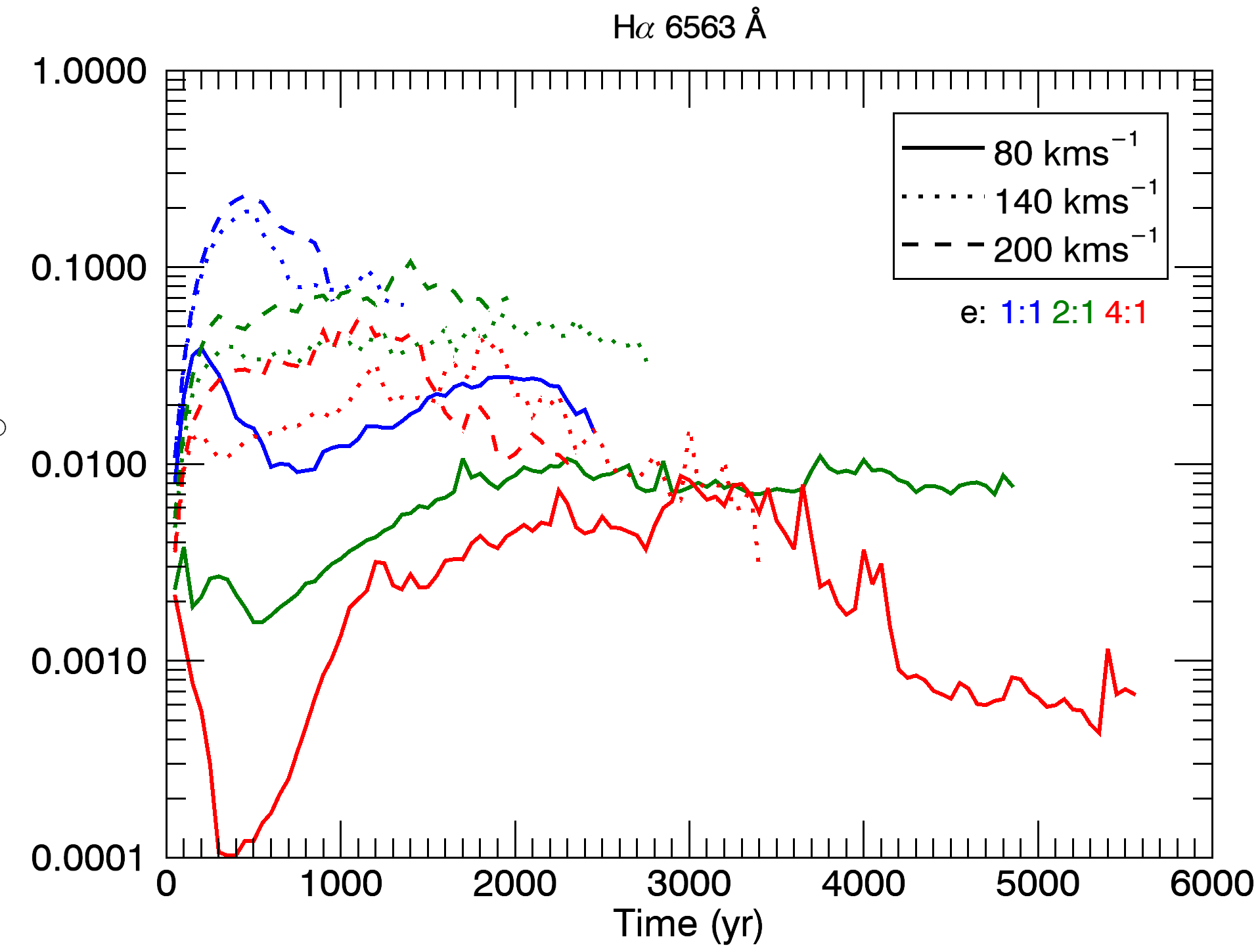}}\\
\caption[Optical $\mathrm { [H \alpha]}$ flux variation]
{\label{atemisHa} Total H$\mathrm \alpha\,6563$ {\AA} emissivities by M2, M3 and M4 outflows at high/low wind velocities for all ellipticities.}
\end{figure*}

\begin{figure*}
   \subfloat[\label{genworkflow}][M2.]{%
      \includegraphics[width=0.3\textwidth,height=0.3\textheight]{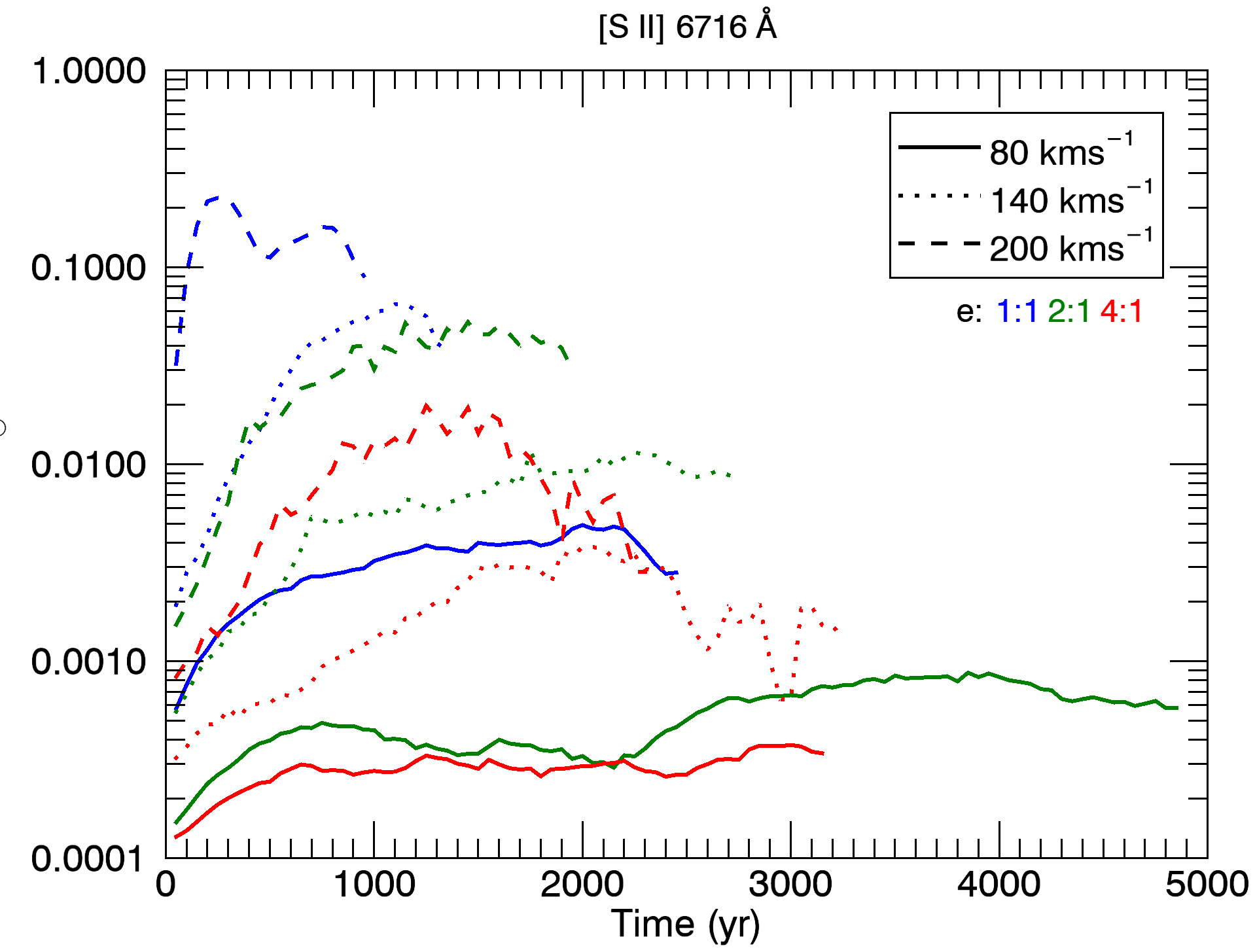}}
\hspace{\fill}
   \subfloat[\label{pyramidprocess}][M3.]{%
      \includegraphics[width=0.3\textwidth,height=0.3\textheight]{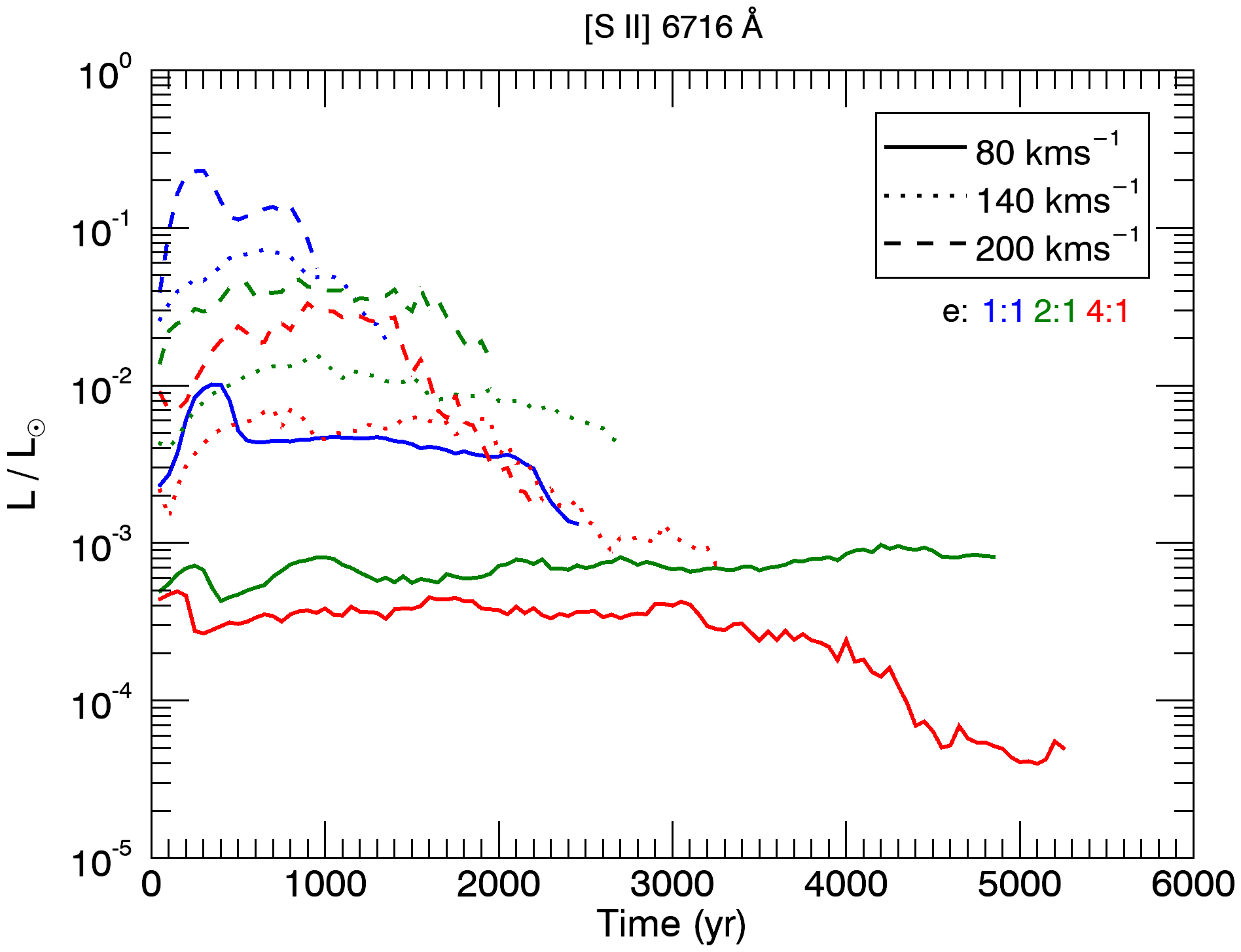}}
\hspace{\fill}
   \subfloat[\label{mt-simtask}][M4.]{%
      \includegraphics[width=0.3\textwidth,height=0.3\textheight]{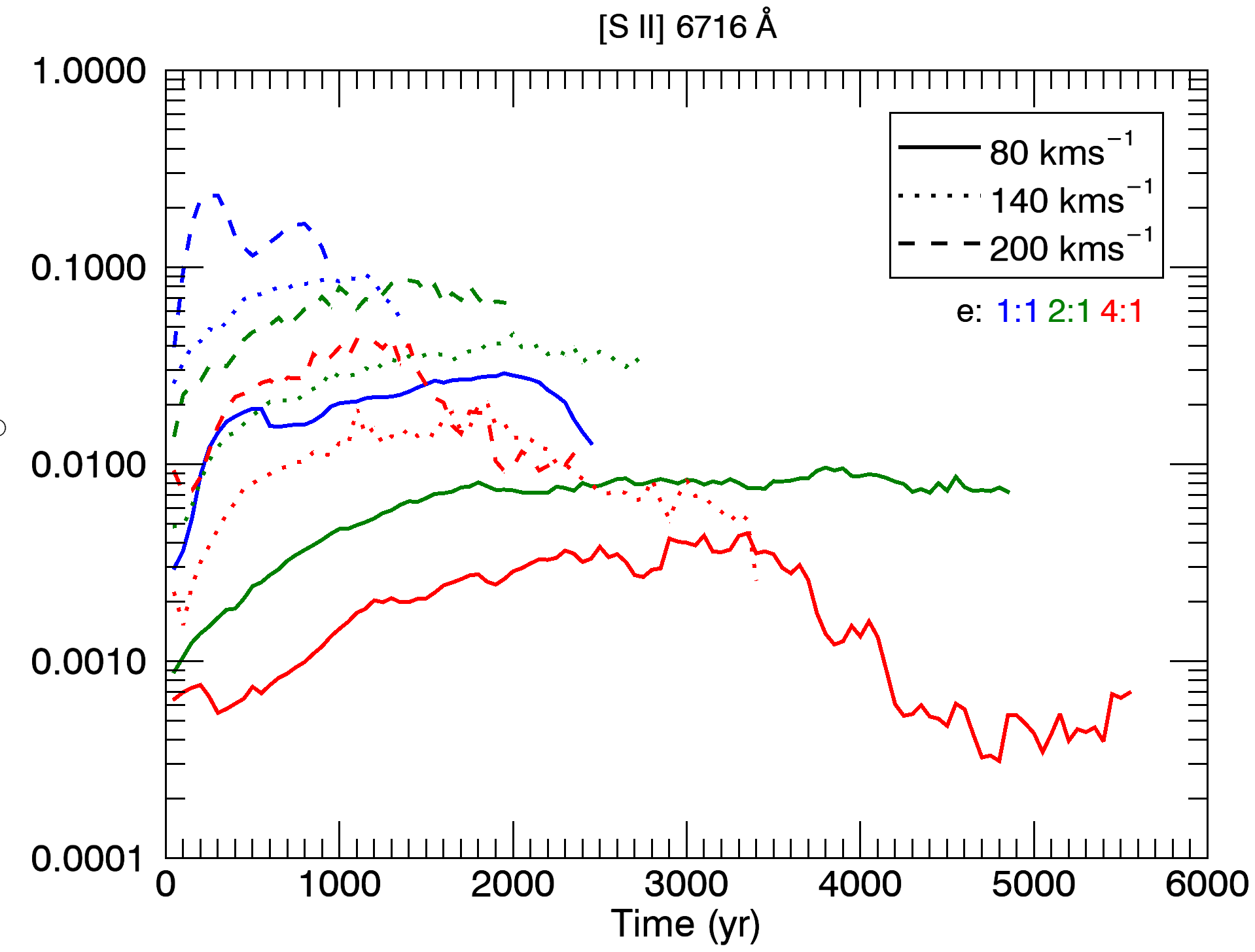}}\\
\caption[Optical $\mathrm{[SiII]}$ flux variation]
{\label{atemisSII} Total $\mathrm{ [S\,{\sc II}] \,6716}$ {\AA} emissivities by M2, M3 and M4 outflows at high/low wind velocities for all ellipticities.}
\end{figure*}
\begin{figure*}
   \subfloat[\label{genworkflow}][M2.]{%
      \includegraphics[width=0.3\textwidth,height=0.3\textheight]{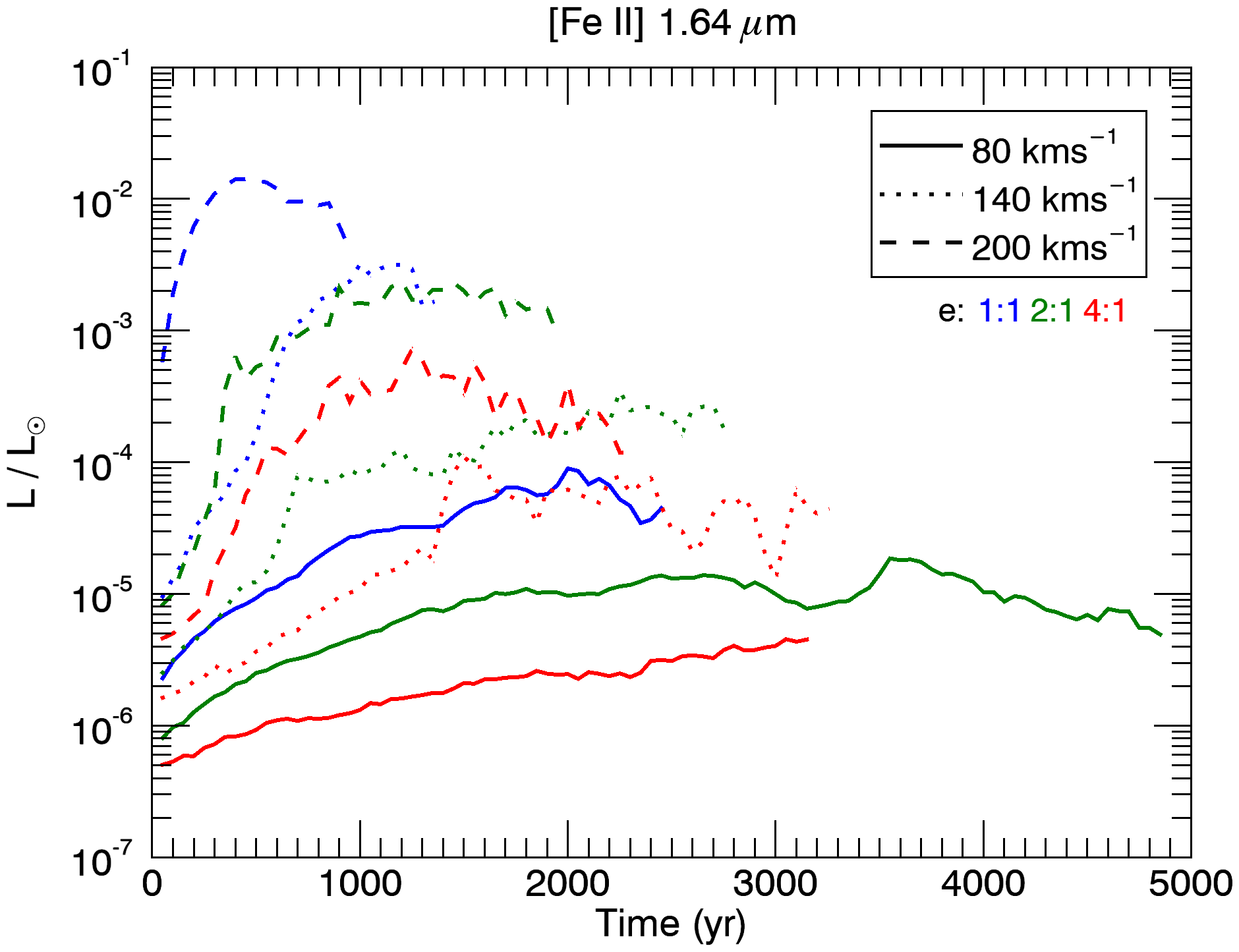}}
\hspace{\fill}
   \subfloat[\label{pyramidprocess}][M3.]{%
      \includegraphics[width=0.3\textwidth,height=0.3\textheight]{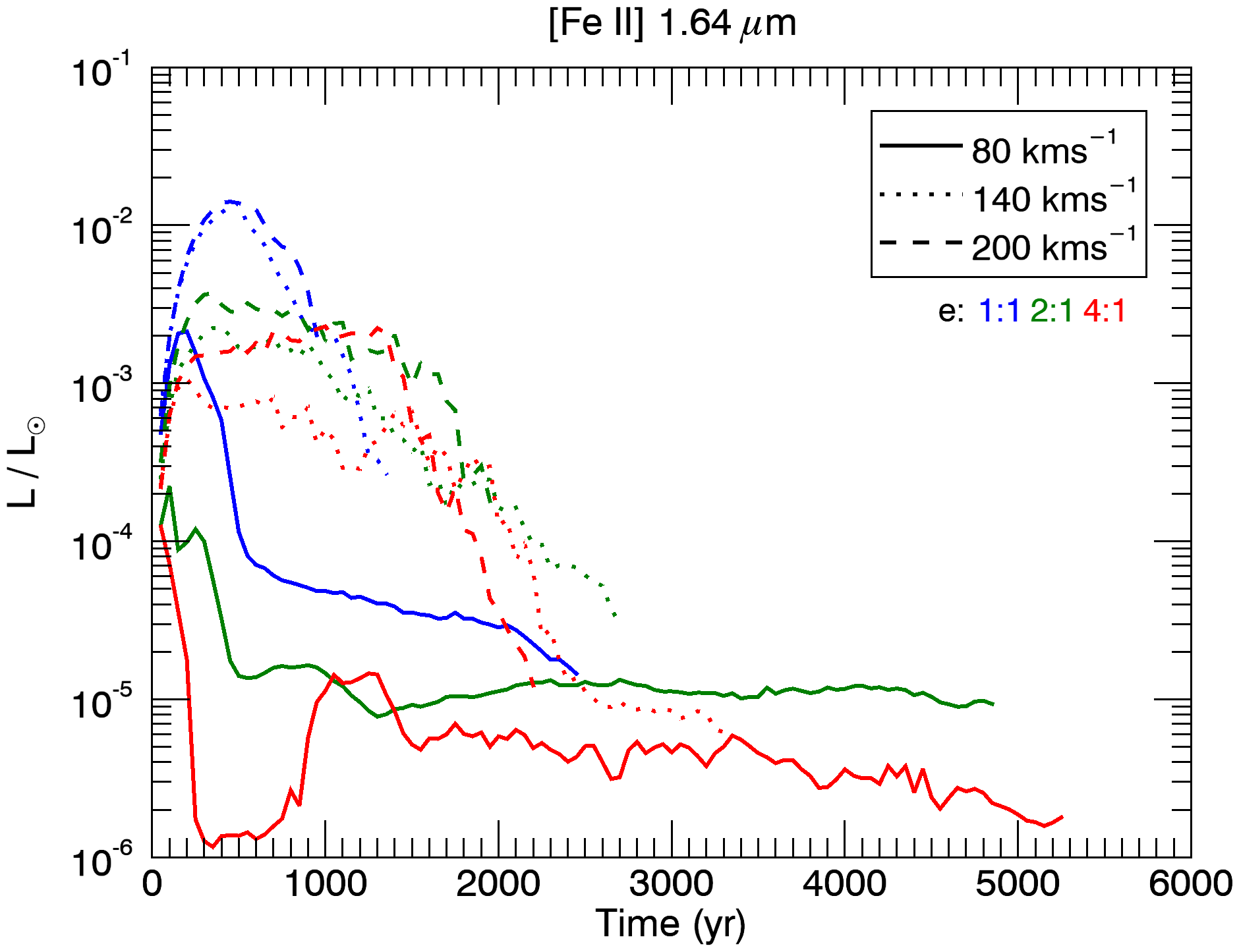}}
\hspace{\fill}
   \subfloat[\label{mt-simtask}][M4.]{%
      \includegraphics[width=0.3\textwidth,height=0.3\textheight]{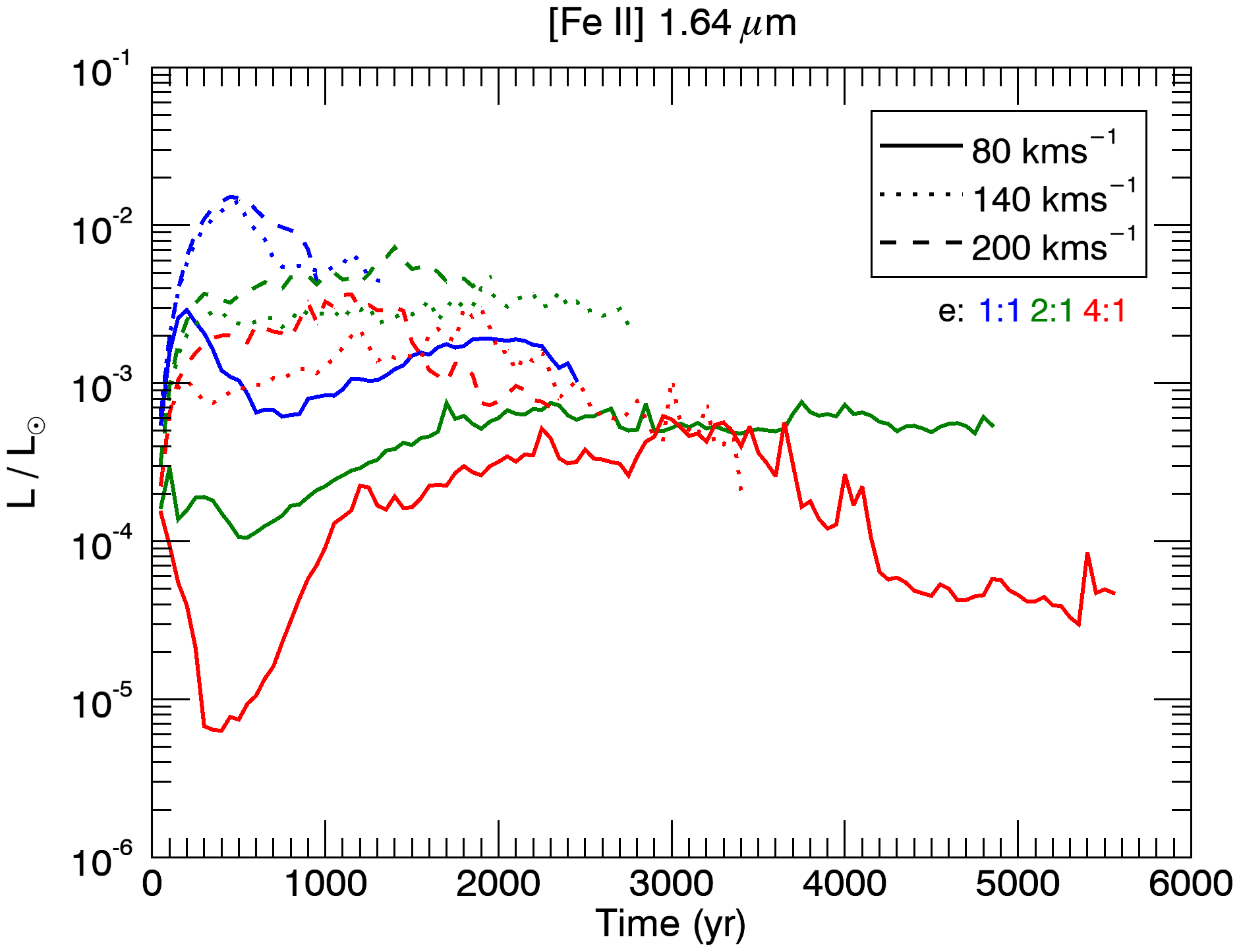}}\\
\caption[Optical $\mathrm {[FeII]}$ flux variation]
{\label{atemisFeII} Total $\mathrm {[Fe\,{\sc II}] \, 1.64\,\mathrm{\mu m} }$ emissivities by M2, M3 and M4 outflows at high/low wind velocities for all ellipticities.}
\end{figure*}

\begin{figure*}
   \subfloat[\label{genworkflow}][M2: (MWAA).]{%
      \includegraphics[width=0.3\textwidth]{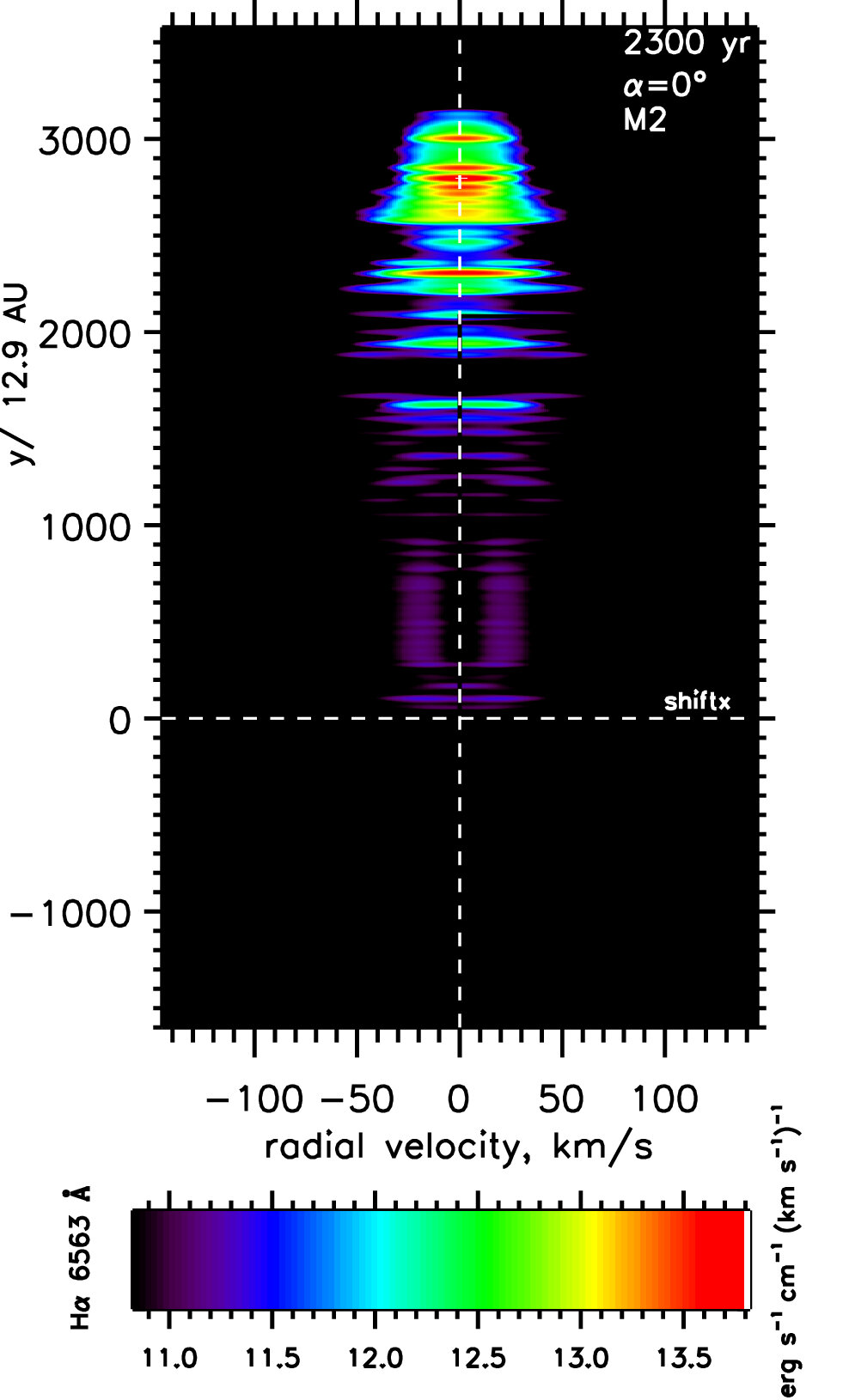}}
\hspace{\fill}
   \subfloat[\label{pyramidprocess}][M3: (AWMA).]{%
      \includegraphics[width=0.3\textwidth]{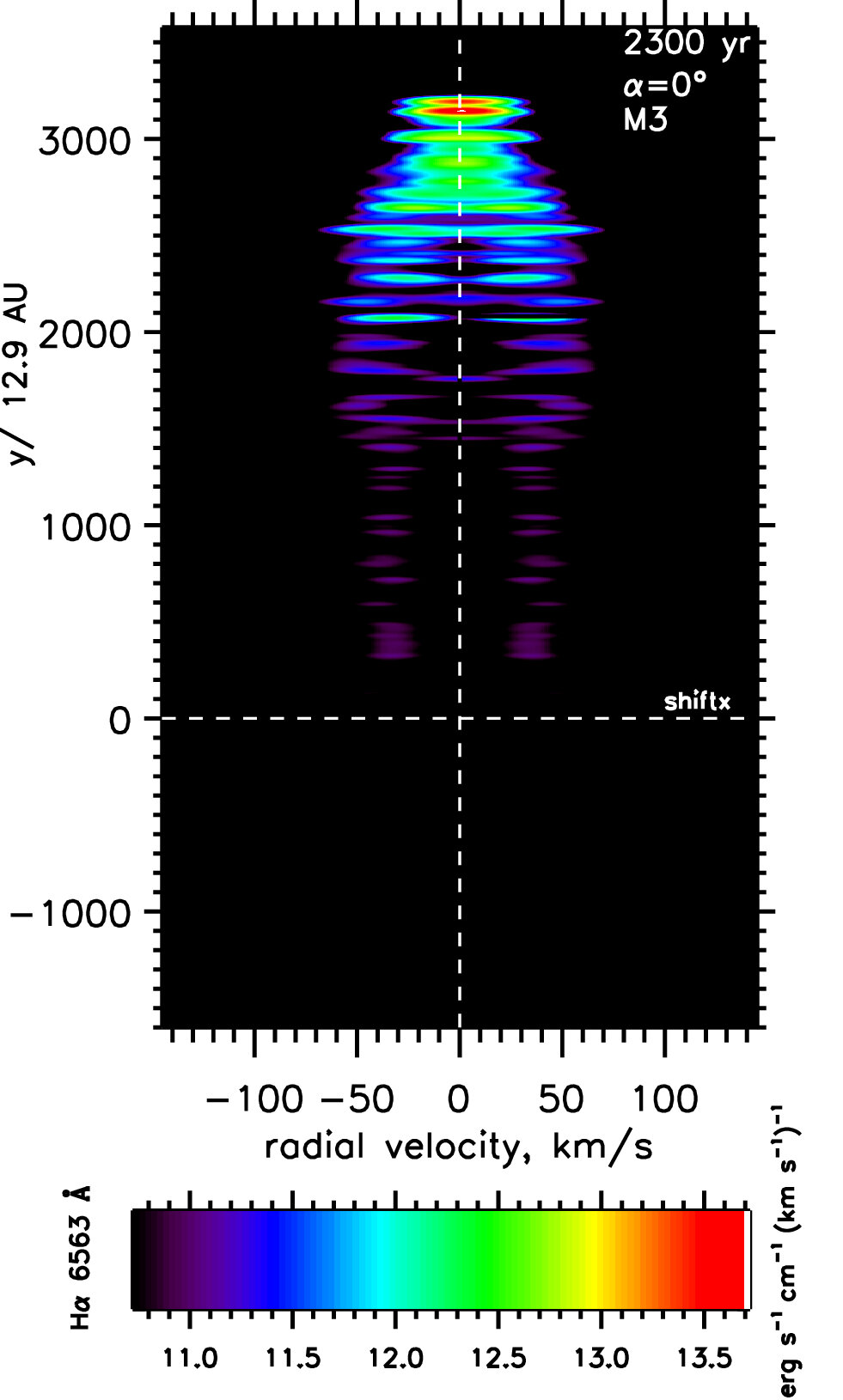}}
\hspace{\fill}
   \subfloat[\label{mt-simtask}][M4: (AWAA).]{%
      \includegraphics[width=0.3\textwidth]{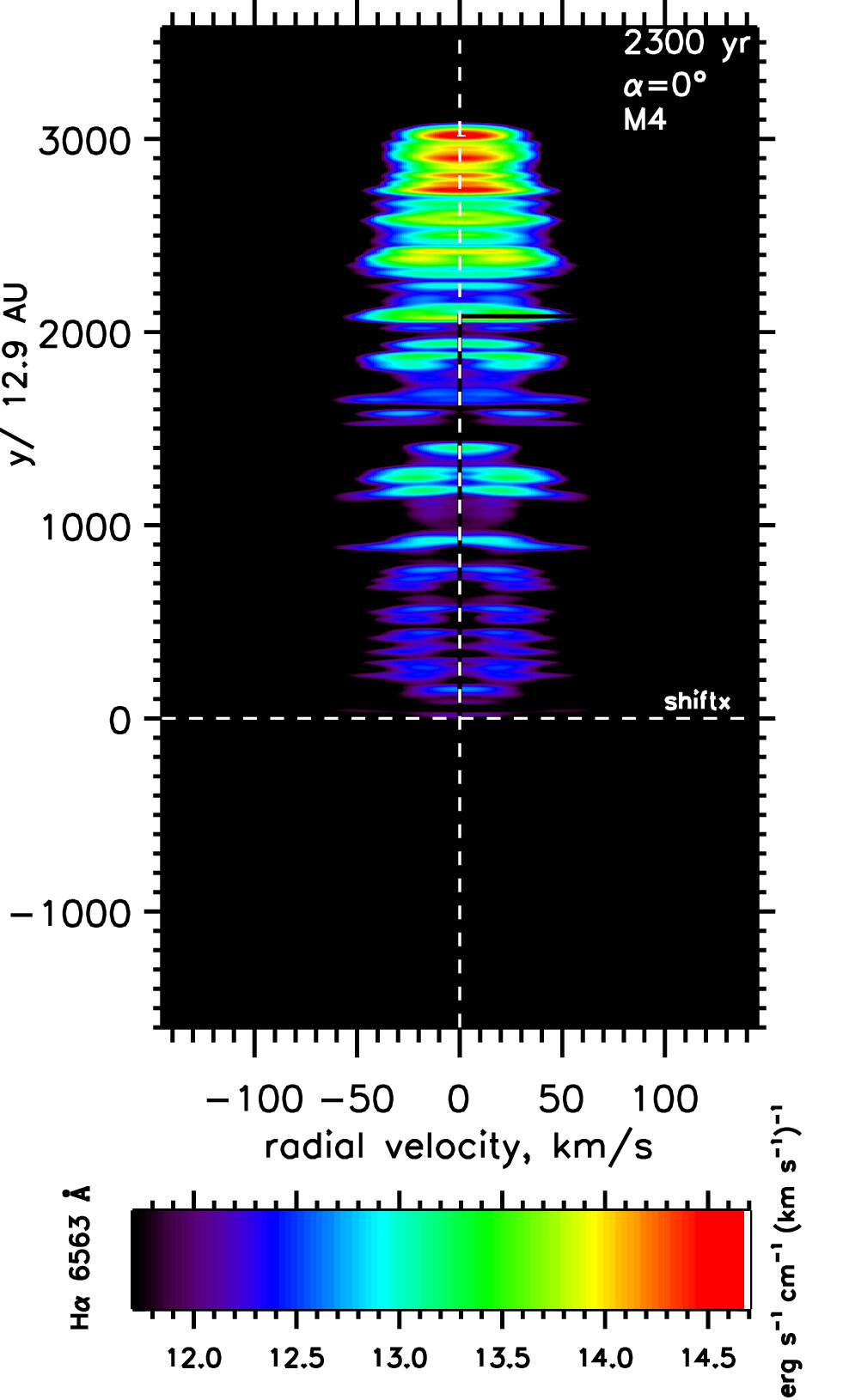}}\\
\caption{\label{2.1atpv0_Ha}Position-Velocity diagrams for the $\mathrm { H \alpha}$ emission from the 2:1 ellipsoidal wind with the long axis in the plane of the sky 
($\alpha = 0^\circ$)  at a late stage of wind expansion. The three composition models are as indicated.}
\end{figure*}

\begin{figure*}
   \subfloat[\label{genworkflow}][M2: (MWAA).]{%
      \includegraphics[width=0.3\textwidth]{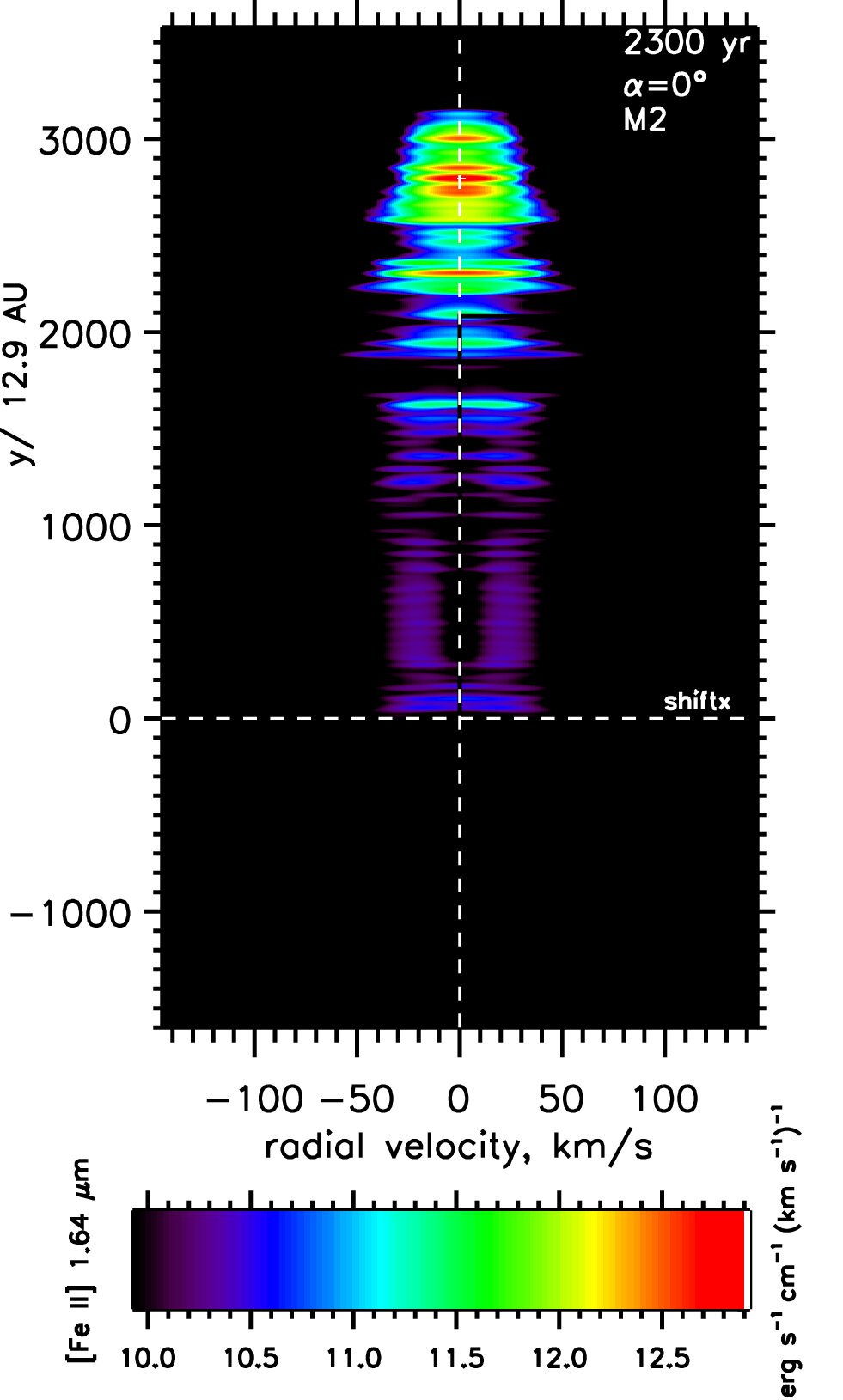}}
\hspace{\fill}
   \subfloat[\label{pyramidprocess}][M3: (AWMA).]{%
      \includegraphics[width=0.3\textwidth]{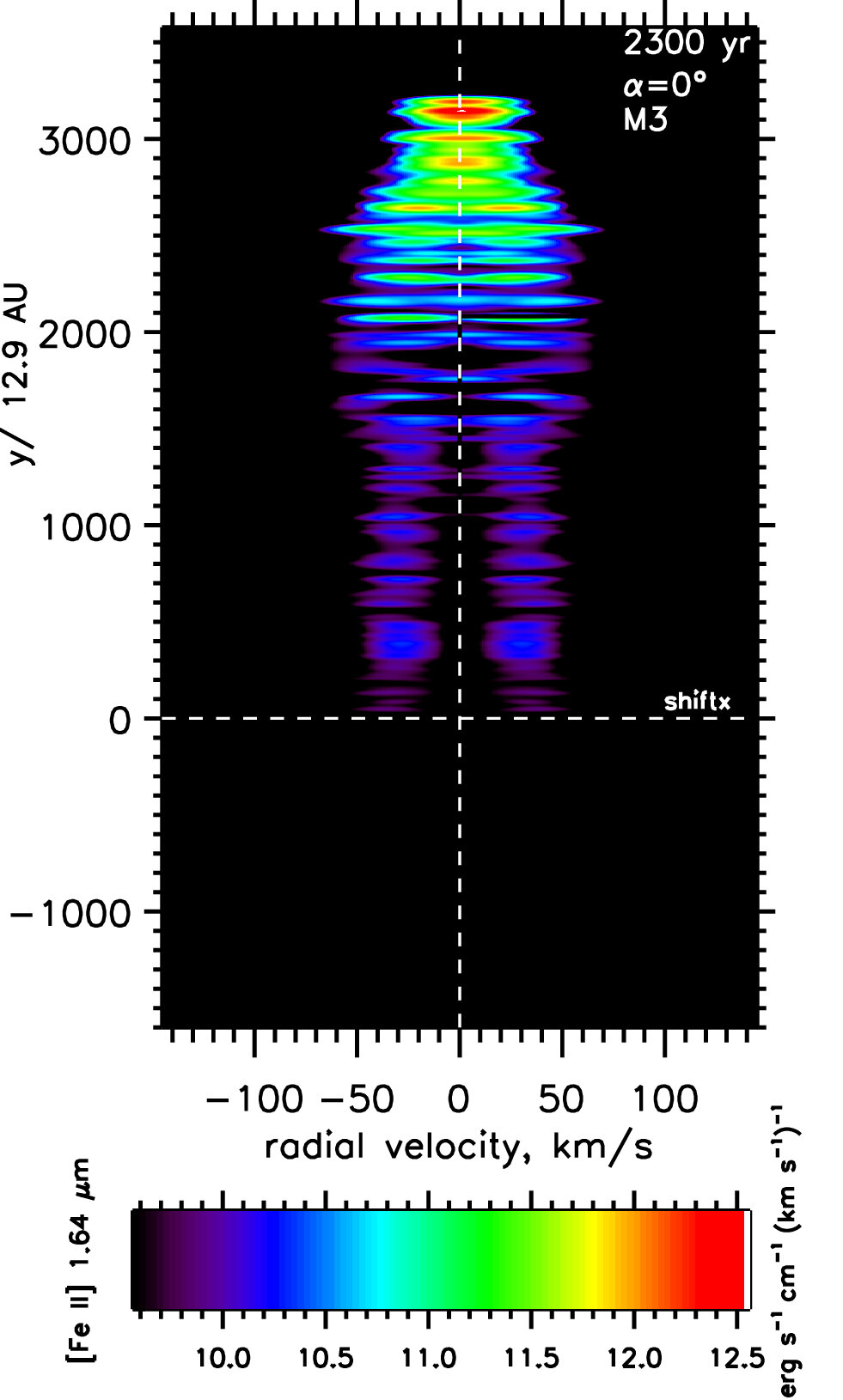}}
\hspace{\fill}
   \subfloat[\label{mt-simtask}][M4: (AWAA).]{%
      \includegraphics[width=0.3\textwidth]{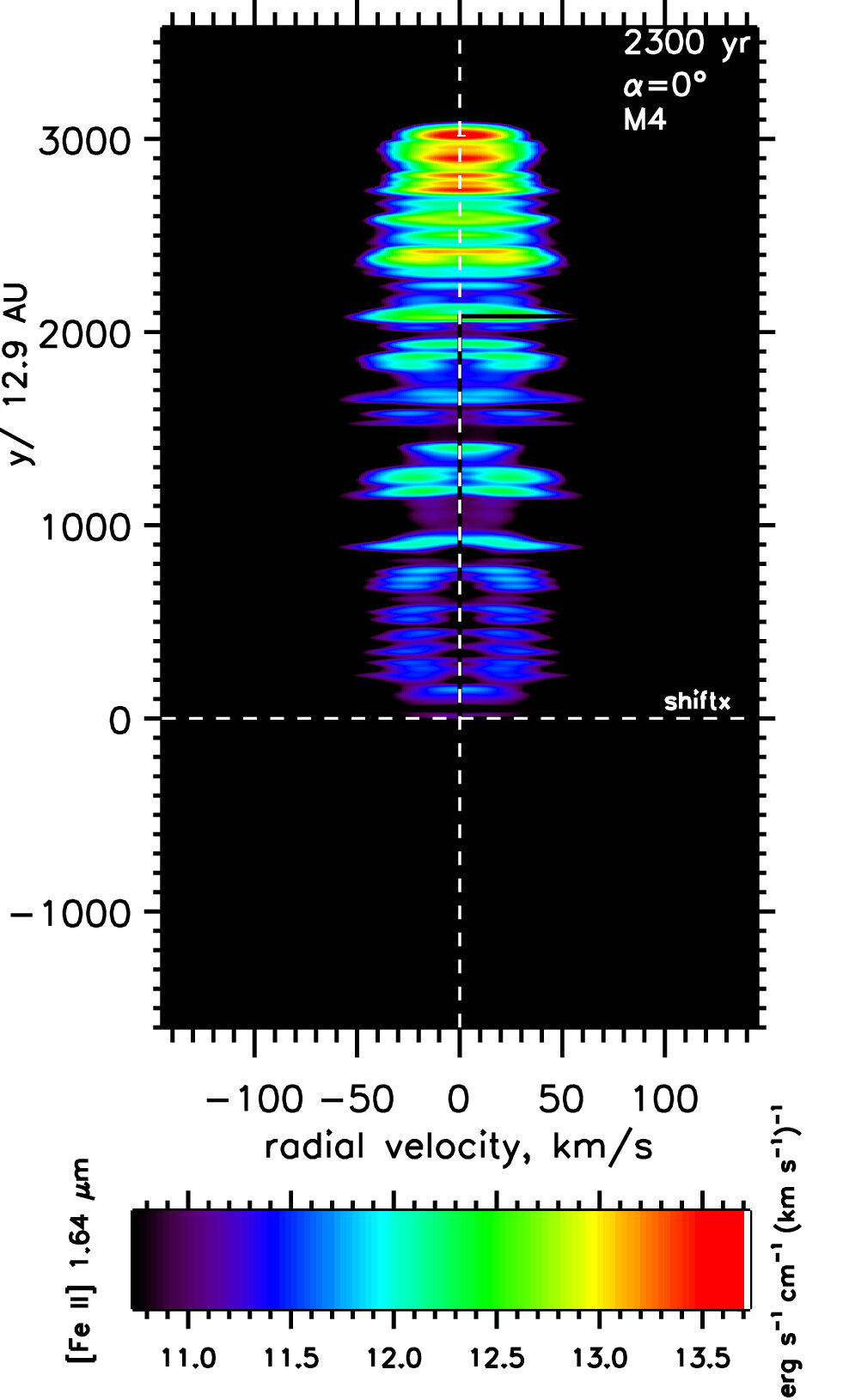}}\\
\caption{\label{2.1atpv0_FeII}Position-Velocity diagrams for the $\mathrm {[Fe\,{\sc II}] \, 1.64\,\mathrm{\mu m} }$ emission from the 2:1 ellipsoidal wind with the long axis in the plane of the sky 
($\alpha = 0^\circ$)  at a late stage of wind expansion. The three composition models are as indicated.}
\end{figure*}
\begin{figure*}
\subfloat[Subfigure 3 list of figures text][M2: 30$^\circ$, 2,300 yr.]{
\includegraphics[width=0.3\textwidth]{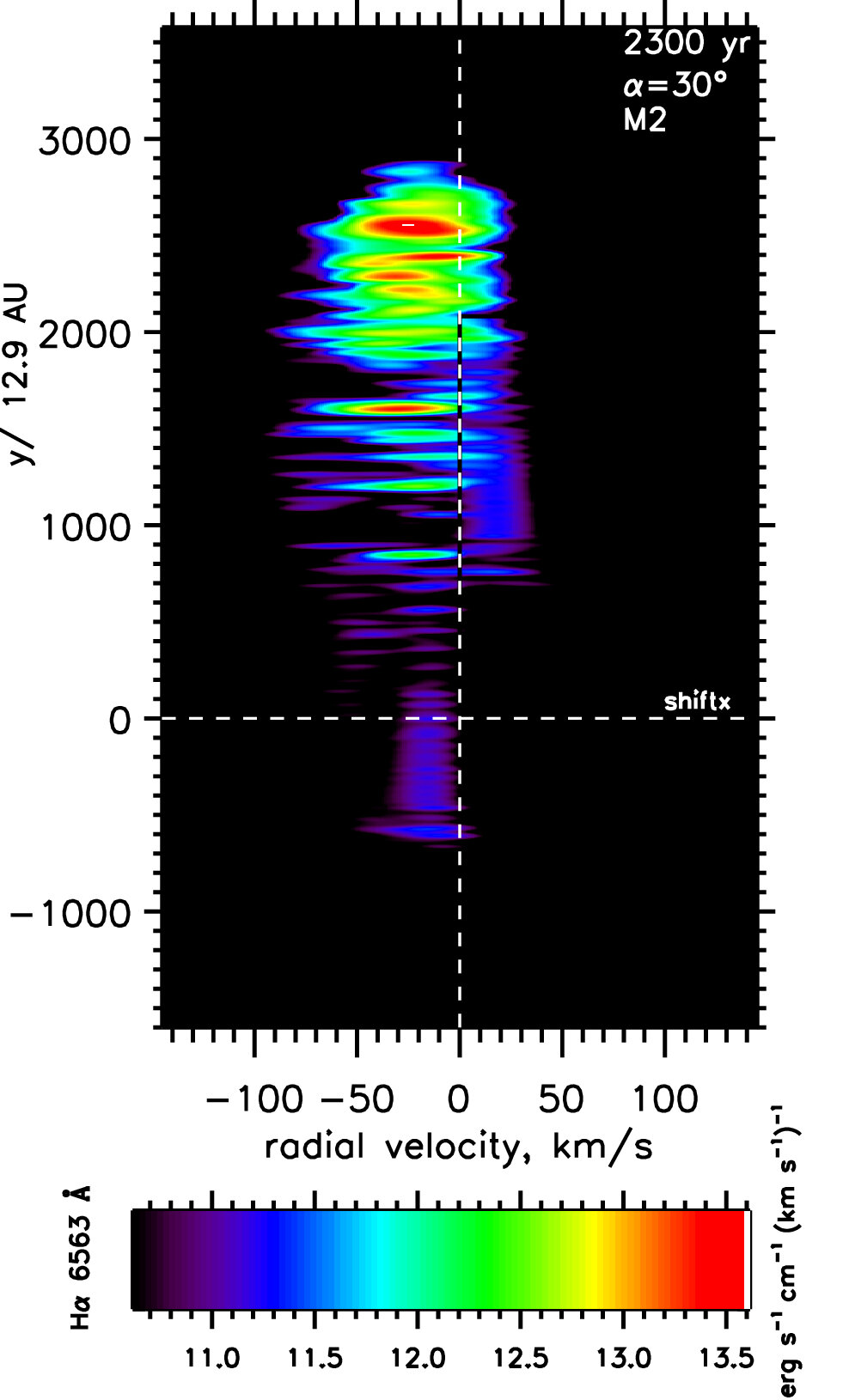}
\label{fig:subfig3d}}
\subfloat[Subfigure 4 list of figures text][M2: 60$^\circ$, 2,300 yr]{
\includegraphics[width=0.3\textwidth]{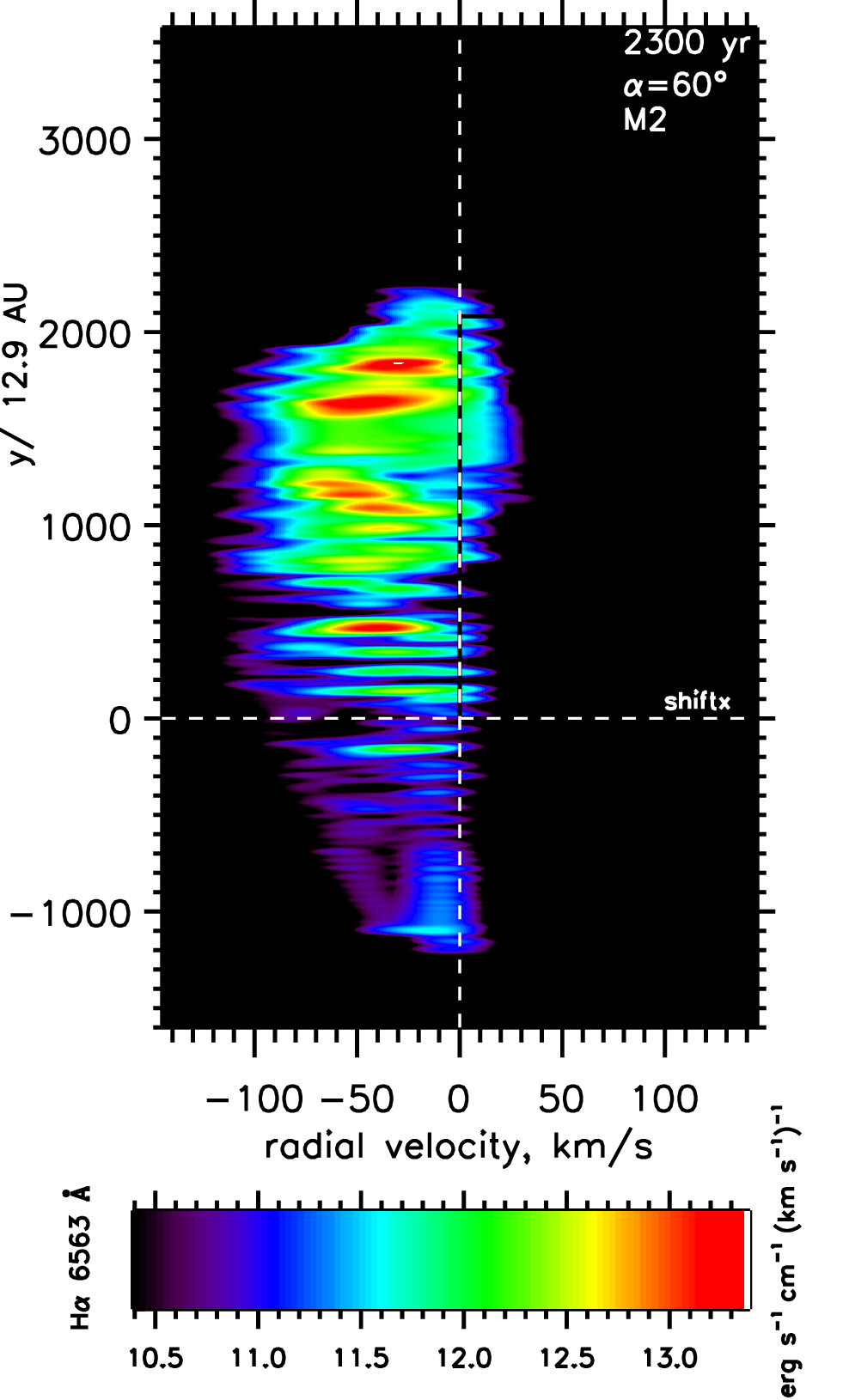}
\label{fig:subfig4d}}
\caption{H$\alpha$ PV diagrams for a molecular wind impacting an atomic ambient gas. This is Model M2 with a 2:1 $V_{w}\sim 140 \,km\, s^{-1}$ wind orientated out of  the plane of the sky at the indicated time}
\label{M2-Halpha-21-4panel}
\end{figure*}
\begin{figure*}
\subfloat[Subfigure 3 list of figures text][M2: 30$^\circ$, 2,300 yr.]{
\includegraphics[width=0.3\textwidth]{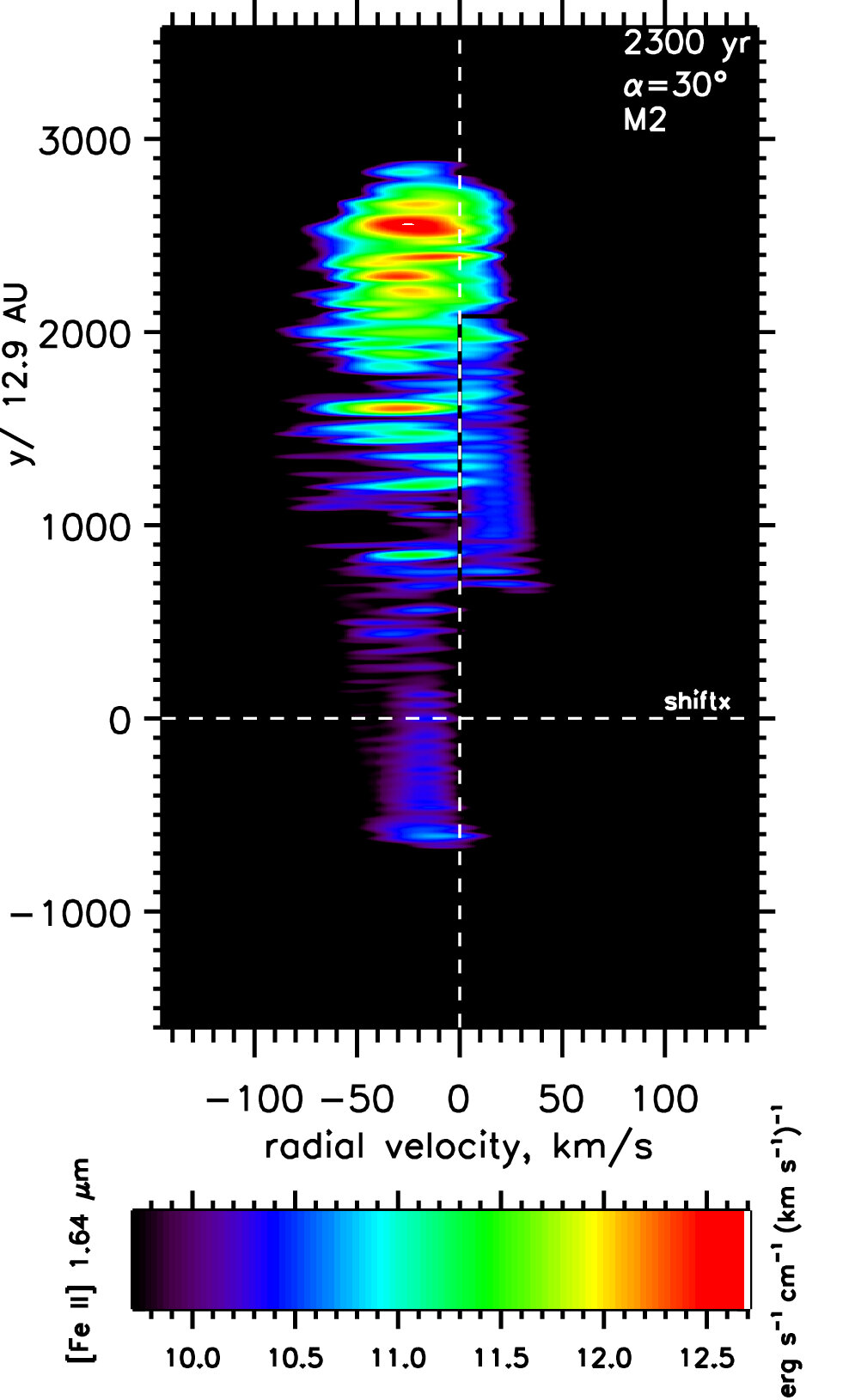}
\label{fig:subfig3d}}
\subfloat[Subfigure 4 list of figures text][M2: 60$^\circ$, 2,300 yr]{
\includegraphics[width=0.3\textwidth]{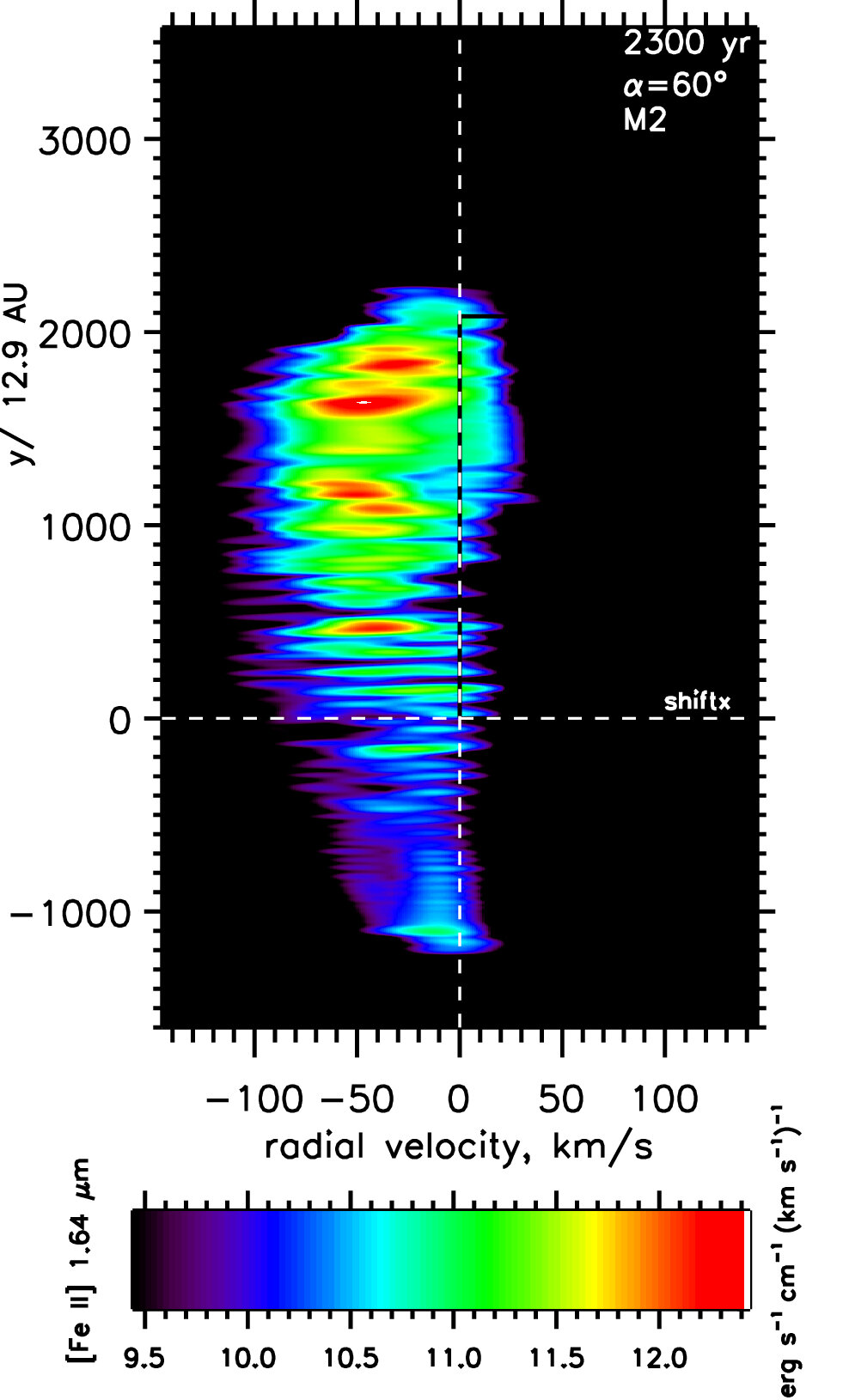}
\label{fig:subfig4d}}
\caption{$\mathrm {[Fe\,{\sc II}] \, 1.64\,\mathrm{\mu m} }$ PV diagrams for a molecular wind impacting an atomic ambient gas. This is Model M2 with a 2:1 $V_{w}\sim 140 \,km\, s^{-1}$ wind orientated out of  the plane of the sky at the indicated time}
\label{M2-FeII-21-4panel}
\end{figure*}
\begin{figure*}
\subfloat[Subfigure 3 list of figures text][M3:  30$^\circ$, 2,300 yr.]{
\includegraphics[width=0.3\textwidth]{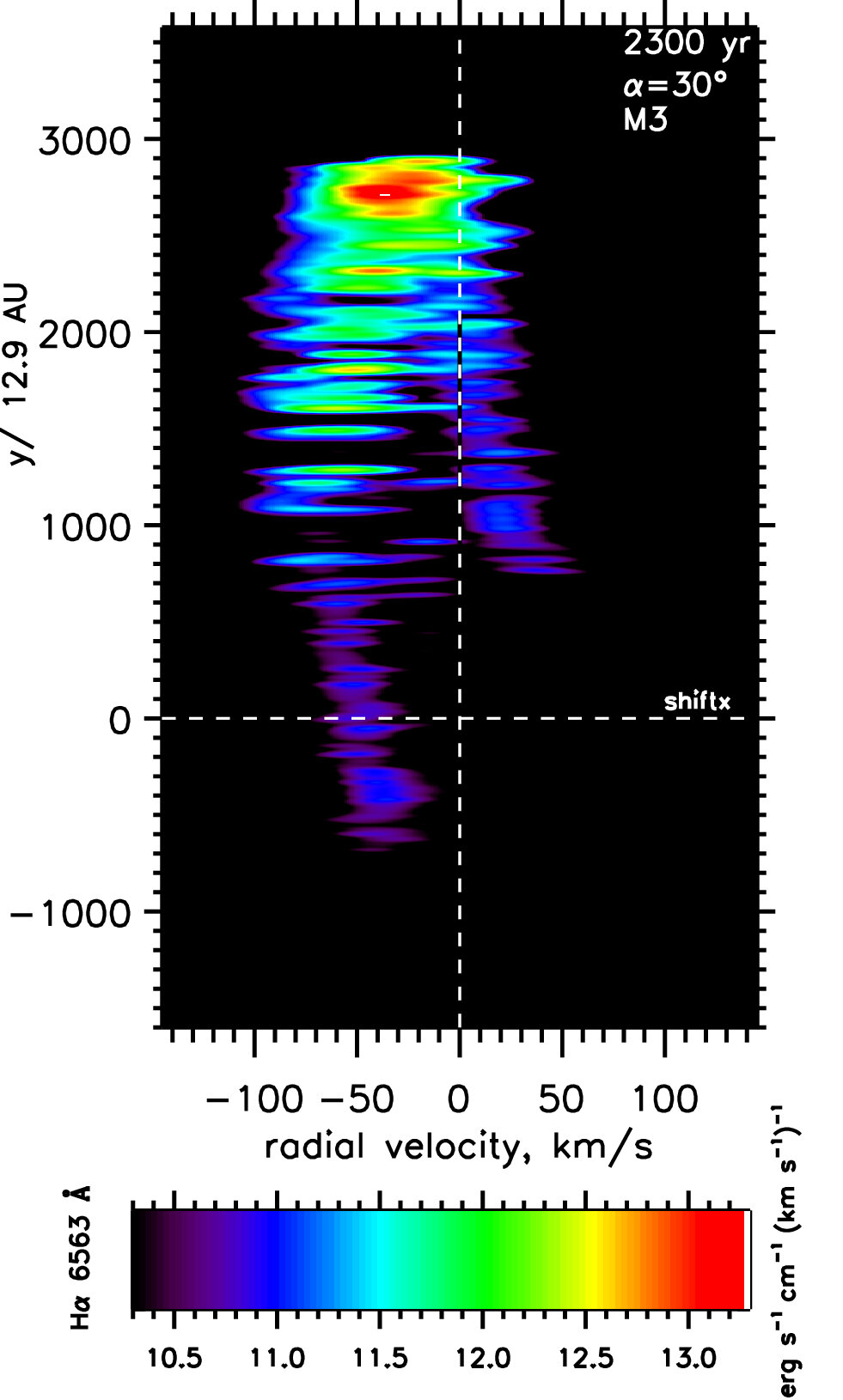}
\label{fig:subfig3d}}
\subfloat[Subfigure 4 list of figures text][M3: 60$^\circ$, 2,300 yr.]{
\includegraphics[width=0.3\textwidth]{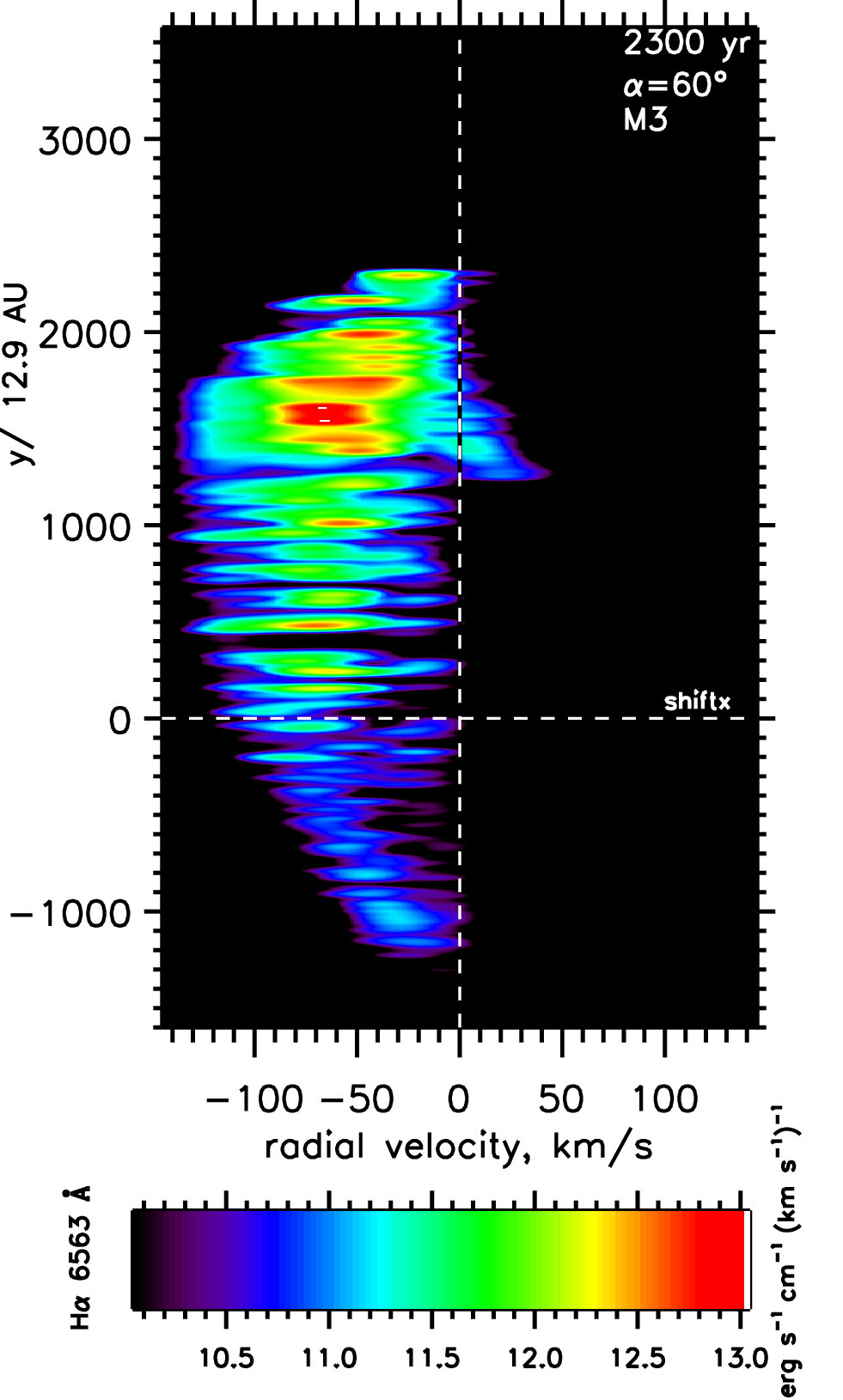}
\label{fig:subfig4d}}
\caption{H$\alpha$ PV diagrams for an atomic wind impacting a molecular ambient gas. This is Model M3 with a 2:1 $V_{w}\sim 140 \,km\, s^{-1}$ wind orientated out of  the plane of the sky at the indicated time}
\label{M3-Halpha-21-4panel}
\end{figure*}

\begin{figure*}
\subfloat[Subfigure 3 list of figures text][M3:  30$^\circ$, 2,300 yr.]{
\includegraphics[width=0.3\textwidth]{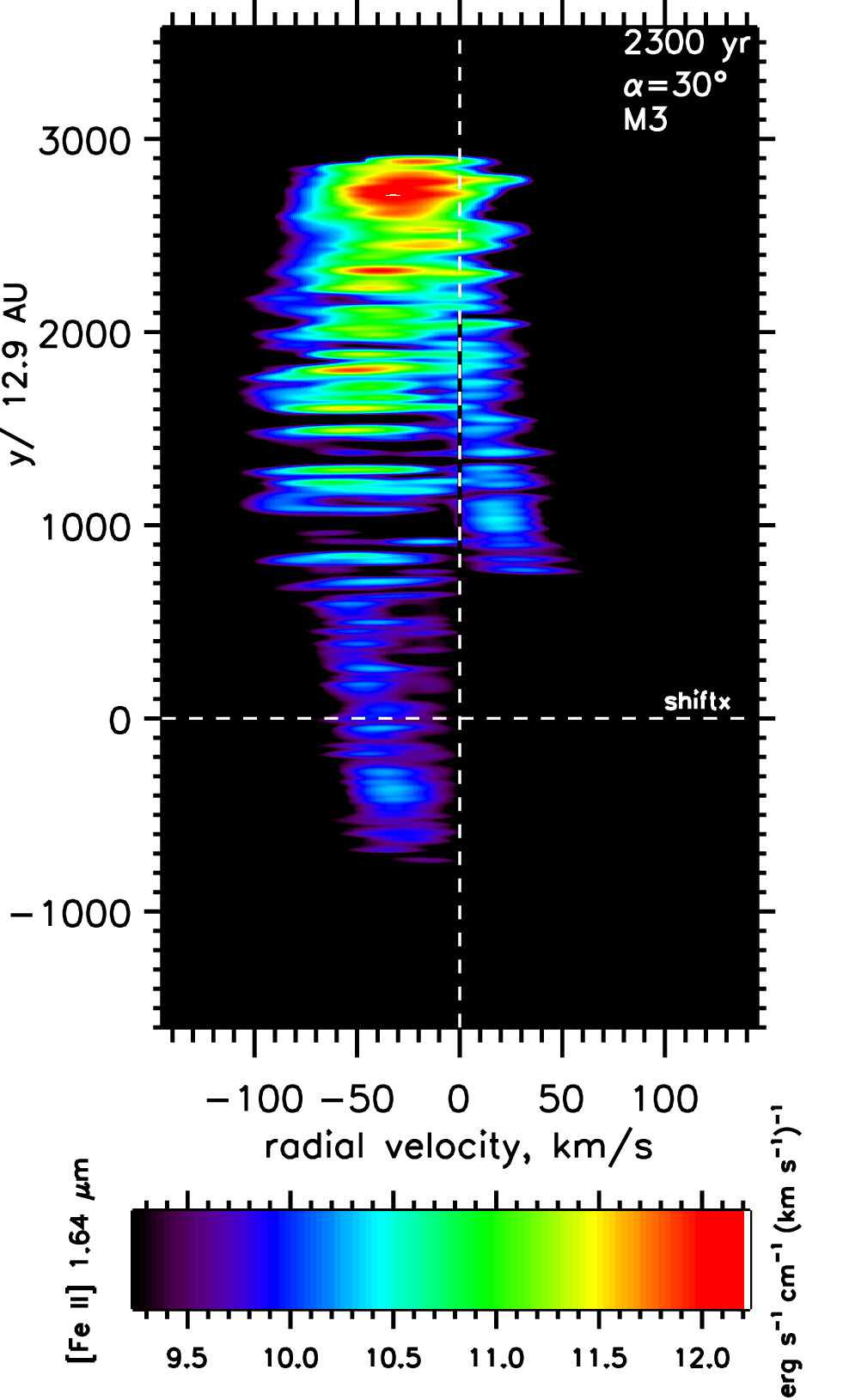}
\label{fig:subfig3d}}
\subfloat[Subfigure 4 list of figures text][M3: 60$^\circ$, 2,300 yr.]{
\includegraphics[width=0.3\textwidth]{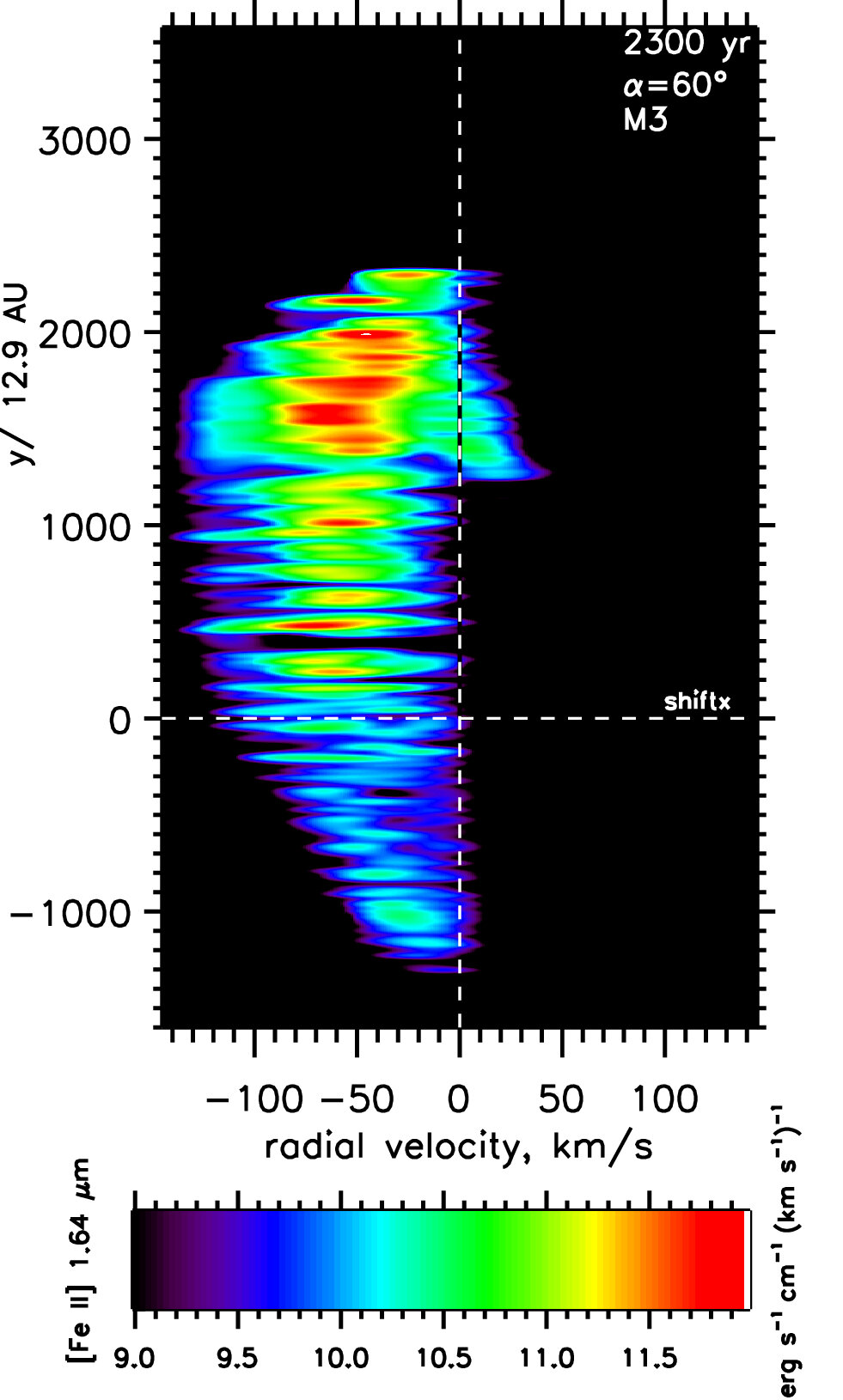}
\label{fig:subfig4d}}
\caption{$\mathrm {[Fe\,{\sc II}] \, 1.64\,\mathrm{\mu m} }$ PV diagrams for an atomic wind impacting a molecular ambient gas. This is Model M3 with a 2:1 $V_{w}\sim 140 \,km\, s^{-1}$ wind orientated out of  the plane of the sky at the indicated time}
\label{M3-FeII-21-4panel}
\end{figure*}

\begin{figure*}
\subfloat[Subfigure 3 list of figures text][M4: 30$^\circ$, 2,300 yr.]{
\includegraphics[width=0.3\textwidth]{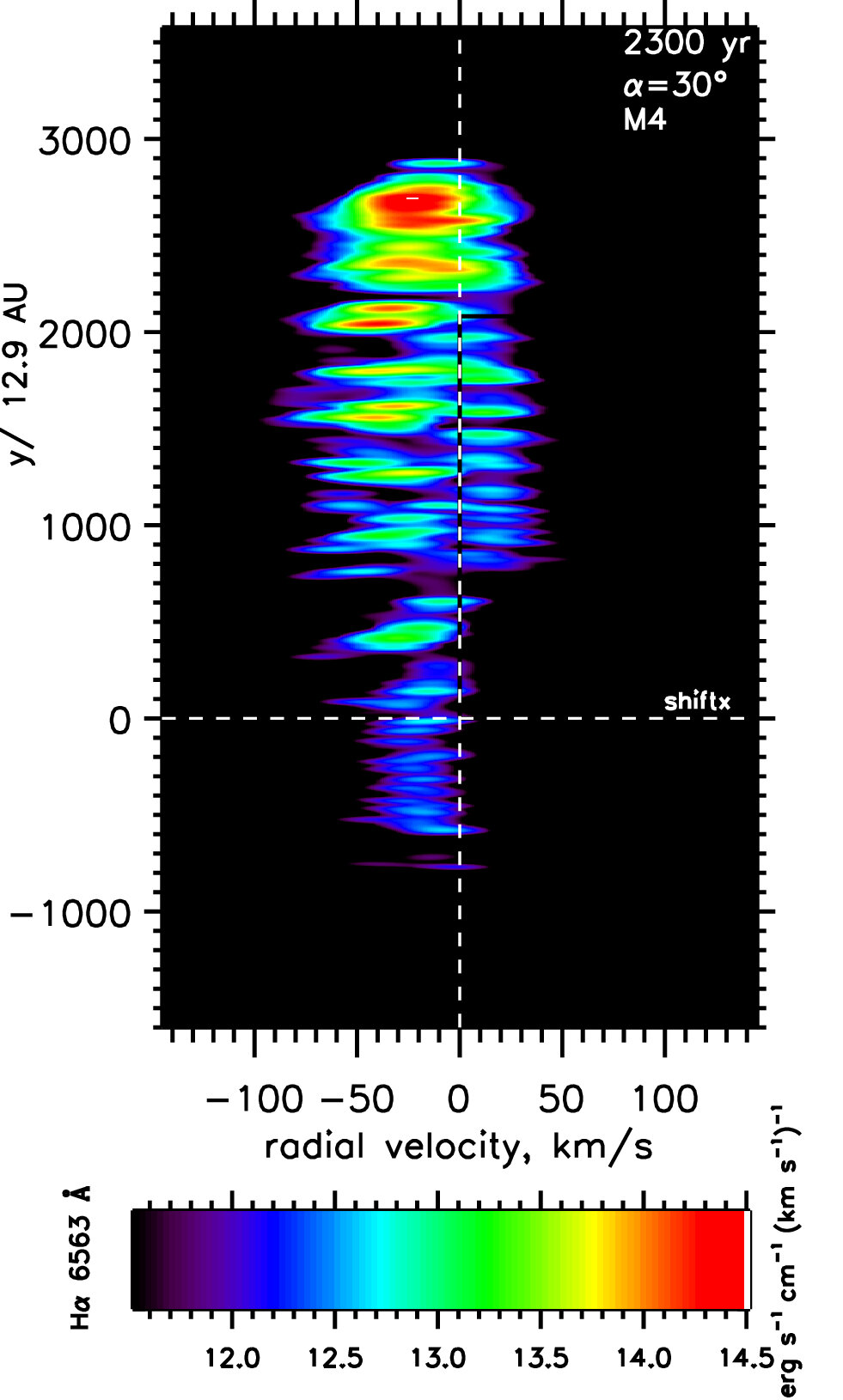}
\label{fig:subfig3d}}
\subfloat[Subfigure 4 list of figures text][M4: late time, 60$^\circ$, 2,300 yr.]{
\includegraphics[width=0.3\textwidth]{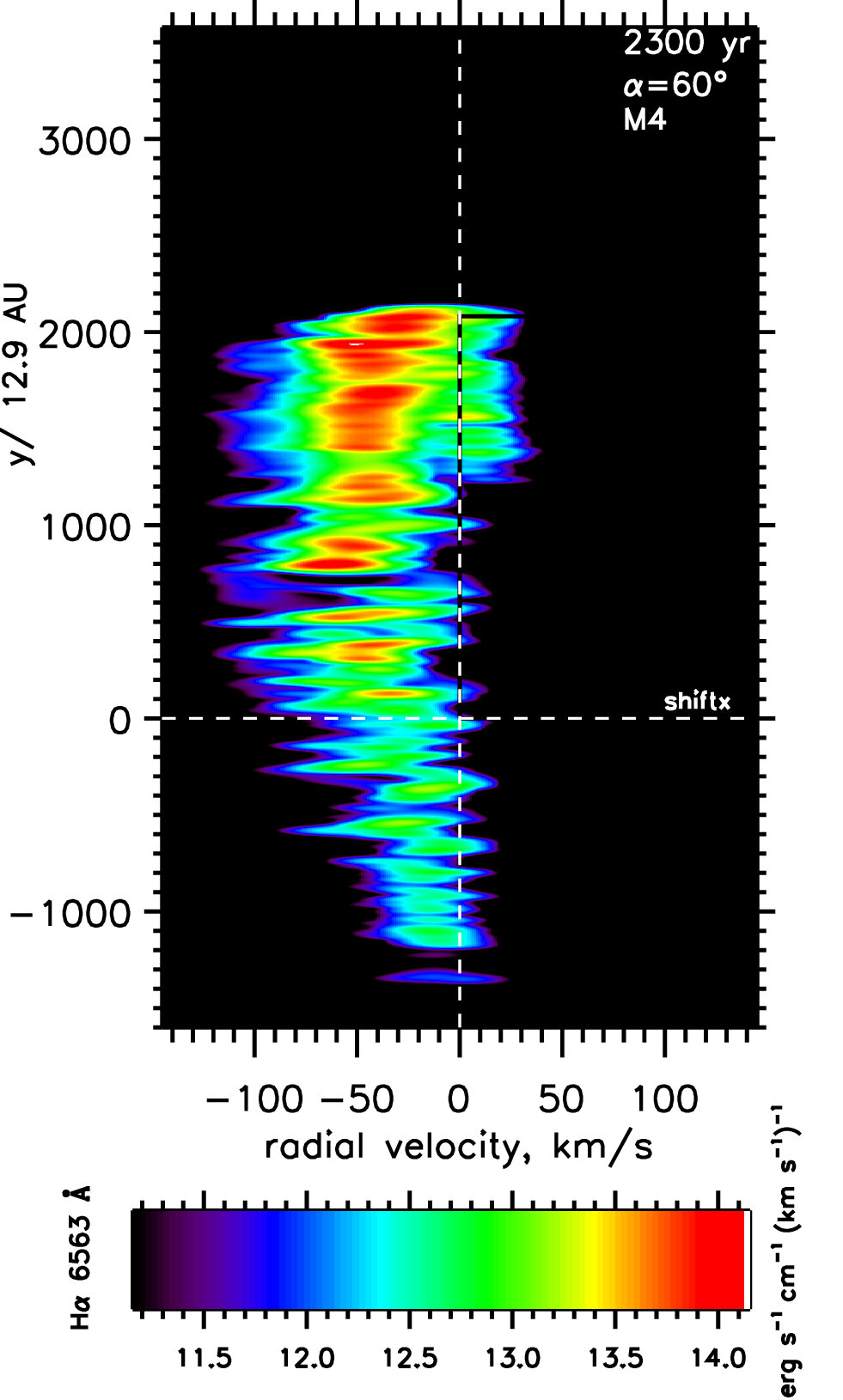}
\label{fig:subfig4d}}
\caption{H$\alpha$ PV diagrams for an atomic wind impacting an atomic ambient gas. This is Model M4 with a 2:1 $V_{w}\sim 140 \,km\, s^{-1}$ wind orientated out of  the plane of the sky at the indicated time}
\label{M4-Halpha-21-4panel}
\end{figure*}

\begin{figure*}
\subfloat[Subfigure 3 list of figures text][M4: 30$^\circ$, 2,300 yr.]{
\includegraphics[width=0.3\textwidth]{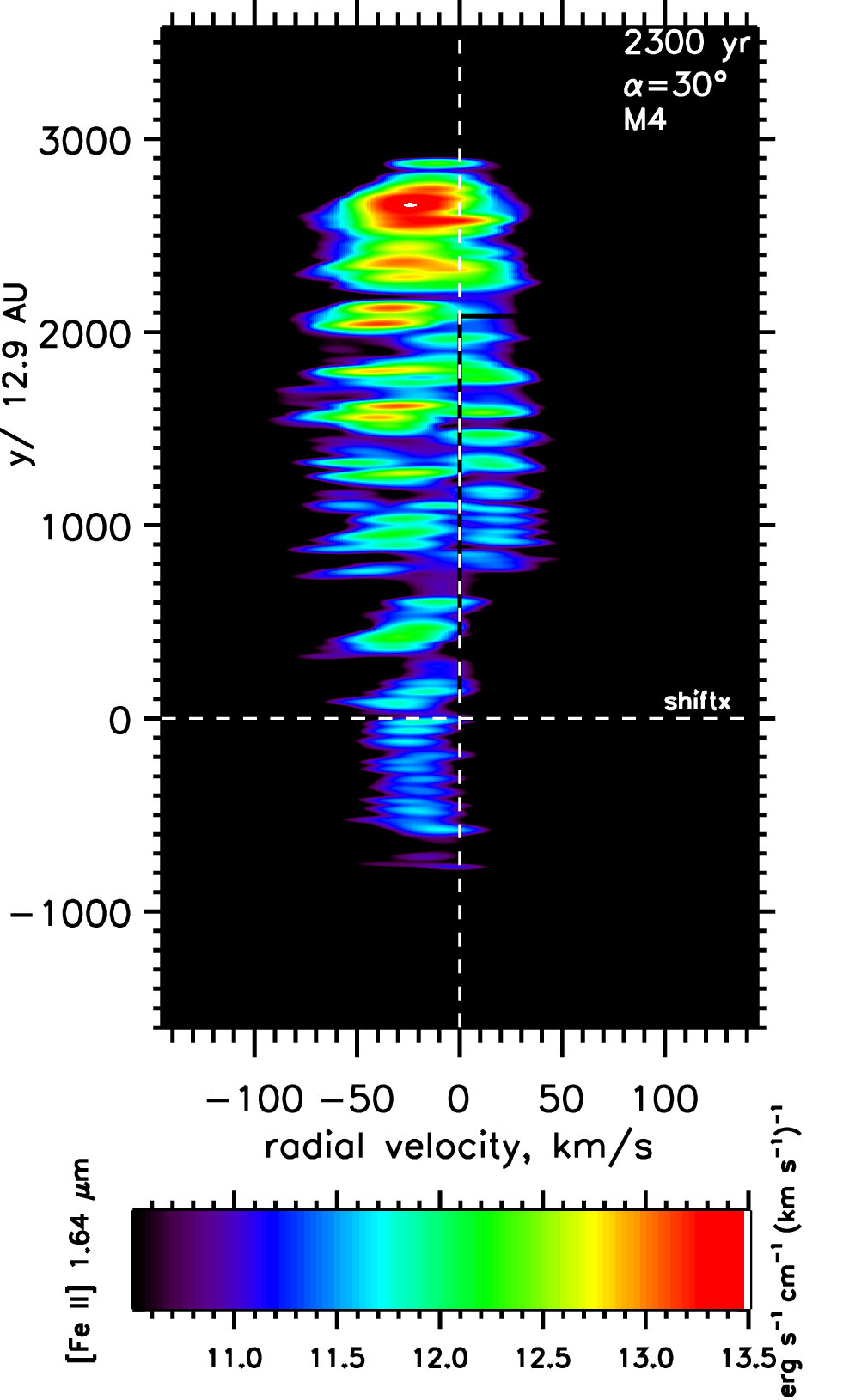}
\label{fig:subfig3d}}
\subfloat[Subfigure 4 list of figures text][M4: late time, 60$^\circ$, 2,300 yr.]{
\includegraphics[width=0.3\textwidth]{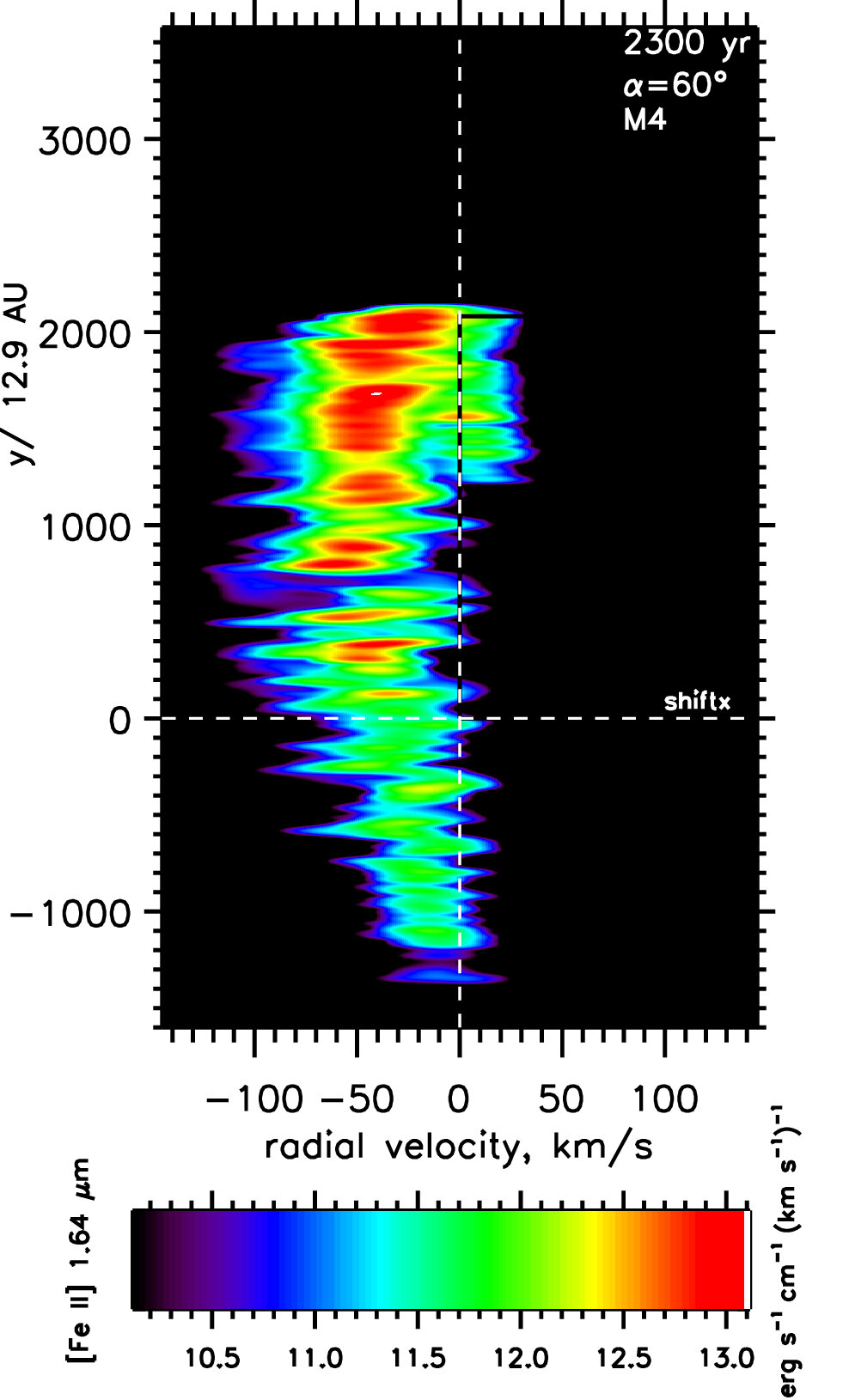}
\label{fig:subfig4d}}
\caption{$\mathrm {[Fe\,{\sc II}] \, 1.64\,\mathrm{\mu m} }$ PV diagrams for an atomic wind impacting an atomic ambient gas. This is Model M4 with a 2:1 $V_{w}\sim 140 \,km\, s^{-1}$ wind orientated out of  the plane of the sky at the indicated time}
\label{M4-FeII-21-4panel}
\end{figure*}

\begin{figure*}
\centering
   \subfloat[\label{genworkflow}][Model M2]{%
      \includegraphics[width=0.3\textwidth,height=0.25\textheight]{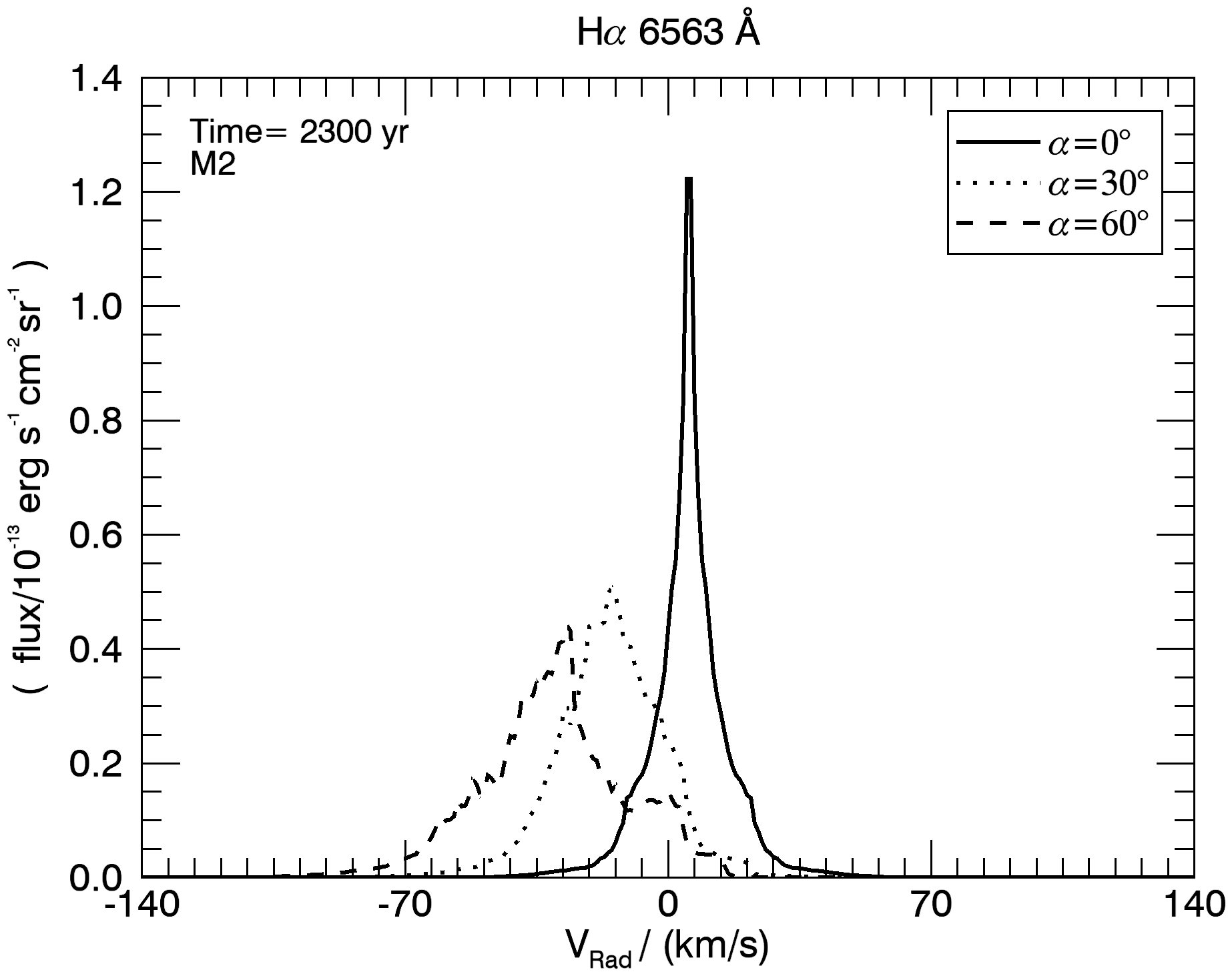}}
\hspace{\fill}
   \subfloat[\label{pyramidprocess}][Model M3]{%
      \includegraphics[width=0.3\textwidth,height=0.25\textheight]{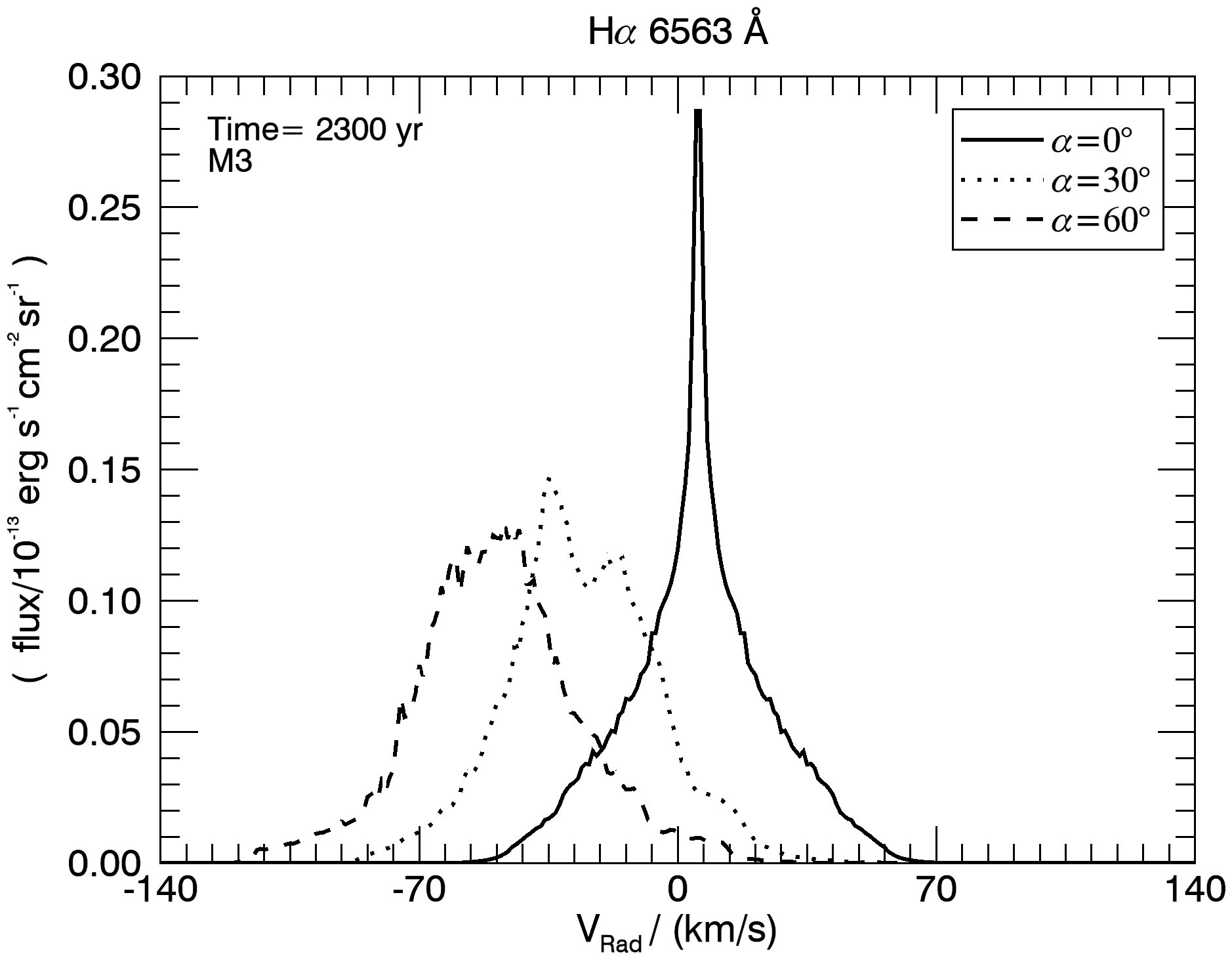}}
\hspace{\fill}
   \subfloat[\label{mt-simtask}][Model M4]{%
      \includegraphics[width=0.3\textwidth,height=0.25\textheight]{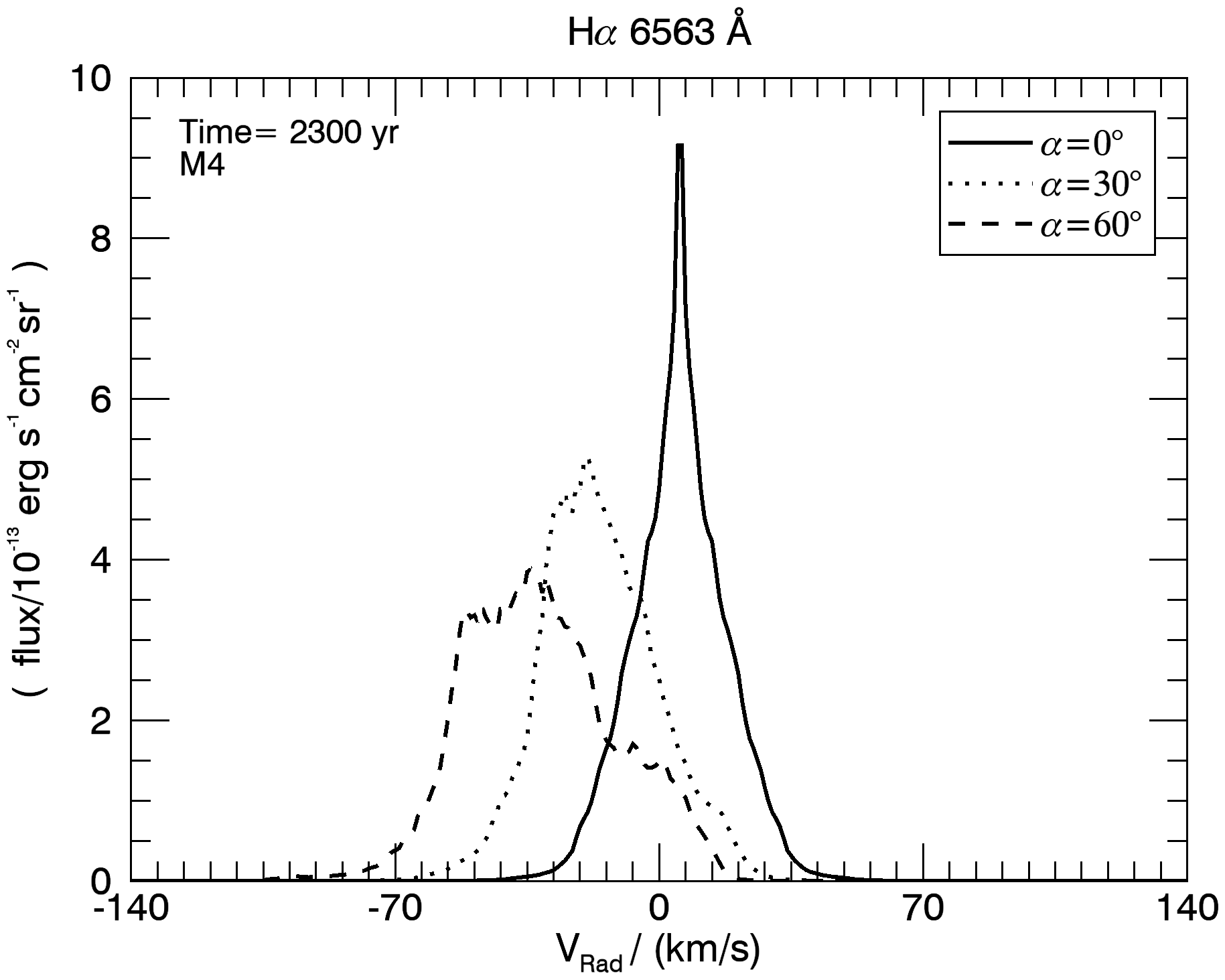}}\\
\caption[$\mathrm { [H \alpha]}$ line profiles by 2:1 elliptical wind]
{\label{lineprofile_Ha}Line profiles of the entire PPN in H$\alpha$ emission from  2:1 elliptical winds with the symmetry axis at  $0, \,30 \,\& \,60^{\circ}$ to the plane of the sky.}
\end{figure*}

\begin{figure*}
\centering
   \subfloat[\label{genworkflow}][Model M2]{%
      \includegraphics[width=0.3\textwidth,height=0.25\textheight]{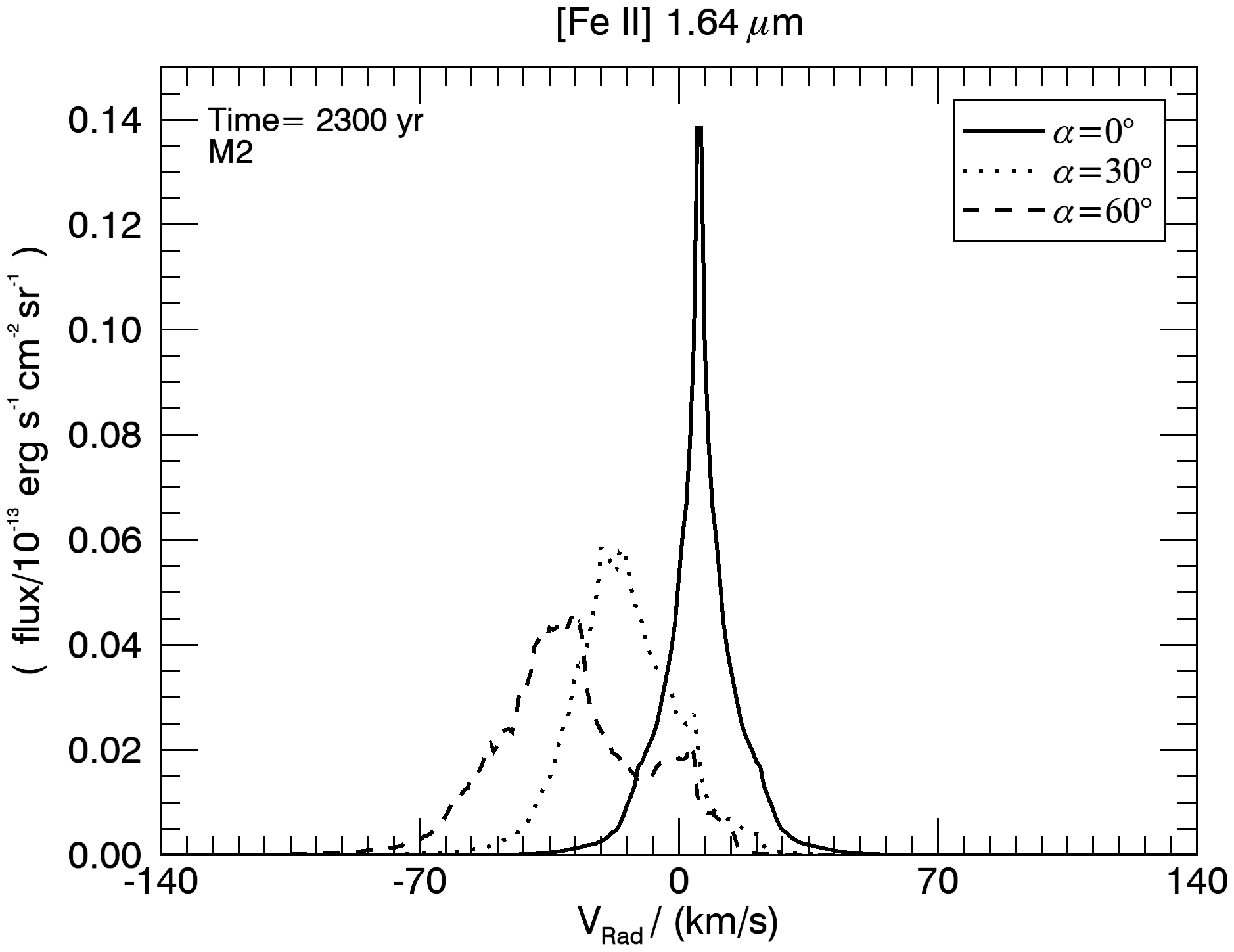}}
\hspace{\fill}
   \subfloat[\label{pyramidprocess}][Model M3]{%
      \includegraphics[width=0.3\textwidth,height=0.25\textheight]{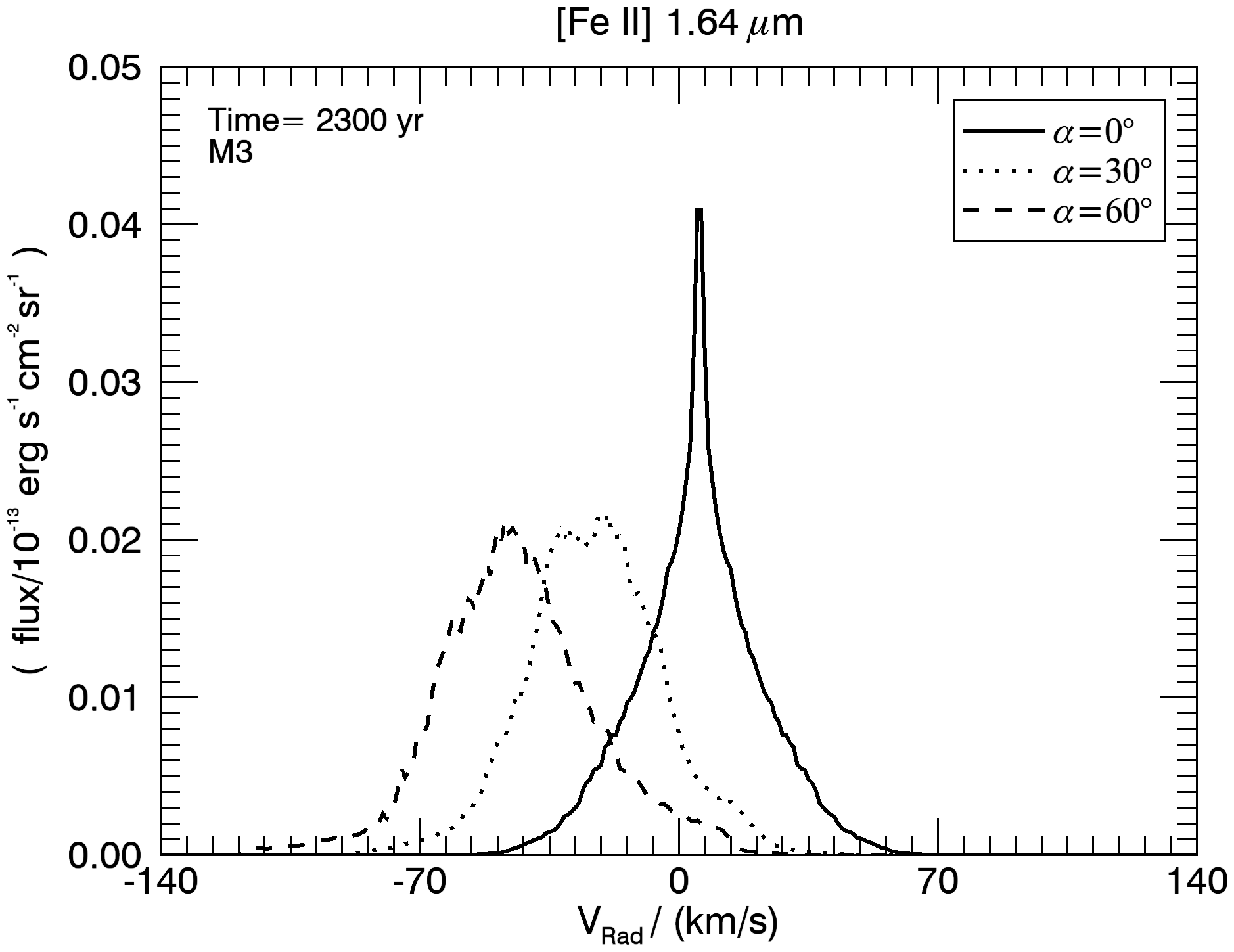}}
\hspace{\fill}
   \subfloat[\label{mt-simtask}][Model M4]{%
      \includegraphics[width=0.3\textwidth,height=0.25\textheight]{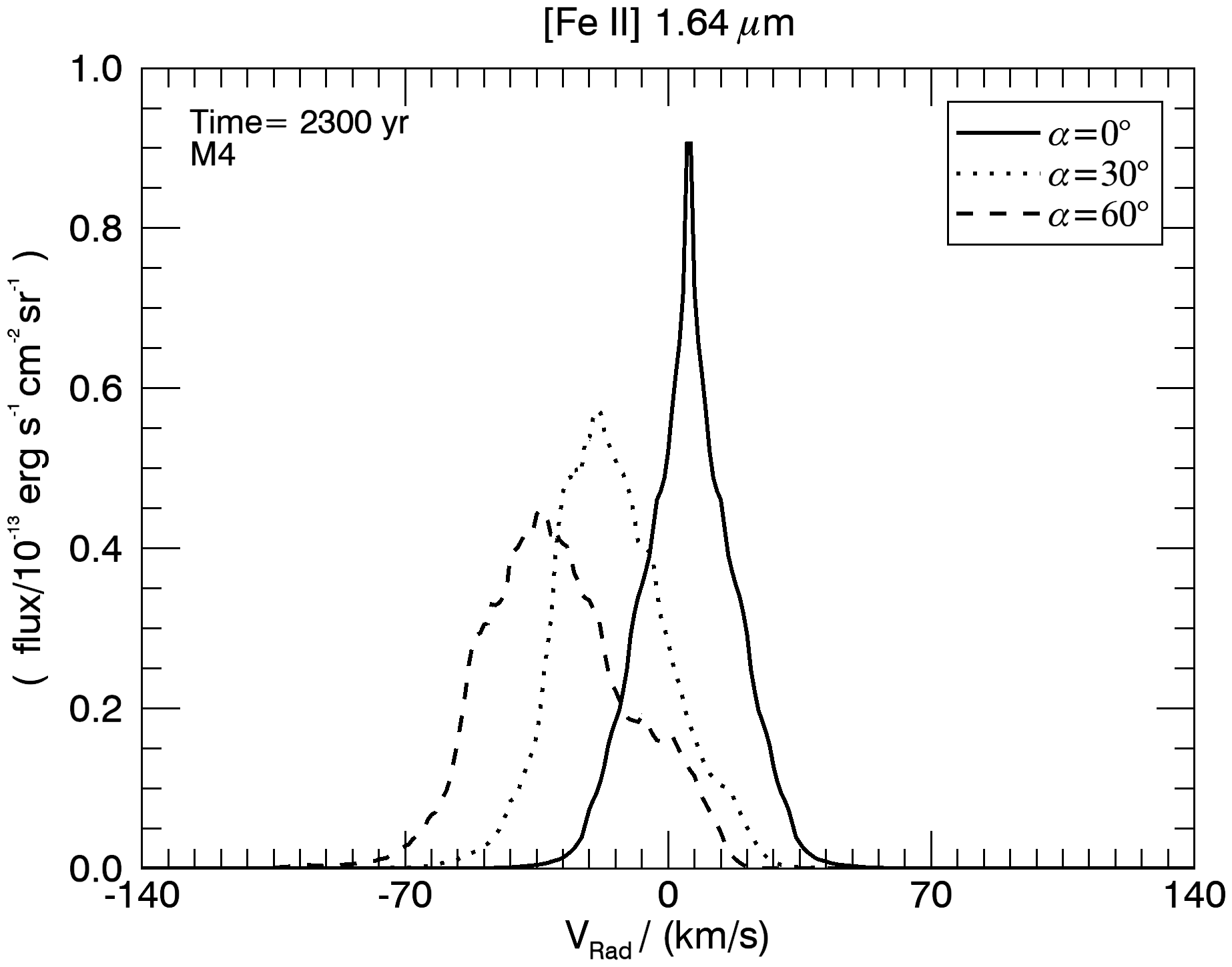}}\\
\caption[$\mathrm {[Fe\,{\sc II}] \, 1.64\,\mathrm{\mu m} }$ line profiles by 2:1 elliptical wind]
{\label{lineprofile_FeII}Line profiles of the entire PPN in $\mathrm {[Fe\,{\sc II}] \, 1.64\,\mathrm{\mu m} }$ emission from  2:1 elliptical winds with the symmetry axis at  $0, \,30 \,\& \,60^{\circ}$ to the plane of the sky.}
\end{figure*}

\section{ EMISSION MAPS } \label{results}
Our aim here is to identify stages of emission from shock tracers such as $\mathrm {[Fe\,{\sc II}] \, 1.64\,\mathrm{\mu m} }$ and $\mathrm{ [S\,{\sc II}] \,6716}$ {\AA} forbidden lines, the $\mathrm{[O\,{\sc I}]\,6300}$ {\AA} airglow line and H$\mathrm \alpha\,6563$ {\AA} emission. The simulated maps resulting from the four chosen atomic transitions are displayed consecutively in Fig.~\ref{2.1atEm_All}, modelling the 2:1 ellipsoidal wind. A 2--D convolution Gaussian smoothing operator is applied for all models to resolve numerical errors at the forward shock with standard deviation of $\mathrm{\sigma=20}$ zones, and flux spread over $\mathrm{4\sigma}$. The above convolution setting enables observation comparison within optimum computational run time.  

These panels can be compared to the 4:1 ellipsoidal winds of Fig.~\ref{OI-60deg-140} - \ref{FeII-60deg-140} at 60$^\circ$ to the sky plane. The resulting emission emanate from three different compositional models that are split into the following panels: (a) a molecular wind into a uniform  atomic medium (Model M2), (b) an atomic wind into a uniform molecular ambient  (Model M3) and  (c) an atomic wind-ambient run (Model M4). In each case of Fig.~\ref{OI-60deg-140} - \ref{FeII-60deg-140} the wind is 4:1 ellipsoidal in velocity with an axial speed of 140 km\,s$^{-1}$ with the axis directed at 60$^\circ$ to the sky plane. The lower elipticity 140 km\,s$^{-1}$ wind in Fig.~\ref{2.1atEm_All}, is displayed in the plane of the sky.   

Regarding Fig.~\ref{OI-60deg-140} - \ref{FeII-60deg-140} the first remark is that the emission occurs in  projected rings due to the axial symmetry. Secondly, most emission occurs in case (c) where both atomic wind and ambient media contribute. The peak emission shows a degree of variability corresponding with chemical composition. 
It is the two higher excitation species which show stronger peaks in Model M4. Hence, it is not only the completely atomic interface but also the more uniform advancing shell which generates stronger shock fronts than the distorted oblique shocks when molecules are involved.

The uniform ISM simulations presented here show that Model M4 displays optical emission from  both the flanks and the leading bow shock: the surface brightness is much smoother. In contrast, the emission is concentrated toward the faster leading edge of the ellipsoid for all four tracers in Model M2 where only the ambient medium is atomic. The peak emission in all cases is set back from the leading edge, independent of forward/backward shock structure, serving as further evidence to suggest that the emissivity distribution is mainly determined by the source geometry, as supported by time-dependent flux animations\,\footnote{ \url{http://astro.kent.ac.uk/~in32/ppn.html} }.

The integrated atomic emission  for each of the four lines are displayed in Fig.~\ref{atemisOI} -- Fig.~\ref{atemisFeII}. The panels show the time-dependent luminosity variation with Model type, axial speed and wind ellipticity. First to note is that there is a drop in flux with increase in ellipticity. This can be expected since the advance speed along the axis is held as constant.

All computed models show a similar correlation between axial velocity and magnitude of resultant emission. The line luminosities displayed in the left-hand panels representing Model M2 all start off with an increase in flux. Clearly, the fast molecular wind is sweeping and heating up atomic ambient gas at an increasing rate. However, the wind is hindered and the speed of advance decreases. Therefore, the flux eventually falls, except for the low-excitation lines in the low wind speed case. 
For Model M3 with the atomic wind, the line luminosities start high but then fall rapidly before reaching steady magnitudes. Overall, Model M3 generates a plateau of emission with time. 
  
At later stages, the wind shows less emission from the wings and lower flux values from the cap, where the dense molecular wind impinges on atomic medium, this set up enables the wind cap to obtain lower flux values of one magnitude, relative to it's inverted counterpart, Model M3. The atomic material surrounding the wind's cap in M2 gets excited and shocked by $\mathrm{H_2}$ molecules in the wind, setting off emission in optical range. 
 
 The line luminosities toward the end of these runs, where the PPN has expanded to the scale of 0.1\,pc, show a consistent pattern. Generally, the M4 fully atomic run generates more emission than the M2 run by a factor of 10, significantly more than one would expect from adding together the atomic wind component of M3 to the atomic ambient of M2. Hence, the production of the thick turbulent shell intensifies the energy channelled into the strong atomic lines.

\section{Spectroscopic Properties}
The radial velocity shift of spectral lines often provides the information on the third dimension for objects beyond our reach. Long-slit spectroscopy can be presented as channel maps or more succinctly as Position-Velocity diagrams. Simulated optical Position-Velocity (PV) diagrams of 140~km~s$^{-1}$ outflows are displayed in Figs.~\ref{2.1atpv0_Ha}-\ref{M4-FeII-21-4panel} . All atomic compositional models are shown at three orientations. 

A direct comparison of the three compositional models is displayed in Fig.~\ref{2.1atpv0_Ha}. This demonstrates that the models can be distinguished. The left panel shows a quite narrow line along the slit due to the relatively small acceleration of the ambient atomic gas. In contrast, if the emission is dominated by an atomic wind, as in the middle panel, then a broad line appears in the flanks where the fast wind is shocked and decelerated at the interface. The full line width in the flanks can reach 100~km~s$^{-1}$.

The PV diagram in the full atomic simulation differs remarkably, consistent with the different dynamical behaviour. The right panel of  Fig.~\ref{2.1atpv0_Ha} shows that the thickening turbulent shell generates much stronger atomic emission that is distributed far back into the flanks. The radial velocities are moderate and quite centre-filled for the wind orientation displayed. Thus, the pure atomic Model M4 (AWAA) generates higher flux densities for both the wind cap and wings of the outflow with emission being contributed by both ambient and wind.

The PV diagrams are equally interesting when the wind axis is orientated toward the line of sight. As shown in Fig.~\ref{M2-Halpha-21-4panel} for Model M2, the PV diagram is not only shifted to the blue but also spread more evenly along the axis with several notable peaks spaced out. At 60$^\circ$ out of the plane, the emission is a distinct arc with the peak well behind the leading edge. The peak is now also blue-shifted relative to the leading edge. 

For the atomic wind, Fig.~\ref{M3-Halpha-21-4panel} for Model M3, the blue-shift is much greater and the arc extends back to the star and across to the other side which would be superimposed on the receding lobe. In a symmetric PPN, this would appear as both blue and redshifted arcs close to the symmetry plane or mid-plane of the configuration.

For the full atomic run, Fig.~\ref{M4-Halpha-21-4panel}, the PV diagram displays an arc which remains bright and strong all the way along. It should also be noted that by adding together two lobes of a double source (i.e. the full ellipsoid) would yield a distorted ellipse, rotated off-axis.

  Many of the above properties are confirmed by studying  the integrated line profiles, as displayed in    
Fig.~\ref{lineprofile_Ha}, where the solid, dotted \& dashed lines correspond to the three indicated wind
orientations, and the models M2, M3  \& M4 are displayed from left to right. It is clear that the atomic wind can create wide lines with shifts approaching the wind speed itself. Even though the wind gets compressed and slowed when passing through the shock, the shock itself is not transverse to the wind, but highly fragmented and distorted into oblique components. Hence the shock-heated wind can maintain a high radial speed. This effect is not prominent in the pure atomic flow because the turbulent shell is not distinctly fragmented. Instead, the energy is channelled into lower-speed turbulence, producing a much higher peak emission, as shown in the right panel.

Finally, it can be noted that many of these properties are apparent in the 4:1 ellipsoidal wind PV and line profile diagrams but not to such an obvious degree although still recognisable.

\section{Discussion}

Near-infrared  observations have broadened our knowledge of PPN in which  H$_2$ emission has been mapped and compared to the atomic emission \citep{2018ApJ...859...92F}. 
It has also been found  that a large fraction of PN may remain undetected in the optical, requiring H$_2$ imaging surveys  \citep{2018MNRAS.479.3759G}. This motivates us to explore the set of simulations in atomic emission lines. Our aim here is to search for structural similarities or differences which may be related to the principle factors of wind speed, wind ellipticity and molecular content. Other factors remain constant or uniform, although we are aware that these limit the applicability. Future  simulations are intended to remove the restrictions, and take off from where these end.

There are many elliptically shaped  planetary nebula that display such rough, broken shell structure, filled with fingers, globules and fragments. It is not clear that such structures could evolve from the earlier wind interaction phase simulated here. Examples include NGC\,1501 \citep{2000A&A...361.1112S}, NGC\,3132 \citep{2000ApJ...537..853M} NGC\,6751 \citep{2010ApJ...722.1260C} and NGC\,6369 \citep{2012MNRAS.420.1977R} which present Hubble Heritage images supplemented with atomic position-velocity diagrams. However, these structures arise in the PN phase where extreme ultraviolet radiation is likely to sculpt any dense clumps into cometary shapes. However, it is difficult to resolve such fine-scale structure in PPN and we resort to lower resolution images and position-velocity diagrams in our analysis.
\subsection{Comparison of hydrodynamic wind-shell interactions}
\label{windyshell}

In Paper \Rmnum{1}, we demonstrated that the shocked interface of a PPN with a molecular medium will distort into corrugations which grow into protruding filaments. On the other hand, the compressed shell between two atomic media will develop into a turbulent layer that thickens and warps. To understand this difference, we now consider the shell dynamics under the various possible outflow conditions.

Persistent supersonic  outflows have been the subject of many hydrodynamic simulations with wide applications. The injected flows are allowed to evolve
 with precession, smooth pulsation and intermittent ejections  often imposed or superimposed. Common geometric forms include ellipsoidal/spheroidal,  partly collimated and  fully channelled into jets. The simulations then follow the interaction between the outflow and the ambient medium which may consist of atomic, molecular and/or relativistic gas. 

 There  are two critical factors which will determine the situation being simulated. The first factor is the cooling time in the shell which forms between the shocked wind and the shocked ambient medium. 
The second factor is the numerical resolution of the cooling layer. Simulations are often not able to resolve the layer and numerical effects rather than the physics can then determine the outcome. To ameliorate this, the simulations often presume very low densities and high speeds, so increasing the cooling length and stability. These simulations are relevant to the fast winds in the PN stage.
We can identify three broad cases from the many outflow simulations, as follows.

In Case 1, the shocked gas of one of the media collapses into a very thin dense shell. The shell breaks up through dynamical instabilities, forming fingers protruding into the opposing medium. Thin-shell instabilities as uncovered in the context of planetary nebula by \citet{1999ApJ...517..767G}   conform to this picture. 

In Case 2, the shocked wind remains hot. A hot cavity forms which will escape through bubbles or de Laval nozzles as jets, depending on the precise configuration \citep{1981Natur.293..277S}. Thus, extremely low density winds have been shown to create hot bubbles under PN conditions \citep{1987AJ.....94..671B,2014MNRAS.443.3486T}.

In Case 3, the thin shell is averted. The cooling is not effective in collapsing the gas into a thin shell because the cooling layer is thermally unstable. In this case, a thick turbulent shell forms and grows without the development of protrusions. Simulations of high resolution are necessary to resolve the compressed shell.
  
High resolution studies corresponding to Case 3 have been presented by \citet{2012A&A...547A...3V}. Here,  a fast wind of 2,000 km~s$^{-1}$ is shocked to provide a hot cavity that drives a dense shell. The shell is highly distorted and  irregular without the formation of dense fingers. The advancing shock front has a speed close to 100\,-\,200 km~s$^{-1}$ which falls into the range where thermal instability and dynamic overstability are present  \citep{1995ApJ...449..727S}.

The specific case studied here corresponds to ellipsoidal winds of high mass outflow into a uniform dense ambient medium. Details are provided in Paper \Rmnum{1}. Both the wind and ambient medium will cool in a time scale significantly shorter than the dynamical time of expansion. The outflow speeds and densities correspond to those expected in protoplanetary nebula and protostellar outflows from low-mass stars. 

With this simulation set-up, we are modelling the very early stages of outflow development before the fast PN wind such as simulated by  \citet{2004A&A...414..993P,2005A&A...431..963S}  is established.
The later fast atomic winds, hot bubbles, photoionisation and radiation pressure of planetary nebulae are not simulated with the present parameters.

We thus find that the simulations presented here are consistent with previous studies. A thin collapsed shell of shocked molecular gas under PPN conditions will degenerate into a regular pattern of roughly radial filaments. On the other hand, if only atomic gas is involved, the shell degenerates into an irregular pattern containing knots and filaments within a slowly thickening turbulent layer.

\subsection{Comparison to observations}

 The literature contains many models of planetary nebula which pertain to finding physical parameter fits to specific images. However, models based on full hydrodynamical simulations are rare and often employ resolutions which cannot adequately explore the shell stability or do not have the resolution to explore the range in temperature and density within the shell. Here, the code contains the major cooling and chemistry  functions which operate between 10 K and $\mathrm{10^{8} \,K}$. A compromise remains in the resolution of the hot atomic post-shock gas for which we approximate the ionisation and temperature structure.
 
The H$\mathrm \alpha\,6563$ {\AA} emission in Fig.~\ref{Halp-60deg-140} shows features similar to the ones computated by \citet{2004A&A...419..991V} for the PPN  Hen 3-1475 with extended bow shock wings produced by a collimated fast wind (CFW) of 400 km\,s$^{-1}$ expanding into uniform ISM. The authors also provide results for the same model expanding into non-uniform $\mathrm{r^{-2}}$ density profile ISM producing high emission values close to the central star and seemingly less bow shock emission due to low density medium parameters at large distances away from the stellar surface. 

Lower ionisation $\mathrm{ [S\,{\sc II}] \,6716}$ emission line from a 200 km\,s$^{-1}$ superwind associated with the PPN CRL\,618 was simulated by \citet{2014ApJ...794..128V} revealing synthetic maps with dynamic flux range of $10^{3} \, \mathrm{erg\,s^{-1}\,cm^{-2}sr^{-1}}$, similar to images displayed in Fig.~\ref{SII-60deg-140}. On the other hand, the predicted PV diagrams line flux for CRL\,618 is dominated by the ejected bullets in contrast with elliptical outflow presented here.
 
The slightly elliptical planetary nebula NGC\,1501 has been modelled as an irregular ellipsoid containing complex filamentary structure. It 
displays inward-pointing tails \citep{2000A&A...361.1112S} which suggests that this structure is formed due to Rayleigh-Taylor instabilities and wind-wind interaction. Position-velocity diagrams of H$\alpha$ show a quite uniform bow-shaped shell with maximum expansion speeds of approximately 55~km~s$^{-1}$. In our case, the corresponding Model M3 shows similar morphological features formed by wind-uniform ISM interaction generating the most well-defined shell structure in the simulated PV diagrams shown in Fig.~\ref{2.1atpv0_Ha}.  However, it is evident that this nebula has evolved well past the PPN phase and other physical processes are likely to have reshaped the environment.
 
 One can also question if ring structures such as those observed in  NGC\,6369 \citep{2012MNRAS.420.1977R} may be related to the ring structures uncovered here in H$\alpha$, although it is not clear that such structure would remain in three dimensional simulations and at the later PN phase. This nebula contains outward-pointing protrusions which would correspond to our Model 2 involving a molecular wind ramming into atomic material. Indeed, \citet{2012MNRAS.420.1977R} found that the protrusions contain molecular and neutral atomic gas.  This is confirmed in H$_2$ images which display considerable molecular material interior to the shell. In addition, PV diagrams do show the distorted ellipses predicted here.
 
  While our PV diagrams exhibit shell structures, our simulated images of 4:1 winds do not, being dominated by the fast leading bow. Thus elliptical shell structures such as for NGC 3132 do not fall into place here. One could propose that the shell has been excavated by a fast ellipsoidal wind but which has now been superseded by a spherical wind which applies a similar pressure around the shell.  Or, in the later stages, photoionisation has taken over as the excitation mechanism.  However, it seems most likely that the elliptical shells are associated with mildly-ellipsoidal winds in which the available shock power is not substantially different. This is also backed up by 2:1 simulations in Fig.~\ref{2.1atEm_All}, where a clear shell structure is observed, produced by less elliptic outflows. In fact, the projected turbulent shell of NGC\.3132 displays an axial ratio of only about 1.25 \citep{2000ApJ...537..853M}, consistent with both shock and photo-excitation. Interacting winds will need to be explored with a possible evolution resulting in hour-glass structure present in some post-AGB nebulae such as MyCn-18 and Frosty Leo. Additionally, an interacting wind structure implies that it is possible to set up an additional sub-volume containing a wind of varying composition included in existing models enabling to compare the effect of uniform ISM vs post-AGB winds structure on the resulting line flux, similar to research conducted by \citep{2004A&A...419..991V}, therefore further defining selection criteria for observational searches.  
 
How far can these simulations of PPN be applied to PN? The major difference is in terms of size with our simulations following the expansion up to about 0.2\,pc whereas PN extend beyond this typically. Nevertheless, H$_2$ has now been detected in a large number  of PN \citep{2015MNRAS.454.2586F} and in almost each case \citep{2017MNRAS.470.3707R}. The PN at this stage appear to have evolved to either ring shaped or waist-shaped. The ring-shaped PN appear to be larger and they emit more in the H$_2$ \citep{2000ApJS..127..125G}. Hence, it is probable that some of the PPN are contained by a dense ambient equatorial ring, reducing shock speed and post-shock excitation; fluorescent excitation then dominates.  On the other hand, where the PPN is not held back, shock excitation can still dominate within the equatorial plane.

 The computed synthetic maps presented here can also be compared with optical emission models of constant and time-varying outflows of CRL 618 by \citet{2003ApJ...586..319L} where the authors run a comparison with high spatial resolution images by the \textit{Hubble Space Telescope} (HST). The optical simulations of time-varying velocity model  has successfully reproduced a series of optical bow-like structures also revealed in CRL 618 HST images. However, the observed optical intensity features show similar magnitude of flux present within the lobes whilst the modelled bowlike structures become progressively weaker with axial distance away from the input radius due to radiative cooling and radial expansion. As a solution, the authors propose a driving jet of constant density with increased axial distance or a less steep radial density profile, hence the observed line emission in some cases can determine the initial set of hydrodynamical parameters chosen to simulate the outflow.
 
 There is a single, leading bow-like structure present in all our models due to constant mass outflow with the peak emission set back from the leading edge, a feature noticed by inspection of the time-dependent flux animations and also consistent with constant mass outflow model of \citet{2003ApJ...586..319L}. An achieved emission/temperature stratification is established due to cold dense outflow layer at the tip outrunning and shocking the ambient prior to the shell. High resolution simulations in this case enable mixing instabilities involving cold material at the tip to mix with the hot shell, significantly lowering the temperature.

\section{Conclusions}
\label{conclusions}

We have applied a hydrodynamic code that includes molecular dissociation and reformation besides molecular and atomic cooling to study protoplanetary nebulae. In this paper, we examined atomic emission line tracers located in the optical and near-infrared bands. This supplements the molecular hydrogen analysis of Paper 1.  We do not include the effects of radiation from the central star; the emission is induced after shock excitation.

These simulations provide the platform for more complex dynamical configurations of the global flow. There are two major extreme models considered in the literature. Firstly, we can envisage that the wind speed picks up abruptly, from a slow wind to fast wind \citep{1978ApJ...219L.125K}, generating a fast expanding inner secondary shell  which eventually hits the outer fragmented shell. Secondly, the wind could switch off completely so that there is no further ram pressure support to drive the shell \citep{2006ApJ...646L..61G}. In this case, the wind cavity is considered to be a hot bubble and the depressurised shell expands back into the cavity causing an intricate fragmented structure \citep{2015ApJ...808..115M}.  Both of these cases should be explored with a molecular code in the future.
 
 The ellipsoidal winds into uniform media investigated here produce giant bow-shaped bipolar nebula. In addition, highly ellipsoidal winds generate emission-line structures that resemble bullets. This is because the low speed of the lateral shock does not ionise the gas. Therefore, we do not generate the elongated ellipses often found as a class of planetary nebula. This is expected since such nebula, while physically created by winds, are then ionised by ultraviolet radiation from the central star. Unfortunately, PPN are quite deeply embedded and optical images remain rare. However, we do find 
 that the later PPN (early development of the PN) phase can be interpreted here especially with position-velocity diagrams.
  
The predicted  systematic variations may help define selection criteria for future searches. We find that the simulations with a molecular component leads to lower atomic excitation, weaker peaks and integrated atomic emission as the PPN grows. This is due to the shell fragmentation in the molecular cases where the shock surface area is increased and oblique shocks are prevalent. 
Position-velocity diagrams for the atomic lines indicate that the  atomic-wind model may be most easy to distinguish with more emission at higher radial velocities.  

Winds and jets have been considered as the drivers behind protostellar outflow, bipolar nebulae and Herbig-Haro flows. We have previously employed a similar code to investigate atomic and molecular jets into molecular environments (e.g. \citet{2004A&A...413..593R}
and \citet{2007MNRAS.378..691S}). We found that the jet-ambient impact region splits up into protrusions, creating small bullets with leading mini-bows in three dimensions.  The difference between a wind and jet is as expected, with the jet  
impact region not extending into the flanks. Position-velocity diagrams for jet-driven flows also show a different morphology with a distinct edge of
broad radial velocity followed by a trail at low radial speed \citep{2014MNRAS.443.2612S}. This is in stark contrast to the arc structures which correspond to wind-driven flows. The arcs are caused by the direct wind impact into the flanks.

We have thus modelled the initial set-up of post-AGB environments that enable us to generate a set of spectroscopic maps that describe PPN emission features. We can investigate further by the addition of physics modules to  approximate wind into wind interaction, i.e the GISW model, in 3--D space for both K-band and optical wavelengths in order to obtain higher resolution of molecular protrusions and atomic turbulent shells.
This should be followed by the inclusion of frozen-in magnetic fields as well as field diffusion which may limit shell compression and shock excitation, respectively.

\section{Acknowledgements}
\label{acks}
We thank the CAPS members at UKC, especially Dr. Mark Price for maintaining the SEPnet funded Forge supercomputer. Special thanks to Justin Donohoe for helping out with IDL code algorithms and Michael Knight for code development. Special thanks to Dr. Alejandro Cristian Raga Rasmussen at UNAM for help with theory of collisionally excited H$\mathrm \alpha$ lines. 

\bibliography{ppn}

\label{lastpage}

\end{document}